\documentclass[9pt]{article}
\usepackage[a4paper, total={7.2in, 10in}]{geometry}

\usepackage{caption,subcaption}
\usepackage{url}
\usepackage{float}
\usepackage{algorithm}
\usepackage{algpseudocode}
\usepackage{graphicx}

\usepackage[utf8]{inputenc}
\usepackage[T1]{fontenc}
\usepackage{lineno}

\usepackage[mathscr]{euscript}
\usepackage{bigstrut}
\usepackage{multirow}

\usepackage{amssymb,amsmath,amsthm}
\usepackage{graphicx}
\graphicspath{{figs/}}

\newcounter{algsubstate}
\renewcommand{\thealgsubstate}{\alph{algsubstate}}
\newenvironment{algsubstates}
  {\setcounter{algsubstate}{0}%
   \renewcommand{\State}{%
     \stepcounter{algsubstate}%
     \Statex {\footnotesize\thealgsubstate:}\space}}
  {}

\makeatletter
\def\ALG@step%
   {%
   \addtocounter{ALG@line}{1}%
   \addtocounter{ALG@rem}{1}%
   \ifthenelse{\equal{\arabic{ALG@rem}}{\ALG@numberfreq}}%
      {\setcounter{ALG@rem}{0}\alglinenumber{Step \arabic{ALG@line}}}%
      {}%
   }%
\makeatother

\newtheorem{problem}{Problem}

\title{Ensembles of Realistic Power Distribution Networks}

\author{Rounak Meyur, Anil Vullikanti, Samarth Swarup, Henning Mortveit, Virgilio Centeno, \\Arun Phadke, H. Vincent Poor and Madhav Marathe}

\begin{document}

\maketitle

\begin{abstract}
The power grid is going through significant changes with the introduction of renewable energy sources and incorporation of smart grid technologies. These rapid advancements necessitate new models and analyses to keep up with the various emergent phenomena they induce. A major prerequisite of such work is the acquisition of well-constructed and accurate network datasets for the power grid infrastructure. In this paper, we propose a robust, scalable framework to synthesize power distribution networks which resemble their physical counterparts for a given region. We use openly available information about interdependent road and building infrastructures to construct the networks. In contrast to prior work based on network statistics, we incorporate  engineering and economic constraints to create the networks. Additionally, we provide a framework to create ensembles of power distribution networks to generate multiple possible instances of the network for a given region. The comprehensive dataset consists of nodes with attributes such as geo-coordinates, type of node (residence, transformer, or substation), and edges with attributes such as geometry, type of line (feeder lines, primary or secondary) and line parameters. For validation, we provide detailed comparisons of the generated networks with actual distribution networks. The generated datasets represent realistic test systems (as compared to standard IEEE test cases) that can be used by network scientists to analyze complex events in power grids and to perform detailed sensitivity and statistical analyses over ensembles of networks.
\end{abstract}

A reliable power grid constitutes the backbone of a nation's economy providing vital support to various sectors of society and other civil infrastructures. 
Power distribution networks are created in a bottom-up fashion connecting small clusters of residential loads to distribution substations, thereby electrifying the entire society.
These bear a structural resemblance to other common networked infrastructures such as transportation, communication, water, and gas networks, and are often interdependent in their operations~\cite{pascal2020}. One may use these resemblances and interdependencies to infer one network from available data about these other networks.

Over the past decade, power engineers have aimed to enhance resilience of power systems through incorporation of distributed energy resources (DERs), by deploying advanced metering and monitoring infrastructures~\cite{Richler2020}, and by performing system vulnerability and criticality assessments thus reinforcing cybersecurity~\cite{onyeji2014}. Furthermore, spatiotemporally variable consumer load demands, such as electric vehicles (EVs), along with an evolving trend towards a distributed operation of the power grid, have posed new challenges to the system planners and operators~\cite{quiroga2019}. 
Network scientists have emphasized the importance of realistic power network data for accurate analysis as opposed to stylized statistical models~\cite{hines2013,anna_2010,zussman_2016}.
In order to address these challenges, there is a pressing need for openly available data containing realistic grid topologies along with available geographic information. For example, in the context of power grid expansion planning, the current grid information in conjunction with geographical knowledge of wind maps and solar trajectories can aid in optimized power grid expansion while introducing DERs in the grid~\cite{You2016}. Similarly, for system vulnerability analysis, a geographic correlation of grid information with cyclone/hurricane paths can help us identify critical sections in the network and raise preparedness levels for natural disasters~\cite{zussamn_2014}.
Further, a detailed knowledge about individual residential load usage and consumer behavior can help address policy-level questions. Examples of such problems include identifying the impact of EV adoption and DER penetration on the current power grid infrastructure as the society moves towards net zero-emission\cite{Popovich2021,Gaete-Morales2021}.

Simulation-based frameworks capable of performing spatiotemporally-resolved simulations can be utilized to analyze the impact of such evolving trends and analyze system vulnerability. Such assessments are useful to system planners aiming to make decisions about infrastructure development and to operators, while handling emergency system conditions. A common drawback of this simulation-based approach is that it requires detailed information regarding the power network and associated components such as locations and capacities of generation and load demands, and line parameters~\cite{hines2010,suchi_centralamerica,brazil_net,suchi_pjm}. Furthermore, since the majority of grid infrastructure advancements are being done at the low voltage distribution level, a high-resolution analysis of the power distribution systems is important. This necessitates a comprehensive knowledge of customer energy-use profiles, customer behavior, and most importantly, the distribution network topology which connects them. Most such data are, at best, partially available, but more typically are not available at all due to their proprietary nature~\cite{review2017}. The lack of such openly available detailed real-world data has been identified as a significant hurdle for conducting research in smart grid technology~\cite{nas2016}. In recent years there has been an increased interest in generating synthetic power network data to address this issue. The synthetic data are not the real-world data; rather, they are generated by mathematical models operating on openly available information, and are designed to ensure the generated data are similar to the real-world data, thus allowing them to be used as a proxy for the actual data. Some examples of synthetic power grid data include synthetic transmission networks~\cite{overbye_101,overbye_102,trpovski_2018,gm2016}, synthetic distribution networks~\cite{rnm_2011,rnm_2013,schweitzer,rounak2020} and synthetic residential customer energy usage data~\cite{swapna_2018,Tong2021,Klemenjak2020}. 

In this work, we focus on constructing a modular framework for generating synthetic \emph{power distribution networks}, that is, networks connecting individual residential customers to the distribution substations. We present a first principles approach where we generate an \emph{optimal synthetic distribution network} connecting all residences in a given geographic region to the HV substations through medium voltage (MV) and low voltage (LV) networks. We use the example of Montgomery county of southwest Virginia (US) to create the synthetic power distribution networks, consider all residences and HV substations within the state boundary, and connect them through the synthetic distribution network. 

In this context, there are two critical questions:~(i) are the created networks the only feasible networks connecting the residences and substations, and~(ii) how similar are the created synthetic networks and the actual power distribution networks? 

To tackle the first problem, we present a methodology for generating an \emph{ensemble} of feasible synthetic power distribution networks for a given region. In the literature related to modeling real-world networks, statistical physics has been used to learn significant structural patterns from an ensemble of networks~\cite{Cimini2019} and thereby to help in network reconstruction from incomplete data. In recent years, statistical aspects of the power networks have drawn the attention of the scientific community for similar reasons. A dataset spanning seventy years
for the electric power grid of Hungary has been studied~\cite{Hartmann2021} for small-world and scale-free properties. Due to a lack of real world power distribution data, ensembles of distribution networks, which have significant resemblance to actual networks, can suffice for a detailed statistical analysis. 

The generated synthetic networks require detailed validation before they can be used as a substitute for actual networks for various applications. To this end, the networks need to be compared against actual distribution networks in terms of their structural properties as well as their power engineering attributes~\cite{validate2020}. Hence the second problem deals with the comparison of synthetic and actual networks using suitable metrics meaningful to power distribution networks which follow a particular structure. In this work, we have compared the created synthetic networks for the town of Blacksburg (in Montgomery county of southwest Virginia, US) against actual networks obtained from a power company operating in the same region. In addition to comparing standard graph attributes such as degree and hop distributions, we compute the difference in geometries between the actual and synthetic networks and provide a measure of deviation.

Our contributions include:~(i)~a holistic modular framework to create synthetic power distribution networks which satisfy structural and power-engineering constraints along with accurate representation of residential load demand profiles.~(ii)~a method to create an ensemble of networks by generating multiple feasible networks for a given region.~(iii)~ an open dataset consisting of ensembles of distribution networks for Montgomery county of southwest Virginia (US). To the best of our knowledge, this dataset is the first of its kind in terms of both size and details. The geographically embedded networks, along with the detailed residential customer usage data, become suitable tools for system-wide planning studies and for addressing policy-level questions.

\section*{Related Work}\label{sec:related}
In recent years, a substantial amount of work has gone into creating synthetic high voltage (HV) transmission networks~\cite{overbye_101,overbye_102}, or combinations of transmission and distribution networks~\cite{overbye_2019,overbye_2020}. The primary focus of these papers is to model the transmission grid with a high level of resemblance to the actual grid. 

For distribution networks, Schweitzer et.al~\cite{schweitzer} were one of the first to analyze real power distribution networks, learn statistical distributions of network attributes from an extensive dataset of actual distribution networks in Netherlands and create synthetic networks which preserve these attributes. The reference network model (RNM) framework~\cite{rnm_2011,rnm_2013,nrel_net} and some of its variants~\cite{anna_naps,highres_net} have proposed heuristics to generate synthetic distribution networks that satisfy structural and power engineering constraints. These heuristics include clustering load groups to identify feeders~\cite{rnm_2013,gm2016}, identifying substation locations from cluster centroids~\cite{rnm_2011,rnm_2013} and construct networks using minimum spanning tree algorithm~\cite{rnm_2011,nrel_net,anna_naps}. Most of these papers do not use individual residence locations and have populated the created synthetic networks with random residences or aggregated loads to zip-code centers. Some of these works used a top-down approach where feeder networks are generated and followed by populating with random loads~\cite{anna_naps,gm2016}. In~\cite{feeder_gan}, the authors use a generative adversarial networks (GANs) to create synthetic power networks. However the approach is inherently data intensive and requires a large number of samples for training, making it practically challenging to use. Here, we propose a rigorous mathematical framework that provides optimality guarantees on the quality of networks created. The ensemble of synthetic networks created can
potentially be used to train GANs.

The RNM~\cite{rnm_2011,rnm_2013} is an important  heuristic-based planning tool for efficient investment options in distribution grid planning. It uses OpenStreetMap and relevant geographic data to create distribution networks in a given region. The comparison of networks generated by the RNM with actual power distribution networks show that the real and the synthetic networks are quite similar~\cite{validate2020}. However, the methods used to compare such networks were somewhat adhoc. In contrast,in~\cite{schweitzer}, the authors performed a statistical fit of the distributions of network attributes and performed a numerical comparison, yielding a more rigorous approach to measuring network similarity.
The RNM framework uses four independent layers (namely logical, topological, electrical, and continuity of supply) to assign constraints while creating the networks. Although, this approach is natural, the set of constraints are not mathematically well-defined to the extent that they can be reproduced. For instance, the framework uses a set of heuristics to satisfy the constraints, which does not always guarantee a feasible solution. Furthermore, several steps in the heuristic-based method involve user-defined parameters, which lead to multiple possible networks for different choices. Furthermore, these papers do not consider the creation of ensembles of networks. 
We present a summary of previous results in Table S1 in the SI. 

Recent works~\cite{eric2017,swapna_2018} have provided detailed synthetic residential demand models along with household geographic footprints. We create synthetic distribution networks connecting substations to these individual residence locations. Our methods differ from the earlier works in the following ways:~(i)~Instead of populating with imaginary demand profiles, we have used behavior-based consumer load modeling which results in an accurate representation of household load demand profiles; ~(ii)~Unlike other heuristic based approaches, the structural and power engineering constraints are mathematically well-defined which makes the framework reproducible for creating other networks with similar constraints;~(iii)~We use actual substation locations obtained from~\cite{eia_substations} and the optimal feeder locations are identified as an output of our optimization framework; and~(iv)~we propose a method to create an ensemble of realistic power networks.

A wide variety of graph comparison methods have been studied in the literature. Tantardini et al.~\cite{Tantardini2019} analyze multiple graph comparison methods, which include comparing whole graphs as well as small portions of the graph known as motifs. Several methods for assessing structural similarities of graphs have also been studied~\cite{simgnn-2018,graphsim-2020}. However, none of the comparison methods consider the node and edge geometries of the graphs. \emph{Edit distance}, or evaluating the minimum number of edit operations to reach from one network to the another, has been widely used to compare networks having structural properties~\cite{xu2015,paben2021,riba2020}. Among these works, Riba et al.~\cite{riba2020} have used Hausdorff distance between nodes in the network to compare network geometries. Morer et al.~\cite{efficiency2020} include edge geometry based comparison and propose an ``efficiency'' metric to measure the distance of a network (where edges have non-straight line geometries between nodes) from its most optimal version (where each edge has a straight line geometry). 




\section*{Methods}
\begin{figure*}[tbhp]
\centering
	\includegraphics[width=0.97\textwidth]{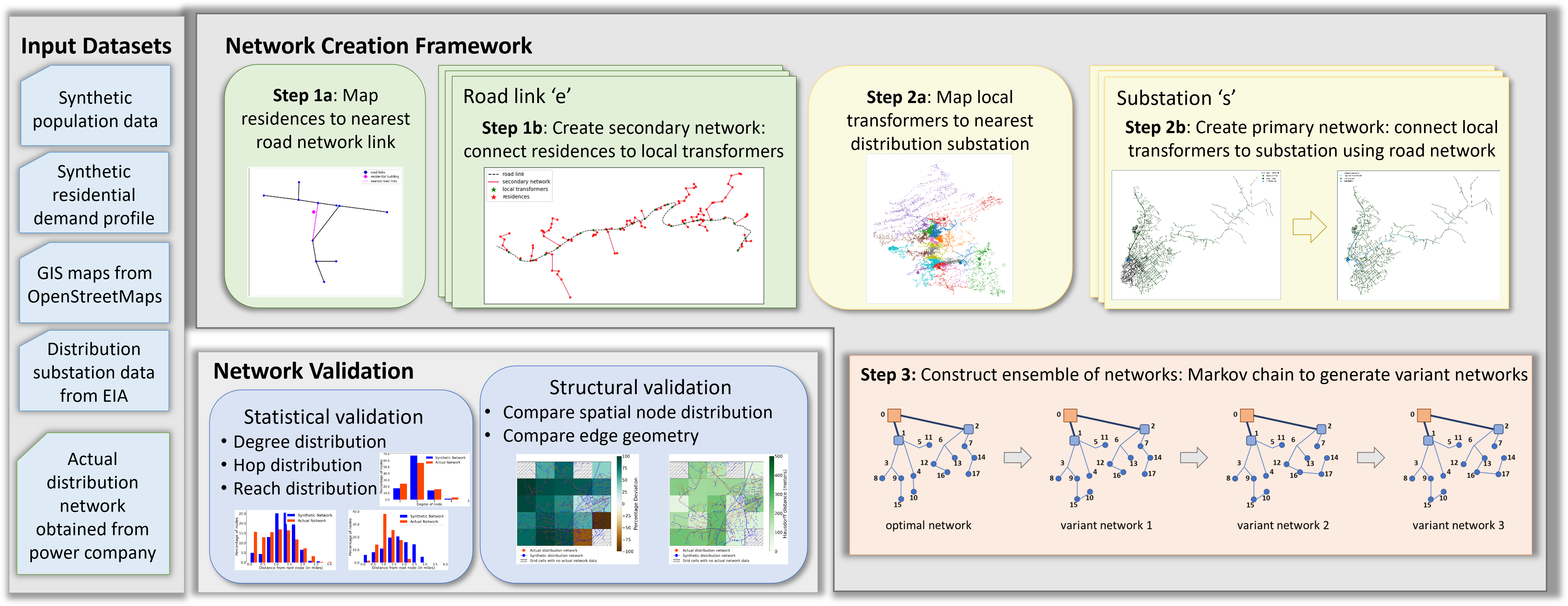}
	\caption{Proposed framework for constructing ensembles of realistic power distribution networks. The framework uses the input datasets and constructs an ensemble of networks using the steps detailed in Algorithm~\ref{alg:total}. The created networks are validated against actual power distribution networks.}
	\label{fig:process-flow}
\end{figure*}
\noindent\textbf{Datasets used.}~We use open source, publicly available information regarding several infrastructures to generate the synthetic distribution networks (more details are presented in Table S3 in the SI): (i) road network data from OpenStreetMap (OSM)~\cite{osm}, (ii) geographic locations of HV (greater than 33kV) substations from data sets published by EIA~\cite{eia_substations}, and (iii) residential electric power demand information developed in earlier work from our research group~\cite{swapna_2018}. We also obtained actual power distribution networks for the town of Blacksburg to validate the synthetic networks.

\noindent\textbf{Approach.}~Algorithm~\ref{alg:total} summarises the steps we use in the paper. The synthetic distribution networks are constructed in two steps using a bottom-up approach. First, we identify local pole-top transformers along the road network and connect the residential buildings to them to create the LV (208--480V) secondary network  (Step 1). Thereafter, we use the road network as a proxy to construct the MV(6--11kV) primary network connecting the local transformers placed along roads to the substations (Step 2). To construct the ensemble of synthetic networks, we propose a Markov chain starting from the already created network to a variant network which is also a feasible distribution network (Step 3). Finally, we add attributes to nodes and edges in each network (Step 4) to create an ensemble of synthetic power distribution networks. Several aspects of the first two and the last steps are similar to the approach taken in earlier papers. The difference lies in the specifics of problem formulation and the resulting algorithmic approach. The third step that creates ensemble of networks has largely not been
explored in the context of distribution networks.

\begin{algorithm}
\caption{Create ensemble of synthetic networks}\label{alg:total}
\textbf{Input} Set of residences $\mathscr{H}$, set of substations $\mathscr{S}$, road network $\mathscr{G}_R\left(\mathscr{V}_R,\mathscr{E}_R\right)$, required ensemble size $N$
\begin{algorithmic}[1]
\State Construct LV secondary network.
\begin{algsubstates}
\State Map residences to nearest road network link.
\State Connect residences to local transformers along road link.
\end{algsubstates}
\State Construct MV primary network.
\begin{algsubstates}
\State Map local transformers to nearest substation.
\State Use road network as proxy to connect transformers to substation.
\end{algsubstates}
\State Construct an ensemble of networks.
\begin{algsubstates}
\State Construct Markov Chain $\mathcal{M}$ to create a variant from an existing network.
\State Run $\mathcal{M}$ to create $N$ variant networks.
\end{algsubstates}
\State Add additional attributes to nodes and edges of each network in the ensemble as follows.
\begin{algsubstates}
\State Assign one of the three phases (A,B,C) to each residence.
\State Assign a distribution line type to each edge.
\end{algsubstates}
\end{algorithmic}
\textbf{Output} Ensemble of $N$ attributed networks.
\end{algorithm}

\subsection*{Step 1. Constructing Secondary Networks}\label{ssec:secnet}
We extract residence and road network data for the geographic region. Let~$\mathscr{H}$ be the set of residences and~$\mathscr{G}_R\left(\mathscr{V}_R,\mathscr{E}_R\right)$ be the road network graph. We evaluate a many-to-one mapping~$\mathscr{F}_M \colon \mathscr{H} \rightarrow \mathscr{E}_R$ such that each residence~$h\in\mathscr{H}$ is mapped to the nearest road network link~$e\in\mathscr{E}_R$. The inverse mapping~$\mathscr{F}_M^{-1}$ defined by~$\mathscr{F}_M^{-1}\left(e\right)=\left\{h\in\mathscr{H};\mathscr{F}_M\left(h\right)=e\right\}$ provides the set of residences assigned to each road link~$e\in\mathscr{E}_R$. 

The secondary network creation problem (denoted by $\mathcal{P}_{\textrm{sec}}$) is defined for each road link~$e\in\mathscr{E}_R$. We provide more details for $\mathcal{P}_{\textrm{sec}}$ in the SI. The objective is to identify local transformers~$\mathscr{V}_T\left(e\right)$ along the link and connect them to the assigned residences~$\mathscr{F}_M^{-1}\left(e\right)$, thereby constructing the secondary distribution network~$\mathscr{G}_S\left(e\right)$ with node set~$\mathscr{V}_S\left(e\right)=\mathscr{V}_T\left(e\right)\cup\mathscr{F}_M^{-1}\left(e\right)$ and edges~$\mathscr{E}_S(e)$. We impose structural constraints to connect residences in chains ensuring tree network structure so that the created networks mimic their physical counterpart.

\begin{problem}[$\mathcal{P}_{\textrm{sec}}$ \textbf{construction}]
Given a road link~$e\in\mathscr{E}_R$ with a set of residences~$\mathscr{F}_M^{-1}\left(e\right)$ assigned to it, construct an optimal forest of trees, $\mathscr{G}_S\left(e\right)$, rooted at points (local transformers) along the link and connecting the residences.
\end{problem}
The problem $\mathcal{P}_{\textrm{sec}}$ is modeled as a mixed integer linear program (MILP) which usually requires exponential computation time. We use different heuristics to reduce the number of binary variables which in turn reduce the overall time complexity. A formal problem statement for the problem has been provided in the SI along with our approach to solve it. The secondary network creation process can be executed simultaneously for different road links~$e\in\mathscr{E}_R$ in the geographic region. In our framework we execute the task sequentially for all edges in a county, with the entire sequence performed simultaneously for different counties. The secondary network generated for the region is
\begin{equation*}
    \mathscr{G}_S=\bigcup_{e\in\mathscr{E}_R}\mathscr{G}_S\left(e\right)=\bigcup_{e\in\mathscr{E}_R}\mathcal{P}_{\textrm{sec}}\left(e,\mathscr{F}_M^{-1}\left(e\right)\right) \;.
\label{eq:create-secnet}
\end{equation*}

\subsection*{Step 2. Constructing Primary Networks}\label{ssec:primnet}
The secondary network results in local transformer nodes~$\mathscr{V}_T=\bigcup_{e\in\mathscr{E}_R}\mathscr{V}_T(e)$ along the road network links. The goal of the primary network construction is to connect these transformers to the set of substation nodes~$\mathscr{S}$ using the road network as proxy. First, we define a many-to-one mapping~$\mathscr{F}_V \colon \mathscr{V}_T \rightarrow \mathscr{S}$ based on a Voronoi partitioning. The details of this mapping are provided in the Appendix. We are interested in the inverse mapping~$\mathscr{F}_V^{-1}\left(s\right)=\left\{t\in\mathscr{V}_T;\mathscr{F}_V\left(t\right)=s\right\}$ which assigns a group of transformers to each substation node.

The primary network creation problem (denoted by $\mathcal{P}_{\textrm{prim}}$) is defined for each substation node~$s\in\mathscr{S}$, and the goal is to crate a minimum length primary network~$\mathscr{G}_P\left(s\right)$ connecting substation node~$s$ to the mapped transformers~$\mathscr{F}_V^{-1}\left(s\right)$ using road network $\mathscr{G}_R$ as proxy, such that the following set of structural and operational constraints are valid:
(i) the network should be a tree rooted at the substation, (ii) all transformer nodes are to be connected, and (iii) all nodes should have acceptable voltages (based on ANSI standards between 0.95 pu and 1.05 pu) when the residential customers are consuming average hourly loads. 

\begin{problem}[$\mathcal{P}_{\textrm{prim}}$ construction]
Given a substation~$s\in\mathscr{S}$ with an assigned set of local transformer nodes~$\mathscr{F}_V^{-1}\left(s\right)$, construct a tree network $\mathscr{G}_P\left(s\right)$ using the road network $\mathscr{G}_R$ as a proxy which connects all local transformers while ensuring acceptable node voltages by power engineering standards.
\end{problem}
We formulate an MILP to solve the problem~$\mathcal{P}_{\textrm{prim}}$ which requires exponential computation time. A formal problem statement for the same has been provided in the SI. We do not use any heuristic to reduce the computational complexity which is determined by the size of underlying road network (used as the proxy). On many occasions, we terminate the optimization program reaching an optimal solution, in order to reduce the running time. This has resulted in the constructed network being a near-optimal solution, but with an acceptable optimality gap of $0-5\%$.
In our framework we execute the task of primary network creation simultaneously for all the substations in the geographic region. The created primary network~$\mathscr{G}_P$ for the entire region is,
\begin{equation*}
    \mathscr{G}_P=\bigcup_{s\in\mathscr{S}}\mathscr{G}_P\left(s\right)=\bigcup_{s\in\mathscr{S}}\mathcal{P}_{\textrm{prim}}\left(s,\mathscr{F}_V^{-1}\left(s\right)\right) \;.
\label{eq:create-primnet}
\end{equation*}

\subsection*{Step 3. Constructing Ensembles of Networks}
In this section, we address the problem of creating multiple realizations of the distribution network which connect the residences to substations. Albeit the modification of user defined parameters in~$\mathcal{P}_{\textrm{sec}}$ and~$\mathcal{P}_{\textrm{prim}}$ can produce different realizations of synthetic networks, the procedure is computationally expensive, since optimization problems of similar order need to be solved. We propose a methodology which uses the already created (near)-optimal primary network for a region and creates an ensemble of synthetic networks by reconnecting the transformer nodes in a different manner from the (near)-optimal primary network, while maintaining the structural and power engineering operational constraints. Thereafter, we connect the residences in the same way as in the optimal secondary network. Thus, we construct an ensemble of networks where each network is a combination of a variant primary network and the optimal secondary network (solution of~$\mathcal{P}_{\textrm{sec}}$). The variant primary networks are ``feasible'' (but not necessarily ``optimal'') solutions of~$\mathcal{P}_{\textrm{prim}}$.


\begin{problem}
Given a near-optimal primary network $\mathscr{G}_{P}^{0}:=\left(\mathscr{V}_0,\mathscr{E}_0\right)$ constructed using underlying road network graph~$\mathscr{G}_R:=\left(\mathscr{V}_R,\mathscr{E}_R\right)$, construct $N$ variants of the primary network~$\mathscr{G}_{P}^{1},\cdots,\mathscr{G}_{P}^{N}$ by identifying respective edge sets~$\mathscr{E}_1,\cdots,\mathscr{E}_N\subseteq\mathscr{E}_R$ such that the networks are feasible solutions of~$\mathcal{P}_{\operatorname{prim}}$.
\end{problem}


We consider the ensemble of networks generation problem for each substation~$s$ and the mapped transformer nodes~$\mathscr{F}_V^{-1}\left(s\right)$. Let~$\mathcal{F}_{\textrm{feas}}$ denote the set of feasible solutions of~$\mathcal{P}_{\textrm{prim}}\left(s,\mathscr{F}_V^{-1}\left(s\right)\right)$. From here on we omit the dependency on~$s$ in our notation. We design a Markov chain $\mathcal{M}$ to create variant networks with each state denoting a feasible realization of the network~$\mathscr{G}_{P}^{t}\in\mathcal{F}_{\textrm{feas}}$. The steps involved in transitioning from the primary network~$\mathscr{G}_{P}^{t}:=\left(\mathscr{V}_t,\mathscr{E}_t\right)$ to~$\mathscr{G}_{P}^{t+1}:=\left(\mathscr{V}_{t+1},\mathscr{E}_{t+1}\right)$ are described below. 

Let $\mathcal{F}_{\textrm{rstr}}(e) = \{\mathscr{G}:=\left(\mathscr{V},\mathscr{E}\right)\in \mathcal{F}_{\textrm{feas}}: e\not\in \mathscr{E}\}$. If  $\mathcal{F}_{\textrm{rstr}}(e)\neq\emptyset$, we select a random edge~$e\in\mathscr{E}_t$ to be deleted with probability~$1/|\mathscr{E}_t|$, and then pick $\mathscr{G}_{P}^{t+1}:=\left(\mathscr{V}_{t+1},\mathscr{E}_{t+1}\right)\in\mathcal{F}_{\textrm{rstr}}(e)$ uniformly at random; else $\mathscr{G}_{P}^{t+1} = \mathscr{G}_{P}^{t}$. The ensemble of synthetic power distribution networks for the region is
\begin{equation*}
    \mathcal{E}:=\mathscr{G}_{S} \bigcup \left\{\mathscr{G}_{P}^{t}: t=1,\ldots, N\right\}
\end{equation*}

\subsection*{Step 4. Post-processing of Networks.}
The final step of our framework involves addition of attributes and labels to nodes and edges to each network in the ensemble. We compute the required distribution line ratings for each edge in the network. We assign a suitable type of distribution line from the catalog of distribution lines~\cite{kersting_book} and add edge attributes accordingly. We include positive sequence impedance (resistance and reactance) for each edge in the network. Additionally we use a framework to assign one of the three phases (A, B or C) to each residence in the network. The phase assignment ensures that the three phases are balanced at each substation feeder. We add the assigned phase as a node attribute to the network. The details of this step has been provided in the SI. The node and edge attributes are listed in Table S8 of SI.

Though we have assigned one of the three phases to each residence in the network, we do not consider three phase circuits with different transformer configurations (wye and delta). Therefore, we limit the created synthetic networks with only positive sequence impedance. To this end, such networks can be useful to perform studies involving balanced loads across three phases.

\section*{Results: Synthetic Network Attributes}
\subsection*{Degree, hop and reach distribution}
In this section, we compare the statistical attributes of the synthetic distribution networks created for rural and urban areas. The degree of a node in a network denotes the number of edges connected to it. The degree distribution gives an idea about the connectivity within the network. The `hop' of a node from substation (root) node is defined as the number of edges lying between them. Hence, the ``hop distribution'' provides an idea about the radial layout of nodes around the root substation node.
Finally, we define ``reach'' of a node as the length of network (in miles) connecting it to the substation. The associated ``reach distribution'' of a network becomes a relevant statistic in the context of networks with associated geographic attributes since it provides a distance metric to the hop distribution.
\begin{figure}[tbhp]
    \centering
    \includegraphics[width=0.32\textwidth]{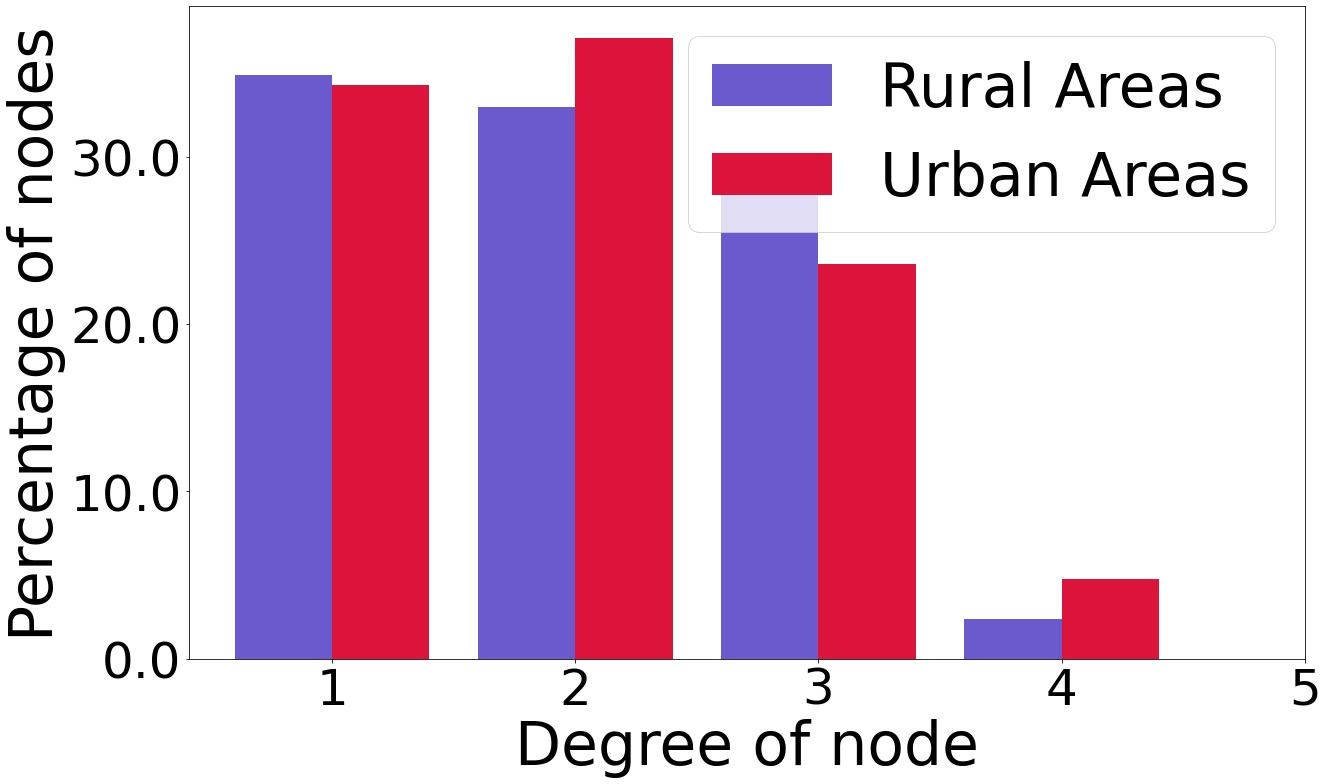}
    \includegraphics[width=0.32\textwidth]{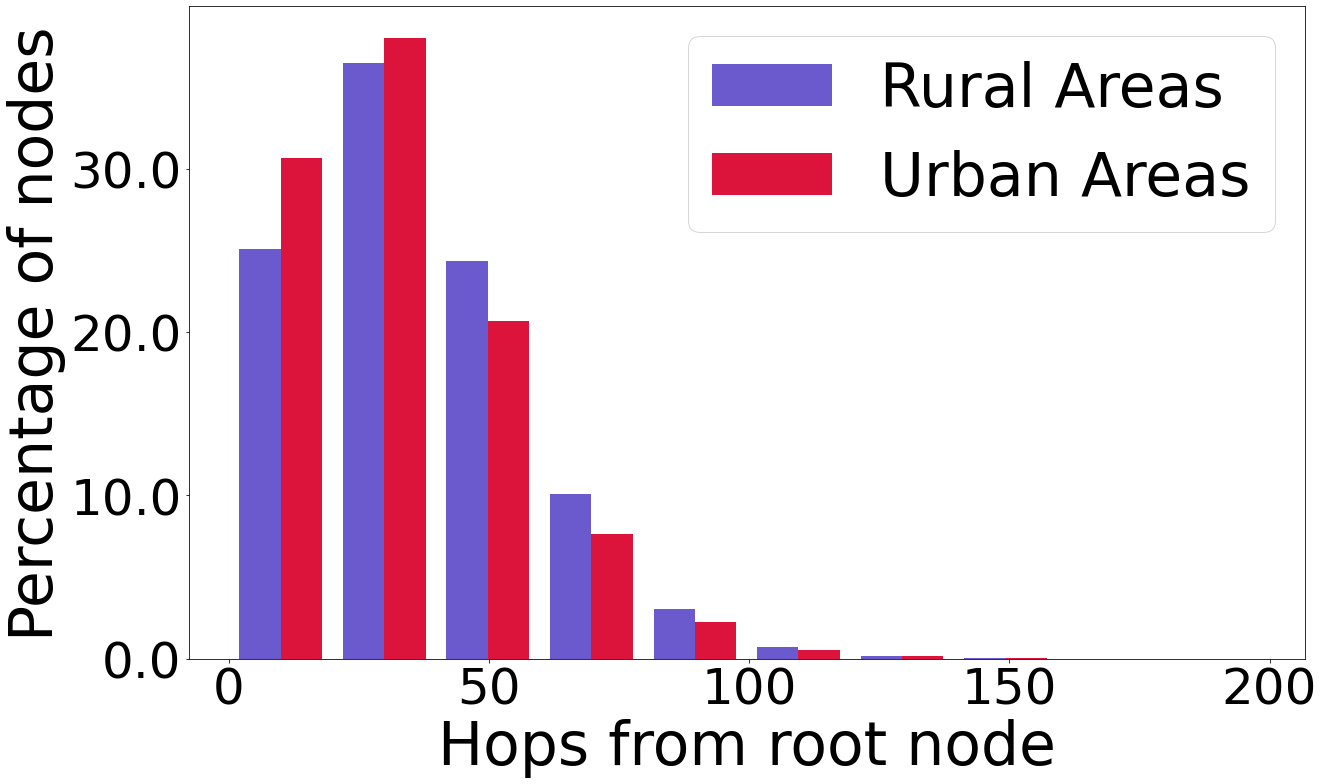}
    \includegraphics[width=0.32\textwidth]{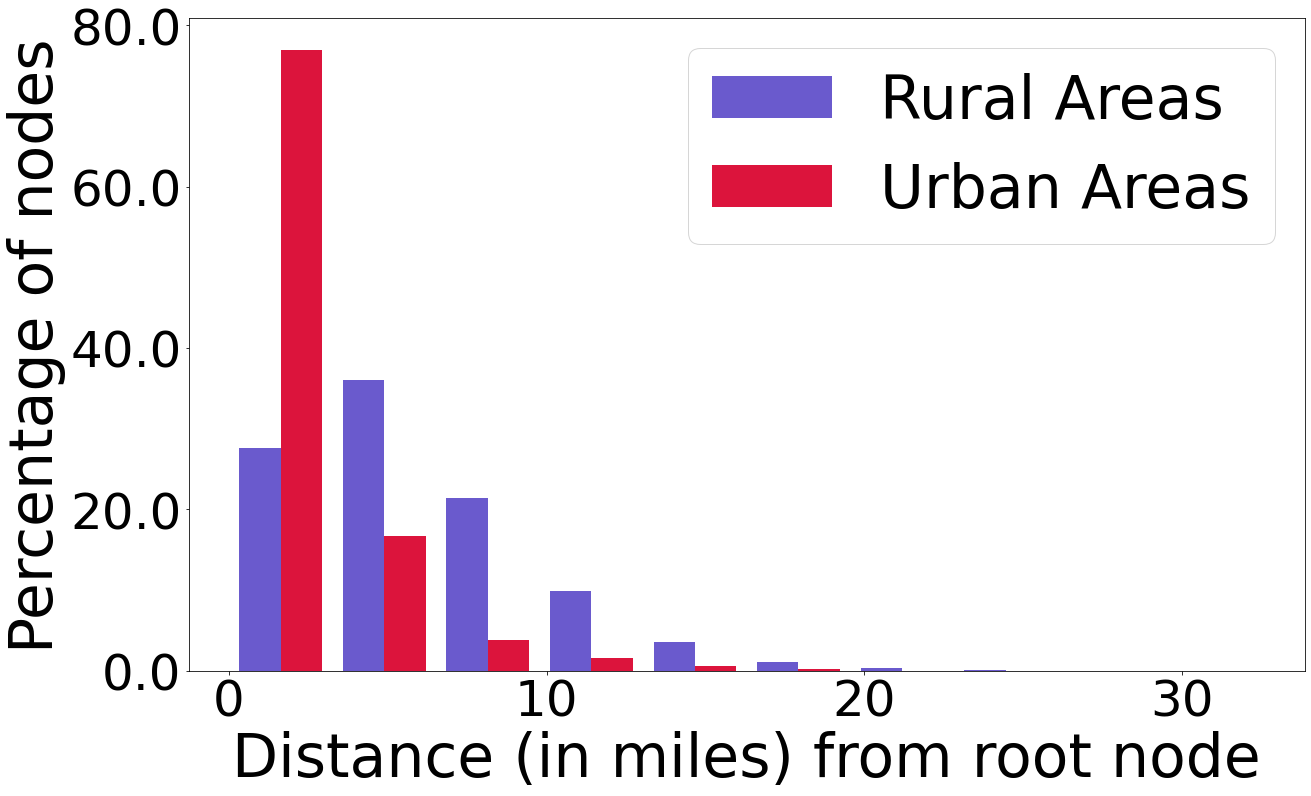}
    \caption{Plots showing degree distribution (left), hop distribution (middle), and reach distribution (right) in rural and urban areas. Colors depict network attributes of urban versus rural areas. The degree and hop distribution are similar for both rural and urban regions. The reach distribution of urban networks peak at small value since the distribution network nodes are more closely placed to the substation than rural areas.}
    \label{fig:rural-urban}
\end{figure}

Fig.~\ref{fig:rural-urban} shows comparison of degree, hop and reach distributions in urban and rural distribution networks. We observe that the degree and hop distributions are fairly similar. However, the reach distribution differs for rural and urban areas. In case of urban areas we notice that a majority of nodes are located very close to the substation whereas rural areas are often characterized by long length network edges. This observation is also consistent with the distribution of residences in urban and rural regions where rural regions have more widely spread out residences than urban areas.

\subsection*{Network motifs}
\begin{figure}[tbhp]
    \centering
    \includegraphics[width=0.48\textwidth]{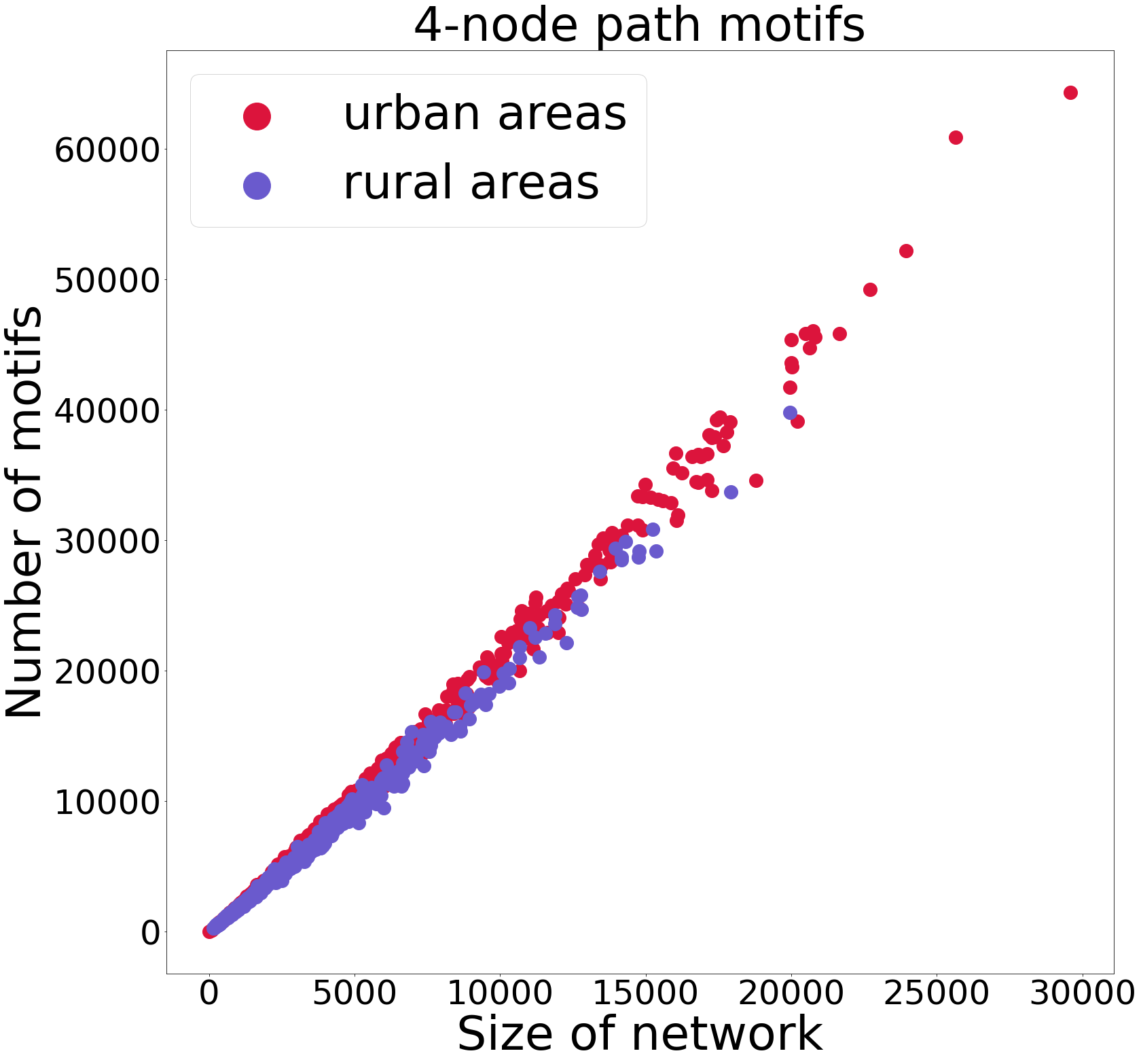}
    \includegraphics[width=0.48\textwidth]{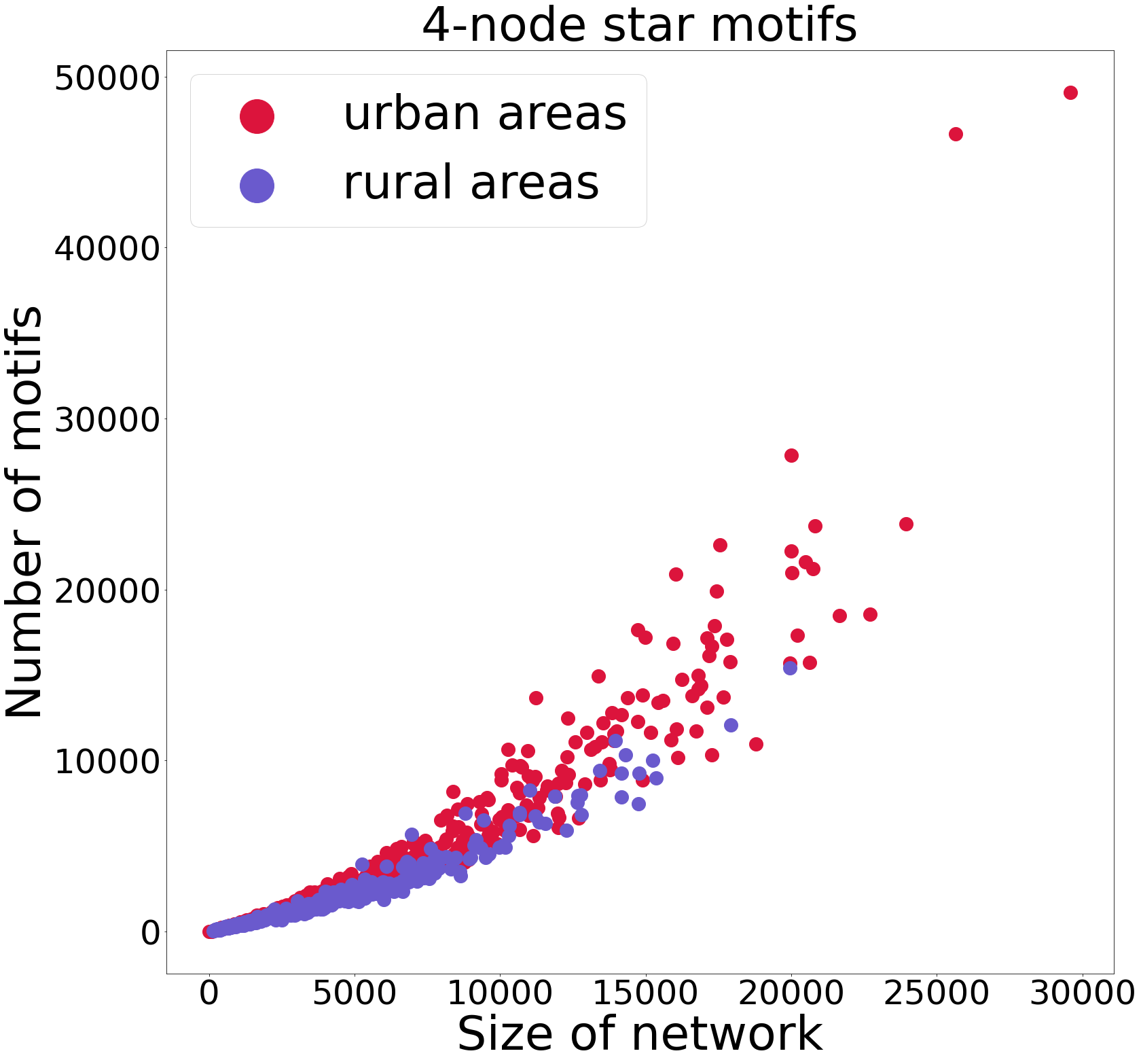}
    \caption{Plots showing number of 4-node paths (left) and 4-node star motifs (right) as a function of network size (measured as number of nodes in the network). Colors depict motif numbers in urban versus rural areas. Urban distribution networks have a larger number of star motifs than rural networks. In contrast, the path motif count does not differ significantly across rural and urban areas.  Urban networks are often larger than rural networks as measured by number of nodes due to larger population size.}
    \label{fig:motif}
\end{figure}
Network motifs are interesting subgraphs which build up the entire network. Network motifs have been used as a metric to understand network resilience in earlier work~\cite{vince_2019}. We focus our attention to small size subgraphs with at most 4 nodes. Since the created distribution networks are tree graphs, we are interested in two types of network motifs: (i) 4-node path and (ii) 4-node star. Fig.~\ref{fig:motif} shows the number of 4-node motifs in the synthetic distribution networks. The two colors show the results for urban and rural networks separately. The star motifs are higher for urban networks as compared to rural networks of similar size. This can be explained from the observation in degree distribution where we notice that urban networks have higher fraction of nodes with degree 4. A single node with degree $4$ results in~$\binom{4}{3}=4$ counts of $4-$node star motifs.

\subsection*{Features in ensemble of networks}
\begin{figure}[tbhp]
    \centering
    \includegraphics[width=0.32\textwidth]{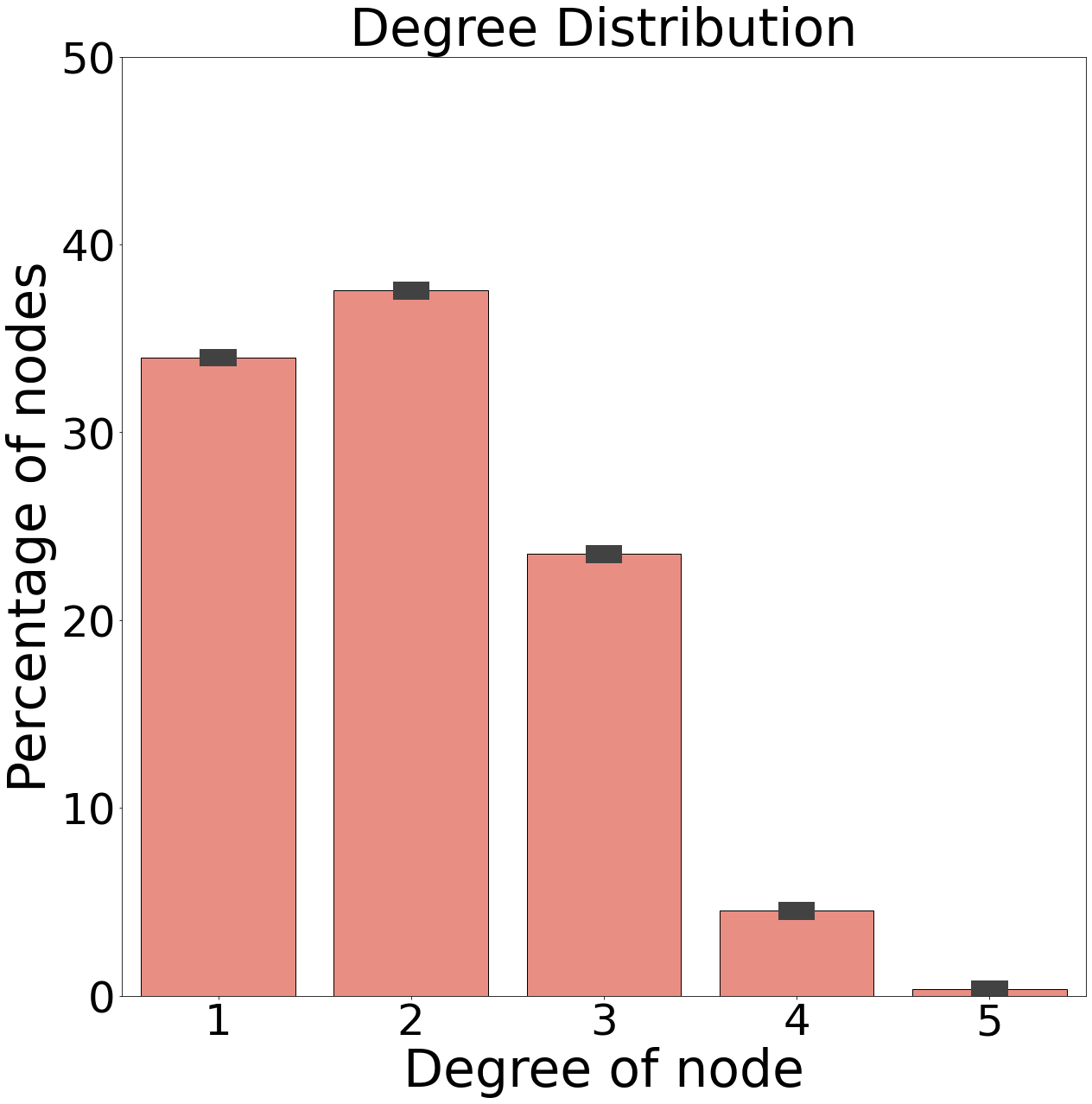}
    \includegraphics[width=0.32\textwidth]{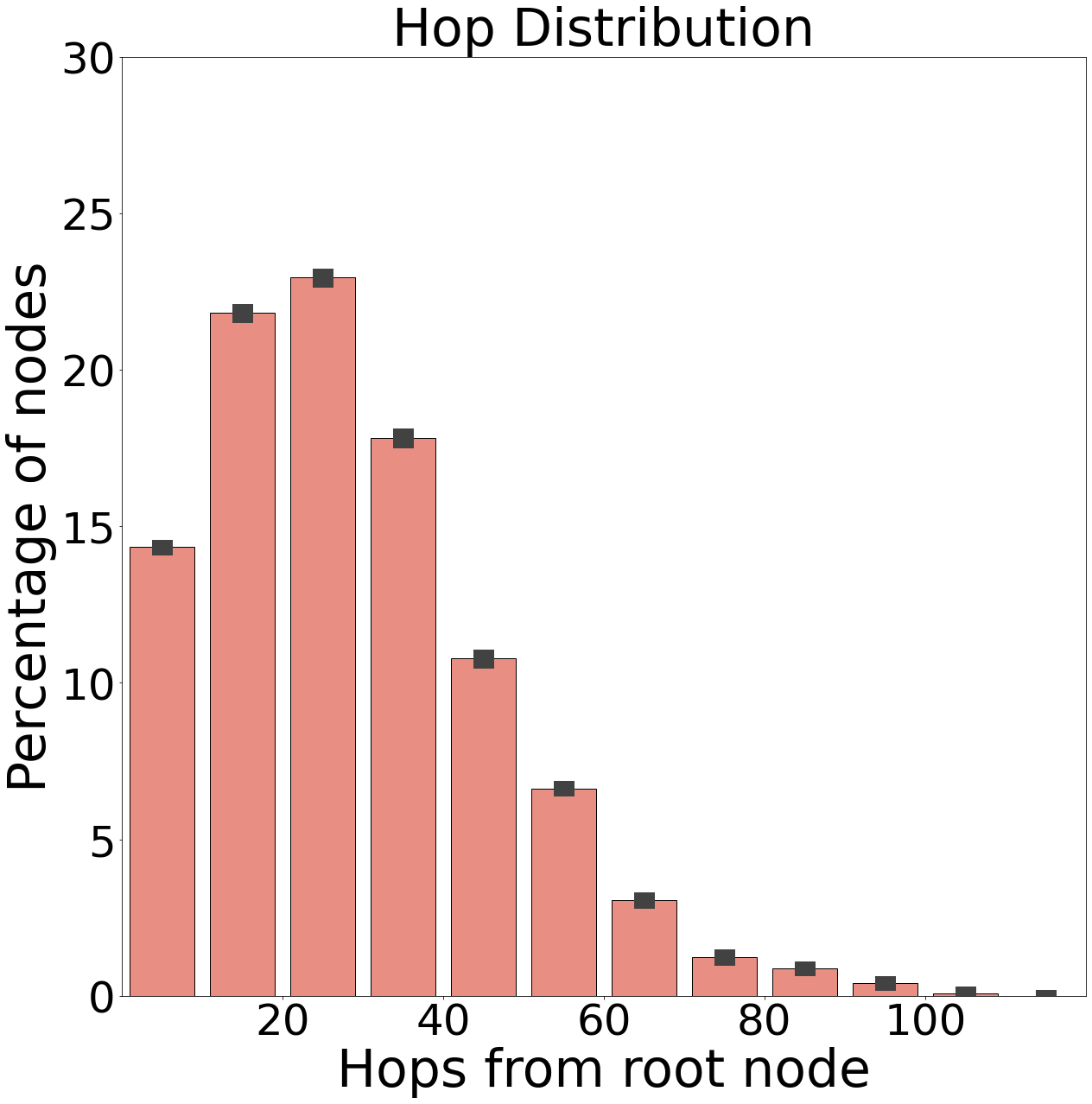}
    \includegraphics[width=0.32\textwidth]{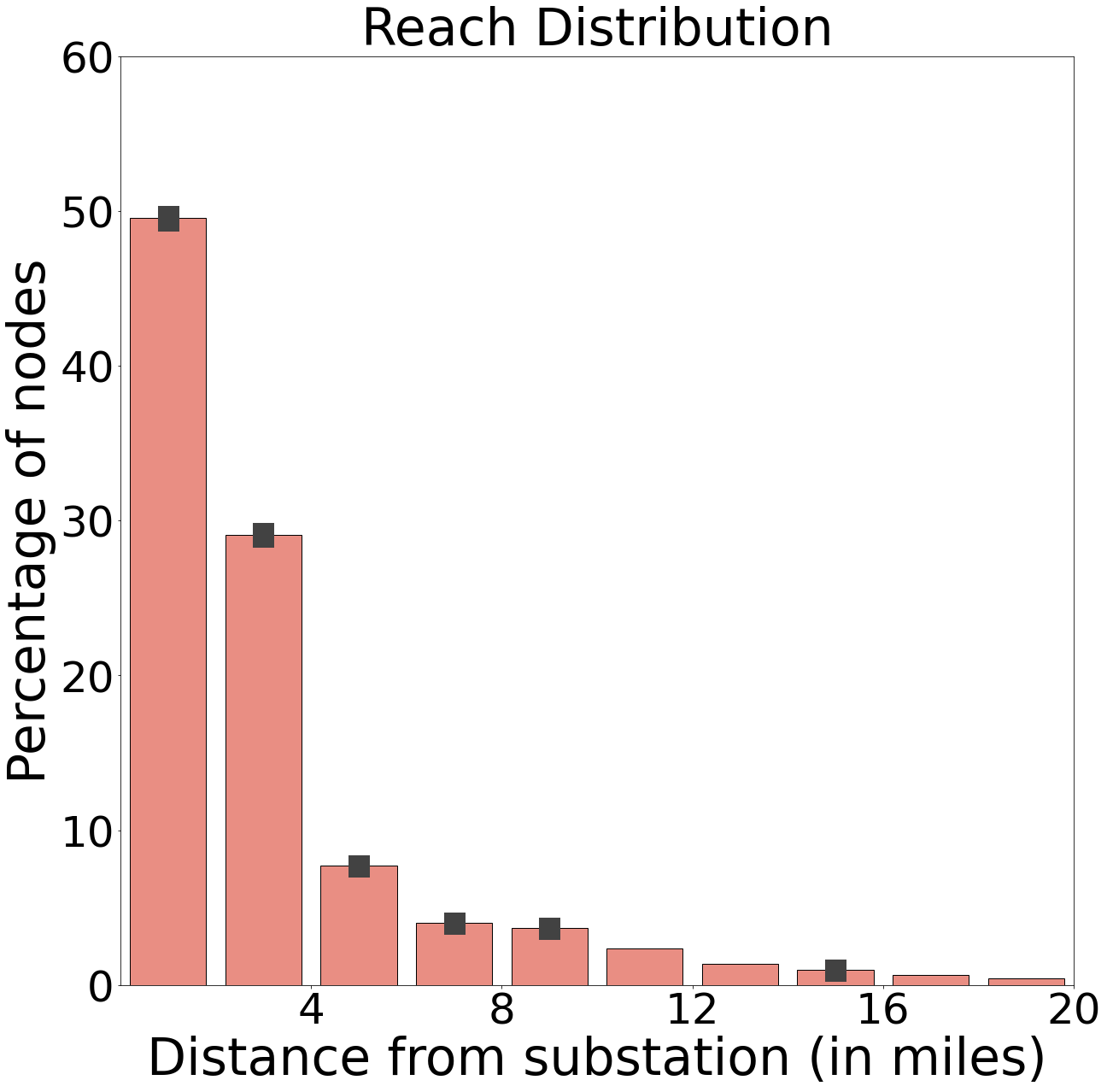}
    \caption{Plots showing variation in degree distribution (left), hop distribution (middle), and reach distribution (right) for the ensemble of distribution networks created for Montgomery county of southwest Virginia. The error-bars in the bar plots show the variation over the networks in the ensemble.}
    \label{fig:ens-sta}
\end{figure}

\begin{figure}[tbhp]
    \centering
    \includegraphics[width=0.47\textwidth]{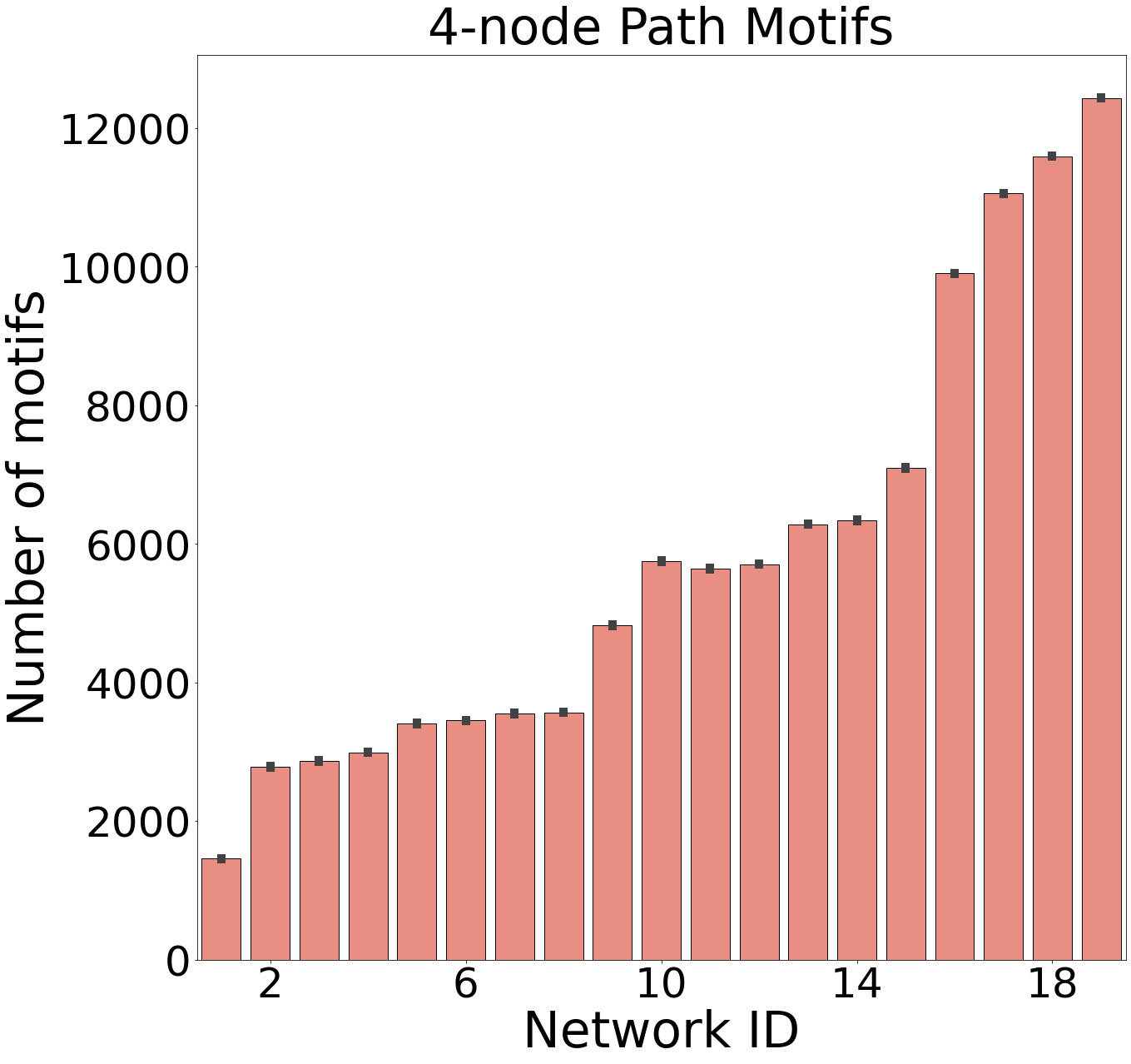}
    \includegraphics[width=0.47\textwidth]{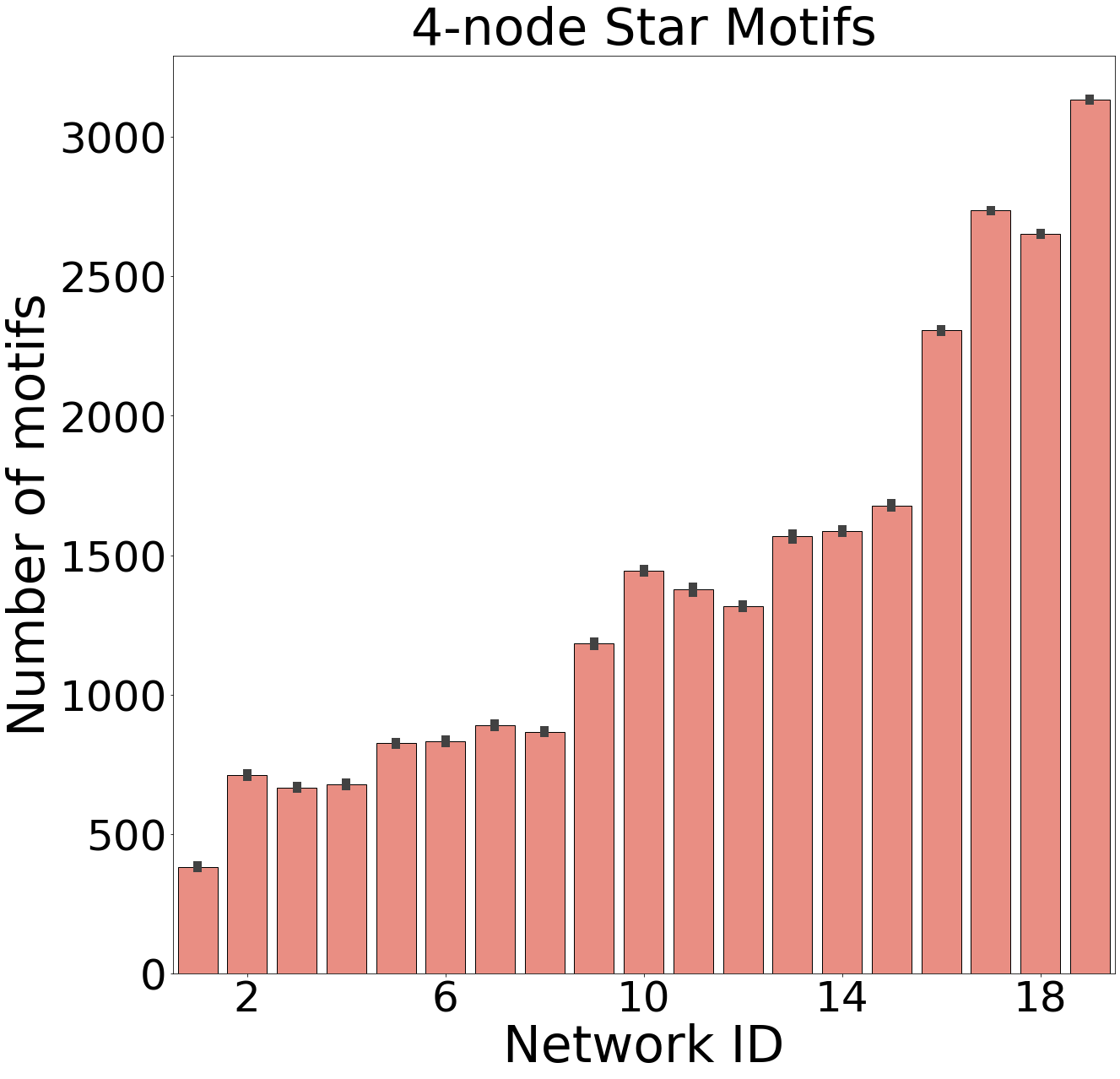}
    \caption{Plots showing variation in number of 4-node path motifs (left), and number of 4-node star motifs (right) for the ensembles of distribution networks created for Montgomery county of southwest Virginia. Results are shown for $19$ ensembles of varying size in the county fed by different substations. Each ensemble consists of $20$ networks. The error-bars in the bar plots show the variation over the networks in each ensemble.}
    \label{fig:ens-motif}
\end{figure}

We create ensembles of distribution networks for Montgomery county in southwest Virginia. The entire network within Montgomery county is composed of $19$ sub-networks (each fed by a different substation). We create an ensemble of $20$ networks for each sub-network and study the variation in network attributes over the ensembles. We plot the variation in degree, hop and reach distributions in Fig.~\ref{fig:ens-sta}. The error bar shows the extent of variation in the ensemble. Fig.~\ref{fig:ens-motif} shows variation in 4-node path and star motif counts for the networks in each ensemble. The bar plots show the motif counts for each ensemble of networks and the error bars (on top of each bar) depict the variation over the ensemble. We observe that the variation of network features over each ensemble is not significant. This shows that the networks are fairly close to each other and each of them can be considered as a digital twin of the actual network. 
Thus our framework is capable of creating an ensemble of synthetic distribution networks which are statistically equivalent to each other. In order to create statistically different networks, the Markov chain in step 3 needs to be altered - deleting multiple random edges, instead of one.

In general, an `ensemble' of networks consist of multiple structurally different networks which connect the same set of residences to the substation. Each synthetic network in the ensemble is a feasible network (has a tree structure and satisfies power engineering constraints) but is not the optimal length network. Therefore, we can consider it as a single random realization of the actual network. This allows us to perform analysis on an ensemble of networks instead of a single network and thereby capture the deviation arising due to different network structure in the ensemble.

\section*{Validation}\label{sec:valid}
In the earlier section, we have presented the proposed framework to create synthetic distribution networks for a geographic region. The aim is to create networks which resemble their actual physical counterparts. We obtained real-world power distribution networks for the town of Blacksburg in southwest Virginia from a distribution company to validate the created networks. This network has been incrementally built over a long period of time with a close dependency on the population growth in the region. In contrast, our proposed framework uses the current population information with no consideration of any historical data. The created synthetic networks are optimal in terms of economic and engineering perspectives. Therefore, it is expected that there would be structural differences between the networks. Furthermore, the comparison methods need to be relevant in the context of distribution networks with associated geographic attributes. 

This section compares the generated synthetic networks with the actual distribution network based on various operational, statistical and structural attributes. The operational validation ensures that we observe similar node voltages and edge flows in both networks. This makes the networks suitable to be used by the scientific community to aid in their research. The methods to compare statistical attributes help us compare the overall connectivity properties of the networks. The comparison of structural attributes involving node and edge geometries enable us to validate the created synthetic networks on a much higher resolution. The results of the comparison show that the created networks bear a significant amount of resemblance to the actual networks.
\begin{figure}[tbhp]
\centering
	\includegraphics[width=0.48\textwidth]{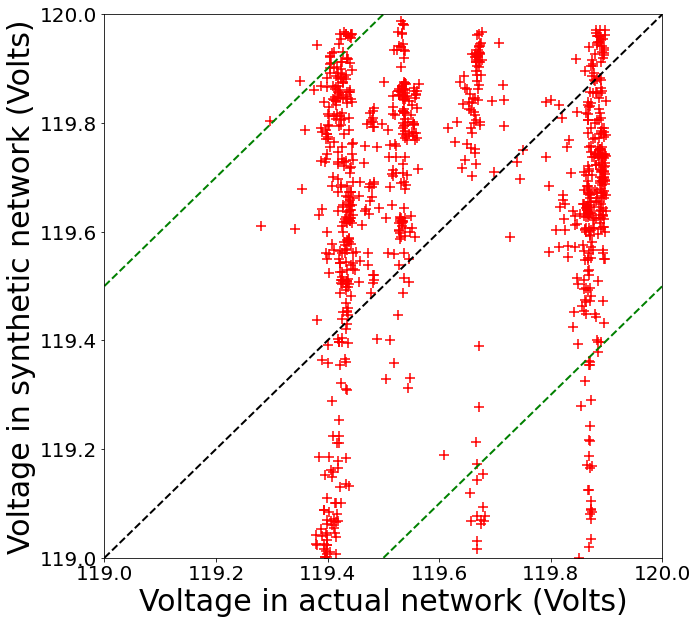}
	\includegraphics[width=0.46\textwidth]{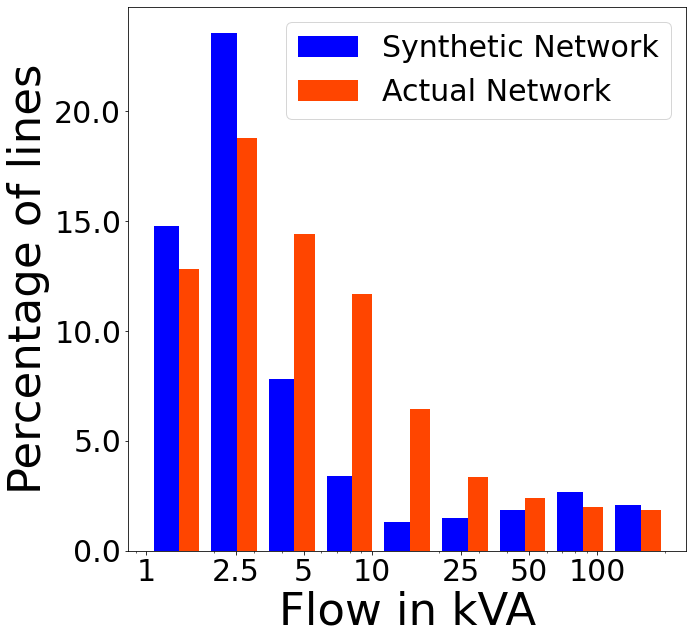}
	\caption{Plots comparing the residential node voltages (left) and edge power flows (right) for actual and synthetic networks. Majority of residence voltages in the synthetic network are within $\pm0.4\%$ voltage regulation of the voltages in the actual network. The edge flows in both network follow similar distributions with a computed KL divergence of $0.15$.}
	\label{fig:validate-voltage-flow-hist}
\end{figure}
\subsection*{Operational Validation}
We compare voltages at the residences when they are connected to the actual and synthetic network in left plot of Fig.~\ref{fig:validate-voltage-flow-hist}. We term this validation as \emph{operational validation}, where the basic idea is that if we substitute the actual network with the synthetic network, we should see minimal voltage differences at the residences connected to either network. Here, the black dotted line denotes the identity line (exact same voltages) and green lines signify $\pm0.4\%$ deviation from the identity line. We observe that majority of residence voltages in the synthetic network remain within this $\pm0.4\%$ regulation. We also compare the edge flows in the two networks through the histogram in right plot of Fig.~\ref{fig:validate-voltage-flow-hist} which also bear a significant resemblance. We performed statistical fit of the flow distributions and the KL-divergence is $0.15$.
\subsection*{Statistical Validation}
\begin{figure}[tbhp]
    \centering
    \includegraphics[width=0.32\textwidth]{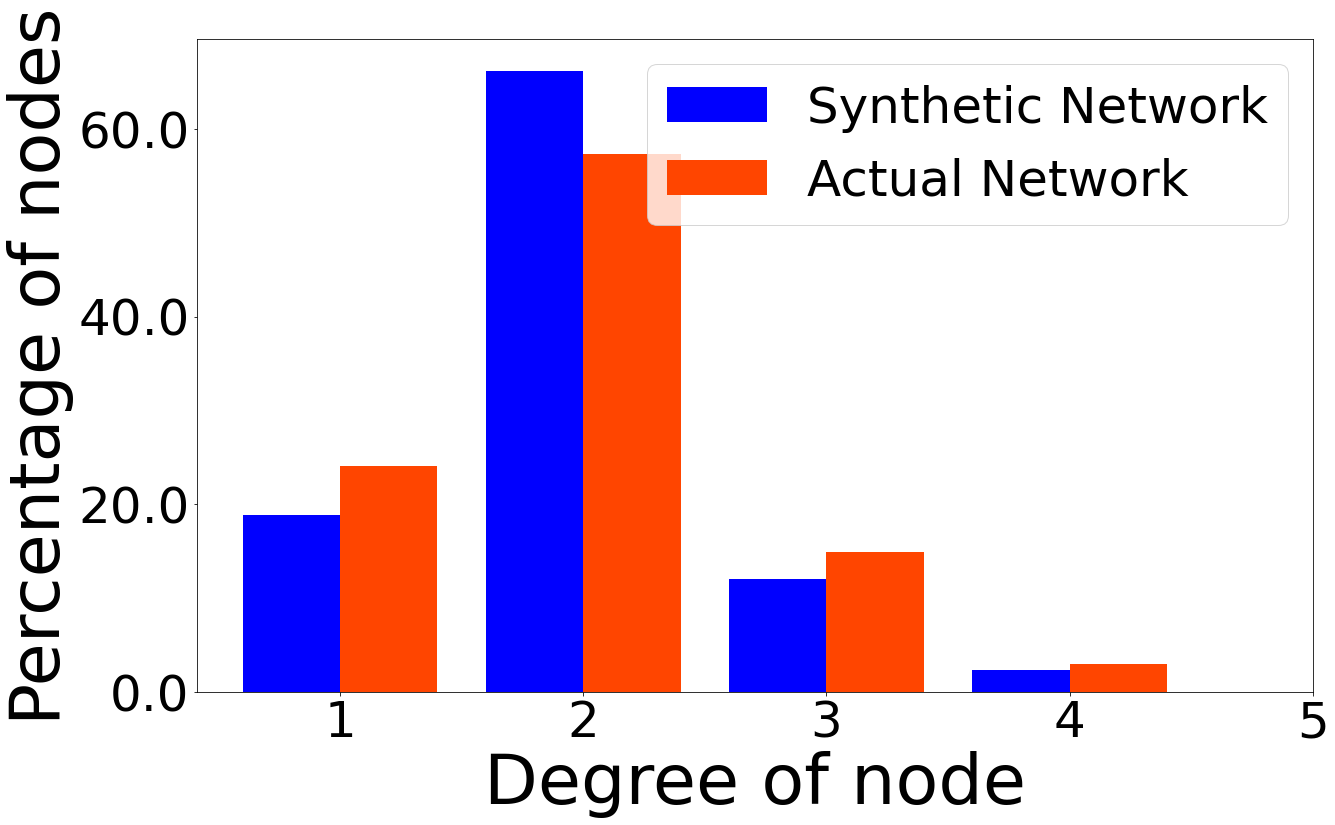}
    \includegraphics[width=0.32\textwidth]{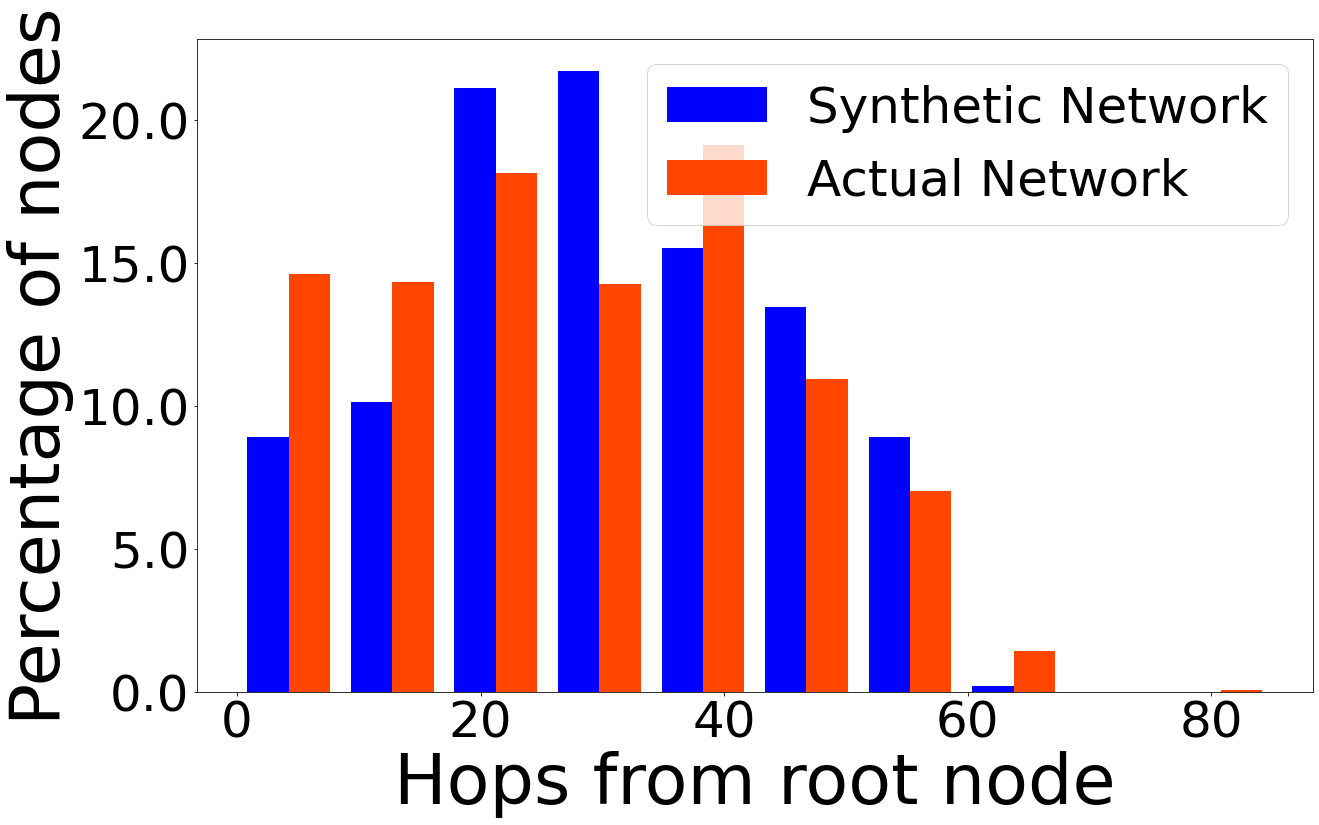}
    \includegraphics[width=0.32\textwidth]{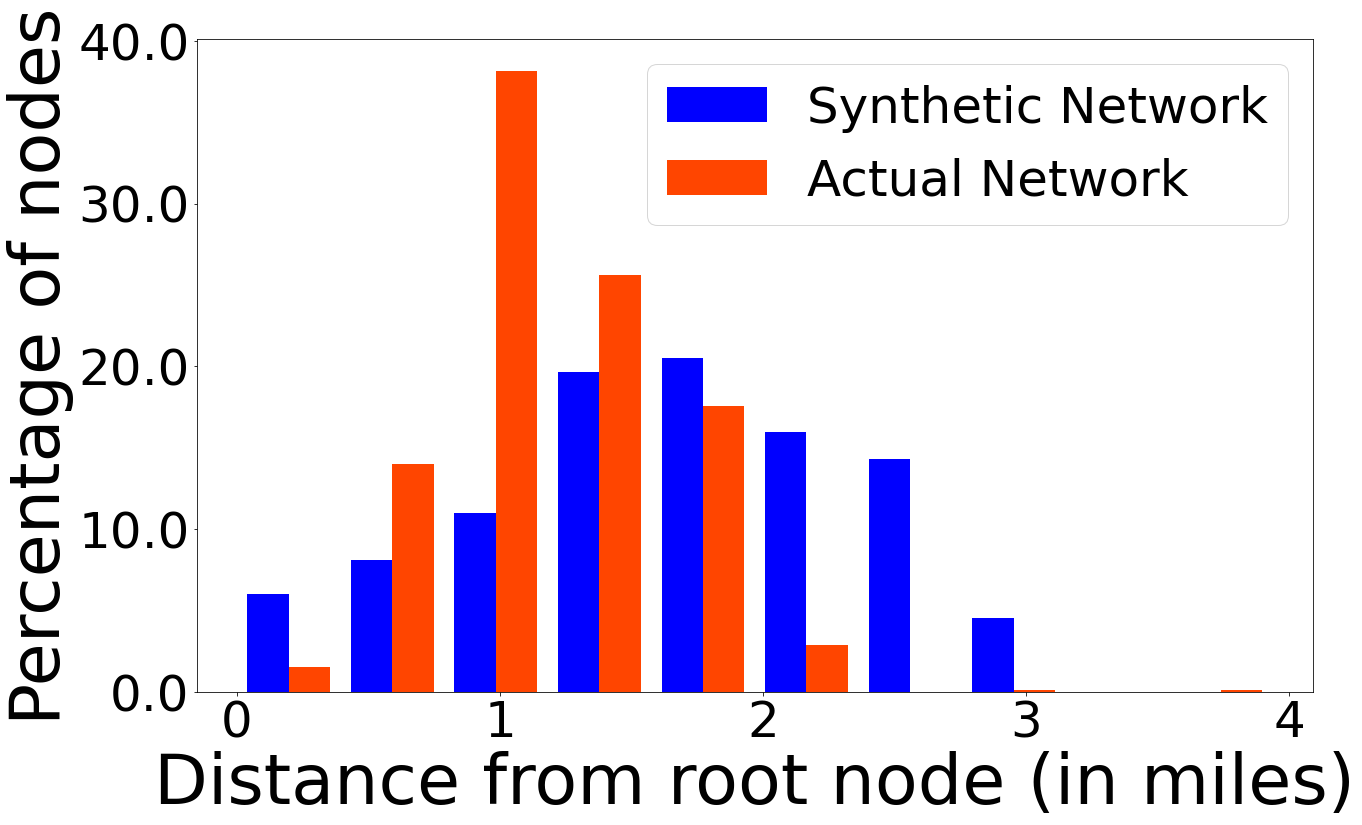}
    \caption{Plots comparing the degree distribution (left), hop distribution (middle), and reach distribution (right) of actual and synthetic distribution networks for town of Blacksburg in southwest Virginia. The degree and hop distributions are fairly close to each other which signifies their resemblance. The reach distribution differs between the networks because of the difference in the way each of them are created.}
    \label{fig:stat-val}
\end{figure}

The created networks are expected to have similar graph attributes to the actual network. We focus on basic graph attributes such as degree and hop distributions and also the newly defined ``reach'' distribution. 
Fig.~\ref{fig:stat-val} compares the synthetic and actual network for the town of Blacksburg in southwest Virginia in terms of these statistical attributes. We use the Kullback-Leibler (KL) divergence to compare each pair of distributions. KL-divergence values for various
structural measures are as follows: (i) degree distributions: $0.0208$;   (ii) hop distribution: $0.0323$; and  (iii) reach distribution: $0.0096$. The small KL-divergence values indicate that the real and the synthetic networks are structurally very similar.
\subsection*{Structural Validation}
One of the important aspects of our work is that the created synthetic networks have a geographic attribute associated with them. Therefore, we need to include network comparison methods which incorporate the geographic embedding while measuring the deviation. In this paper, we use a metric for geometry comparison, i.e., how the edge geometries in the networks deviate from each other. Due to unavailability of actual network information for the entire region, we propose an effective way to compare the structural attributes of the networks. We divide the entire geographic region into multiple rectangular grid cells and perform comparison in each cell separately. In this way we can omit the cells for which network data is missing.

\begin{figure}[tbhp]
    \centering
    \includegraphics[width=0.48\textwidth]{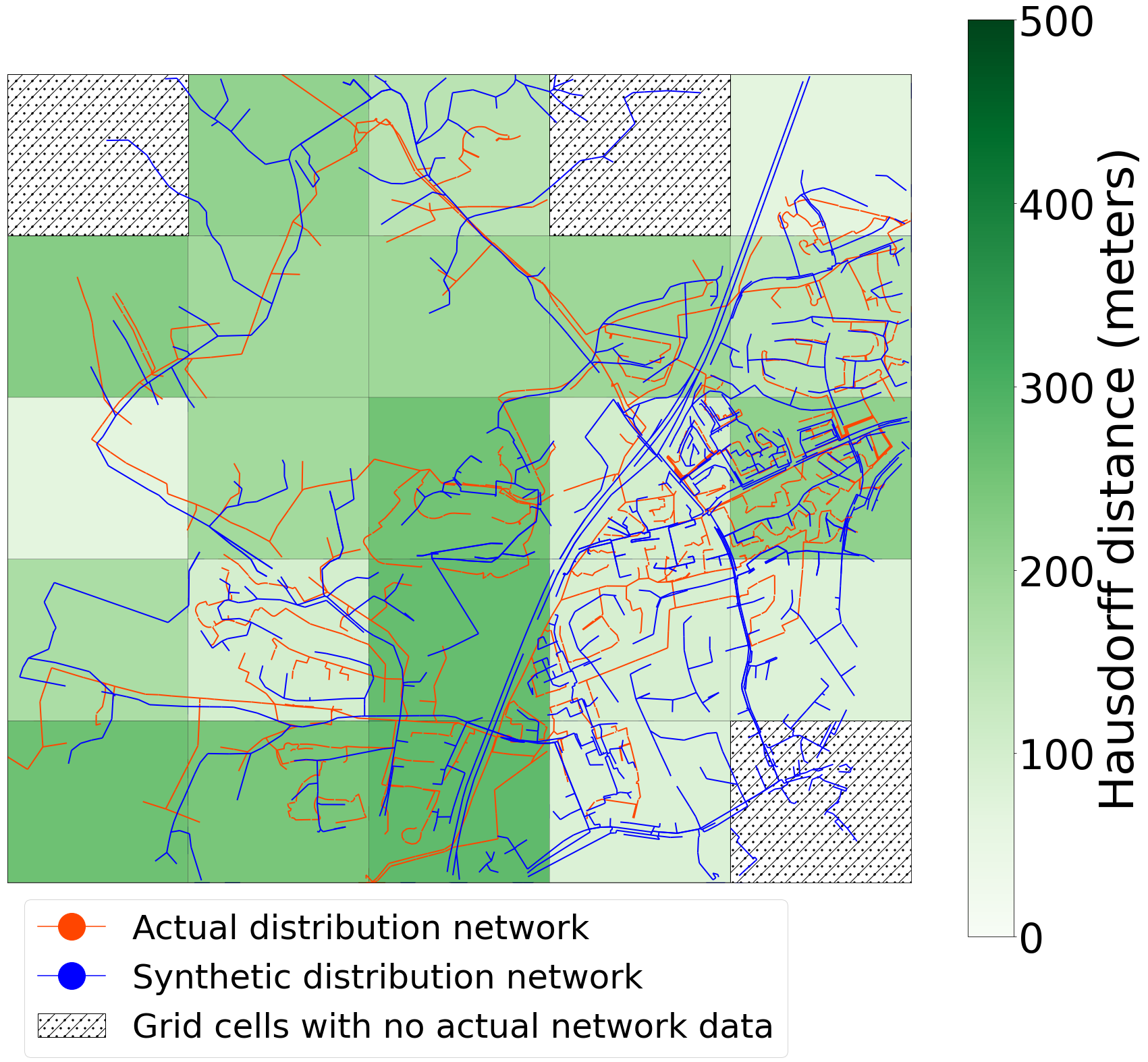}
    \includegraphics[width=0.48\textwidth]{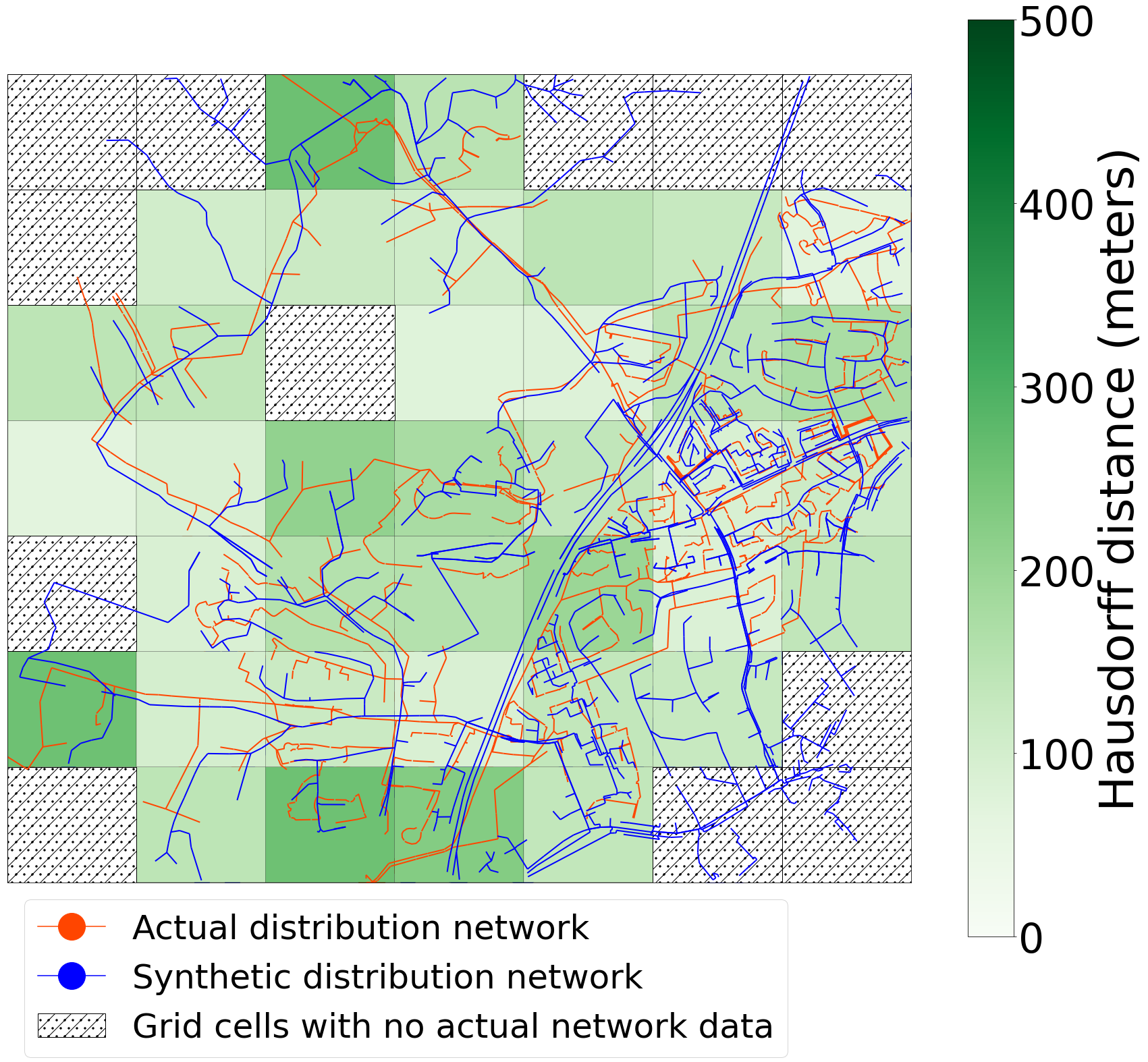}
    \caption{Plots showing Hausdorff distance based geometry comparison of actual and synthetic networks for the town of Blacksburg in southwest Virginia. The geometry comparison is performed for grid cells with two different resolutions: low resolution of $5\times5$ grid cells (left) and high resolution of $7\times7$ grid cells (right). Color in each grid cell denotes the magnitude of deviation in meters. Grid cells with no available actual network data are shaded with black dots.}
    \label{fig:hausdorff-comp}
\end{figure}

We compare the Hausdorff distance between the edge geometries of the two networks in each rectangular grid cell. Let $\mathscr{P}_{\textrm{act}}$ and $\mathscr{P}_{\textrm{syn}}$ represent the set of points representing the geometries of the actual and synthetic networks. We define the Hausdorff distance between networks $\mathscr{G}_{\textrm{act}}$ and $\mathscr{G}_{\textrm{syn}}$ for a rectangular grid cell as,
\begin{equation*}
    \mathsf{D}^{\textrm{CELL}}_{\textrm{H}}\left(\mathscr{G}_{\textrm{act}},\mathscr{G}_{\textrm{syn}}\right):=\max_{x\in\mathscr{P}_{\textrm{act}}}\min_{y\in\mathscr{P}_{\textrm{syn}}}\mathsf{dist}(x,y)
\end{equation*}
The above metric of geometry comparison allows us to measure a degree of proximity for edge geometries which are non-overlapping, yet close to each other. Fig.~\ref{fig:hausdorff-comp} shows the edge geometry comparison between actual and synthetic networks for uniform rectangular grid partitions of two different resolutions. Note that network geometries in certain regions show a significant deviation when compared with low resolution, while comparing with a higher grid resolution shows small deviation. This shows that the networks are fairly close to each other.

\section*{Case Study: Impact of Photovoltaic Penetration}\label{sec: usage-example}
We now present a representative study where we analyze the impact of PV penetration on the system node voltages. We compare the PV penetration in multiple levels of the network (MV primary network or LV secondary network). We consider the following two cases: (i) PV penetration in LV network where PV generators are installed on residence rooftops and (ii) PV penetration in MV network where a single PV generator is installed at a location in the MV network. In the first case, we randomly identify a group of residences (for example, $50\%$ of all residences) and assign PV generation to them. The penetration level indicates the rating of the PV generation installed on these residences. For the latter case, a single node PV penetration represents a `solar farm' which is connected to the distribution grid and the penetration level indicates the PV generation rating as a fraction of the total load.

\begin{figure}[tbhp]
    \centering
    \includegraphics[width=0.48\textwidth]{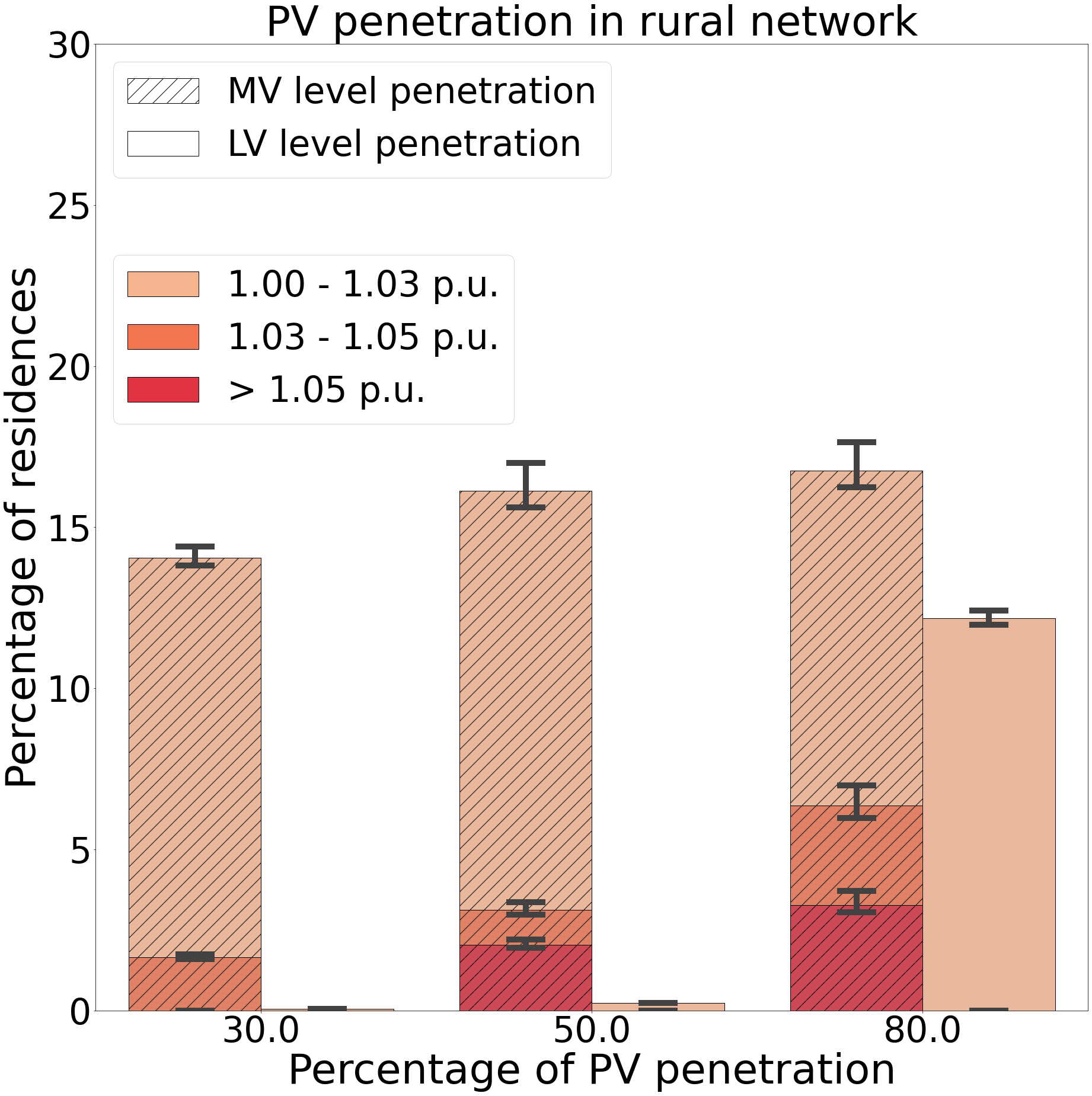}
    \includegraphics[width=0.48\textwidth]{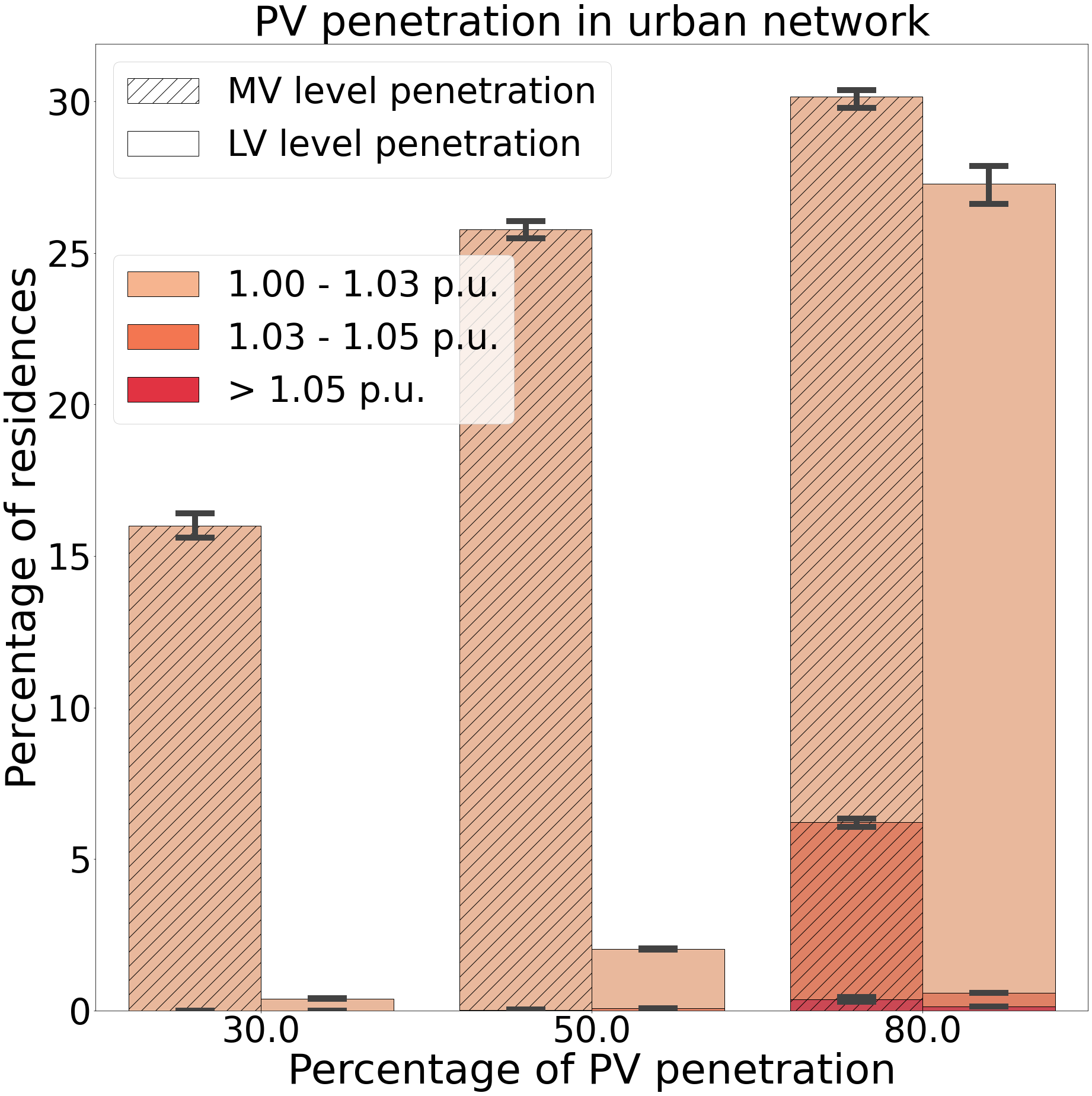}
    \caption{Plots showing impact of PV penetration in rural and urban networks. Colors depict the percentage of nodes with various levels of overvoltages. Shaded and non-shaded bars denote MV and LV-level penetration. LV-level penetration is less likely to cause severe overvoltages as compared to MV level penetration. PV penetration in rural networks is more likely to cause overvoltage issues (greater than 1.05 p.u.).}
    \label{fig:volt-outlimit}
\end{figure}
We perform the comparison on two different synthetic feeders: urban and rural. An urban distribution network is characterized with shorter lines as compared to rural networks where remote nodes are connected by long lines. Fig.~\ref{fig:volt-outlimit} compares impact of LV-level and MV-level PV penetration for two networks. Here, we focus on the percentage of nodes which face overvoltage issues due to different levels of PV penetration. We observe that for either case, the percentage of nodes with overvoltage increases with higher penetration level. Further, we see that LV-level PV generation is less likely to cause overvoltage issues as compared to a single node MV-level PV integration. Additionally, in the case of rural feeders, the percentage of nodes experiencing severe overvoltage (around~1.05 pu, which is the extreme limit of acceptable overvoltage) is higher as compared to urban feeder networks. Therefore an optimal placement of PV generators is required for the rural feeders so that they do not suffer from overvoltage issues.

\section*{Discussion and Limitations}
Although the synthetic power distribution network dataset produced by our framework is comprehensive, it is not without its limitations. In this work we generate networks with only positive sequence parameters. The ensemble of synthetic networks can be used as a tool for performing planning studies or addressing system-wide policy level questions. We can also perform short circuit analysis with symmetrical three phase faults.

However, distribution systems are networks of mixed phase order and mixed network configuration. They are usually three phases in the primary network and the secondary network consists of mixed single and three phase circuits. We have provided a framework in the SI to partition the residences into three phases and thereby create a three phase network. A complete three phase network requires inclusion of zero sequence line parameters and transformer configurations (wye-wye, delta-wye, wye-delta and delta-delta). Therefore, in their current version, these networks might not be suitable to be used for performing dynamic stability analysis or studying detailed transient responses to power grid contingencies.

Further, shunt compensation is used in the primaries for maintaining voltage level within engineering standards. These comprise of capacitor banks which elevate voltage level along the network. Hence, they can be optimally placed in the network to avoid severe undervoltage issues at remotely located residences. We can consider critical sections of the network and design necessary shunt compensation to maintain a high degree of reliability of the network. Additionally, the proposed framework creates a network to connect only the residential buildings in a geographic region. In order to connect heavy load centers, networked secondaries with pad-mounted transformers are used in some large urban areas. These additions can be made to our existing synthetic networks and would be a direction for future research.

\subsection*{Acknowledgements}
This work is partially supported by University of Virginia Strategic Investment Fund award number SIF160, $3$Cavaliers grant number 166914-WMS5F-LC00295-30015 and NSF grants: EAGER 161070 \& CINES 160369. We also thank Pete Sauer (University of Illinois, Urbana-Champaign) and Anna Scaglione (Cornell Tech) for their valuable suggestions.


\bibliographystyle{unsrt}
\bibliography{references}

\end{document}



\maketitle


\section*{Preliminaries of Power Distribution System}
\begin{figure}[tbhp]
\centering
	\includegraphics[width=0.48\textwidth]{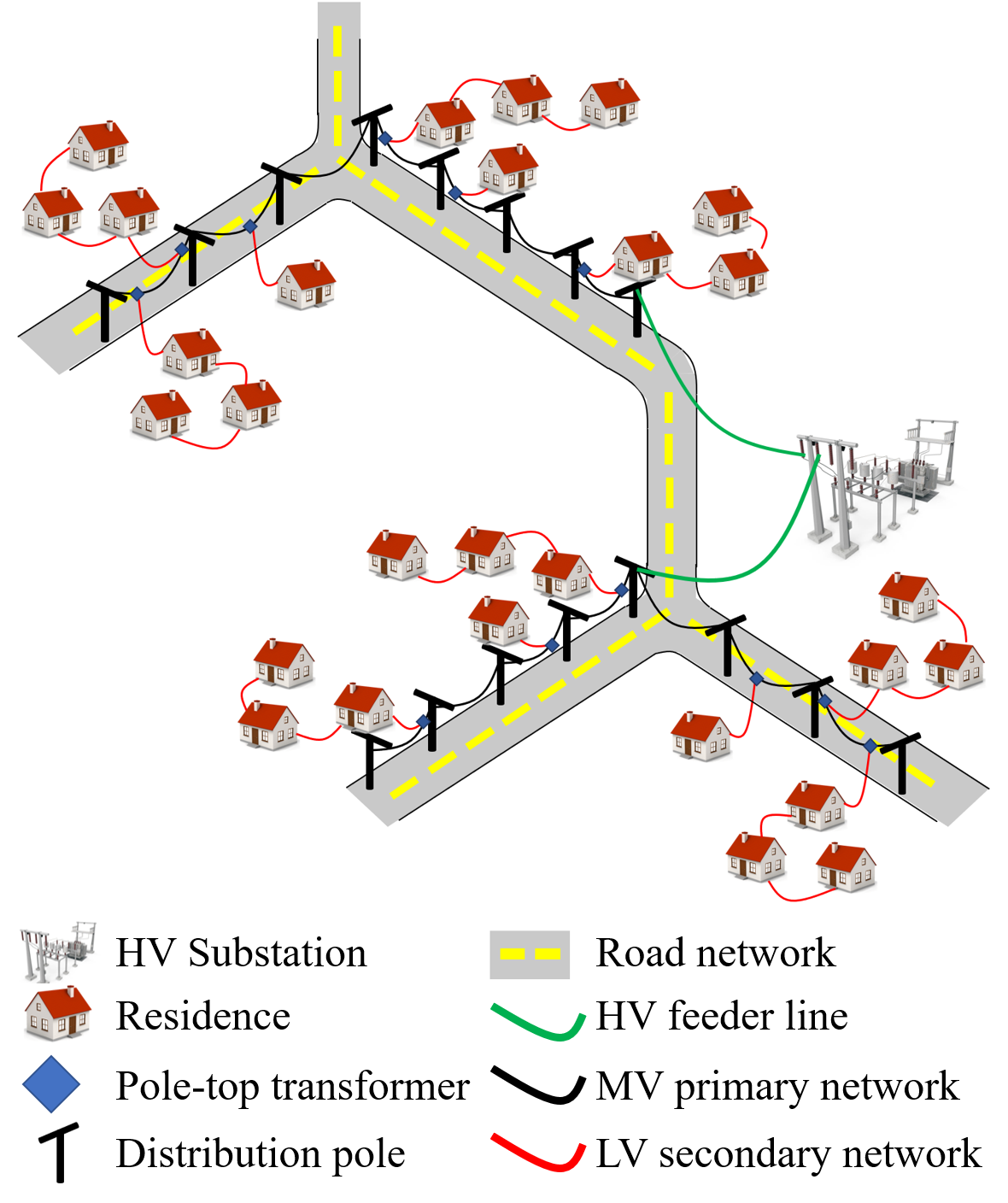}
	\includegraphics[width=0.48\textwidth]{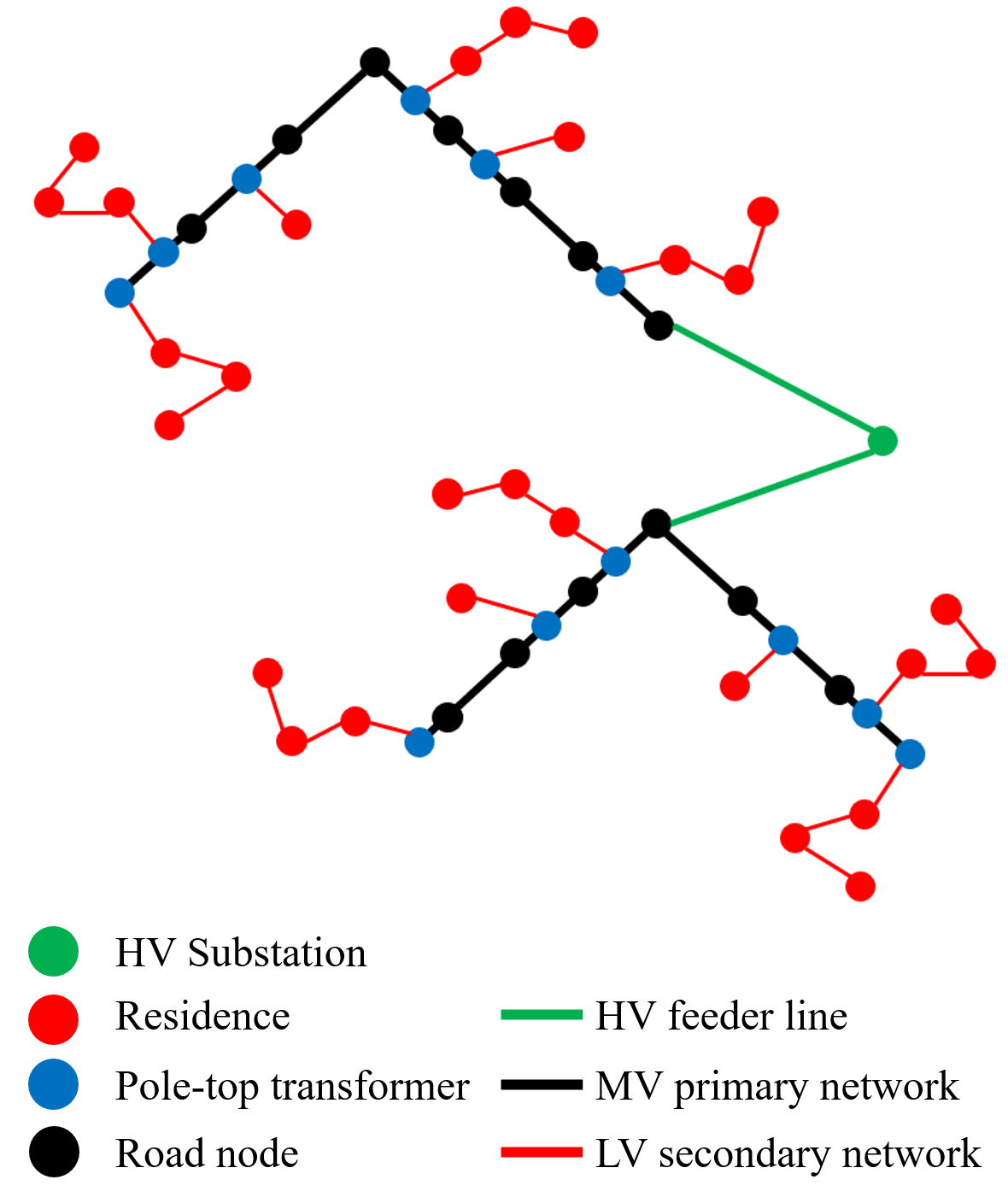}
	\caption{Schematic of power distribution network (left) and corresponding graph structure (right). The proposed approach first creates the secondary network (red) connecting the residences to local transformers. Thereafter, the road network (black edges) is used as a proxy to connect the transformers forming the primary network.}
	\label{fig:approach}
\end{figure}

The power distribution network connects the HV (greater than 33kV) distribution substation to LV (208-480V) residential consumers. In most practical networks, this is done through two sets of networks: (i) the MV (usually 6-11kV) primary network connects a step-down transformer at the substation to local pole-top transformers along the road network 
and (ii) the LV secondary network connects the residences to the local pole-top transformers. 
Additionally, most distribution networks are operated in a \emph{radial} structure (with no cycles or loops) to facilitate protection coordination~\cite{rad_prot}. Residences are primarily connected in chains (the degree of a node is at most 2) to avoid branching and thereby maintaining a healthy voltage level. We term this configuration a \emph{star-like tree} since the edges emerge from a single root transformer node and connect residences without further branching. The schematic of a typical power distribution networks is shown in Fig.~\ref{fig:approach}.
\FloatBarrier

\section*{Related Work}
Over the past decade, researchers have developed methods to create synthetic power networks of different sizes. These attempts include interdependent networks created using simple logical models~\cite{Souza2014} to models created based on statistical attributes of actual networks~\cite{zussman_2016}.
The simple logical network models allowed network scientists to carry out a linear analysis which reveals big picture results for large networks. However, neglecting rich system data (line parameters and residence load demand) often makes these simple models less useful in certain kinds of applications~\cite{hines2013}. More recently, models with actual system parameters have been proposed in~\cite{overbye_101,overbye_102,overbye_2019,overbye_2020}. However, most of these networks are limited to the transmission system and provides little importance to the distribution power networks.

For a long time, power system researchers were dependent on the standard IEEE test systems~\cite{ieee_feeder,epri_feeder} to carry out experiments and validate control/optimization algorithms. These test systems were considered a replica of the actual power grid and therefore, used as a test-bed before deploying a methodology on the actual power system. These models either consider the residential customers as passive consumers of electricity or assume an aggregated version of multiple consumers as system load demand. These provided an almost accurate system analysis as long as we do not consider the role of consumers. However, with the advent of distributed energy resources (DERs) and electric vehicles (EVs),  residential customer behavior has become an integral part of power system analysis~\cite{Bistline2021,Sepulveda2021,Joshi2021}. Therefore, modern power network models need to include detailed residential consumer attributes for accurate representation of the actual power grid. Table~\ref{tab:dist-net-lit} lists the different power network models present in the literature and a summary of the network features.

The literature dealing with the creation synthetic distribution networks can be broadly classified into two categories~\cite{review2015,review2018}. The first category comprises  methods that use one or more heuristics to synthesize the networks. The second category of methods uses mathematical programming to encode various physical and structural constraints that real-world networks usually obey. The methods in the second category involve solving one or more optimization problems that often require large computation time~\cite{nrel_net}. 
The work by Schweitzer et al.~\cite{schweitzer} is one of the earliest papers in the second category. The author analyze a large dataset of actual distribution networks in Netherlands. Then they statistically fit these parameters to create a class of stochastic network models. Random networks are then sampled from using the stochastic model. This method is reminiscent of ``configuration models'' used to construct random social networks~\cite{configmodel}. The generated synthetic networks have random interconnections, yet the network attributes follow similar distributions as the actual networks. In~\cite{feeder_gan}, the authors have used an extensive dataset of actual distribution networks and trained a generative adversarial neural network (GANN). Thereafter, random networks are generated from the trained generative model and hence multiple synthetic networks can be generated which resemble actual distribution networks. However, these methods are dependent on the availability of an extensive amount of actual distribution network dataset. The lack of availability of such information uniformly from all geographic regions makes the statistical models `biased' and hence cannot be used as a generalized framework to create synthetic networks.

On the other end of the spectrum, we have multiple heuristic methods~\cite{rnm_2011,rnm_2013,nrel_net,anna_naps,gm2016,highres_net} to create random feeder networks with or without using the road network information. The heuristics are primarily used in the initial steps to assign substation locations and construct a feeder network. The heuristics include clustering residential loads to assign substation location at the cluster centroids~\cite{rnm_2013}, distribute an aggregated residential load to certain load points in a hierarchical fashion~\cite{gm2016} and use minimum spanning tree algorithms~\cite{rnm_2011,rnm_2013,anna_naps}. Some of these methods construct an imaginary road network~\cite{rnm_2011,rnm_2013}, while others have used openly available information~\cite{anna_naps}. Neither of these methods include the power engineering constraints in the network synthesis phase, rather the cables and shunt compensators are chosen such that power engineering constraints are satisfied.

The mathematical programming methods are mostly restricted to distribution system expansion planning~\cite{quiroga2019,You2016,trpovski_2018,pascal2020} or distribution system restructuring problem~\cite{manish2018OptimalDS,manish2019,lei2019radiality}. We have followed similar works to construct the mathematical framework for our approach (Steps 1 and 2). However, one of the main difference is that in all these works the feeder nodes are known beforehand. In our approach, the feeder nodes are identified as an output of the proposed optimization framework. The development of such a framework allows us to use it in Step 3 of our algorithm to construct an ensemble of synthetic networks which are statistically similar to each other.

\begin{table}[tbhp]
\scriptsize
\centering
\caption{Summary of power system network data available in literature}
\label{tab:dist-net-lit}
\begin{tabular}{ll}
\hline
Authors/Datasets & Description/Comments \\ \hline
EPRI Test Circuits~\cite{epri_feeder} & A set of test feeders, representative of small, medium, and large-scale distribution networks. \\
IEEE Test Feeders~\cite{ieee_feeder} & A repository for small and large-scale distribution feeders with no associated geographic attribute. \\
GridLAB-D Feeders~\cite{pnnl_feeder} & A set of 24 actual radial distribution feeders from 5 climate regions in the US. \\
D'Souza et.al~\cite{Souza2014} & Random graph models to create synthetic interdependent infrastructure networks. \\
Zussman et.al~\cite{zussman_2016} & Synthetic transmission networks with geographic attributes created using a statistical distribution of actual network attributes. \\
Birchfield et.al~\cite{overbye_101,overbye_102} & Large scale synthetic transmission networks for the state of Texas, US with synthetic loads aggregated at the zip-code level. \\
Overbye et.al~\cite{overbye_2019} & Stochastic geometry-based approach was used to place transformers in the distribution network without physical constraints. \\
Trpovski et.al~\cite{trpovski_2018} & Optimization framework to create distribution networks connecting zipcode level aggregated consumer loads to substations. \\
Kadavil et.al~\cite{gm2016} & Creates a feeder network and populates it with stochastic load profiles with no associated geographic attribute. \\
Schweitzer et.al~\cite{schweitzer,anna_naps} & Distribution networks populated with random loads created to mimic statistical distribution of actual network attributes. \\
Mateo et al.~\cite{rnm_2011,rnm_2013,nrel_net} & Heuristic-based approach to connect consumers with no explicit structural and operational constraints. \\
Li et.al~\cite{overbye_2020} & Combined synthetic transmission and distribution networks created for Texas, US by adding models of~\cite{overbye_101,overbye_102} with RNM. \\ 
Liang et.al~\cite{feeder_gan} & Generative adversarial networks (GAN) based framework to create an ensemble of synthetic networks. \\
Bidel et.al~\cite{highres_net} & Synthetic networks were created after learning the statistical distribution of load and network published by network operators. \\ \hline
\end{tabular}
\end{table}
\FloatBarrier

Our work differs from prior works in the following ways. The points are further summarized in Table~\ref{tab:comp-relworks}.
\begin{enumerate}
    \item We propose an optimization framework which creates a minimum length network which satisfies standard structural and power flow constraints. Earlier papers, e.g.~\cite{highres_net,anna_naps,nrel_net,feeder_gan} often do not use such well-defined mathematical models that ensure all the constraints are satisfied while creating the synthetic networks.
    \item Although the heuristics used in the methods proposed in~\cite{anna_2010,anna_naps,rnm_2011,rnm_2013,nrel_net,validate2020} are able to create a distribution network as a minimum spanning tree, they do not ensure that the power engineering voltage constraints are satisfied as per ANSI standards~\cite{ansi}. Therefore, the synthesized networks might be topologically realistic but need not be be realistic
    from the perspective of power engineering. The power engineering constraints are satisfied only after adding shunt capacitors and voltage regulators, which improves the voltage profile.
    \item Most heuristics described in previous papers~\cite{nrel_net,validate2020,feeder_gan,highres_net} deal with the issue of finding locations of substation feeders by clustering residences. In contrast, we start with given locations of substations~\cite{eia_substations} and residences~\cite{swapna_2018}  based on real-world data and construct the network connecting these points. Further, in our primary network creation method (Step-2), we have included the optimal substation feeder selection problem effectively in our optimization framework, which has not been used in any prior work.
    \item The power flow constraints used in our methodology are similar to the constraints used in distribution feeder planning problems,  as in~\cite{trpovski_2018,pascal2020,lei2019radiality,mv_2011}. These constraints define the fundamental physics of distribution network operation and hence are similar to related works. In addition, we include two novel components in our optimization framework: (i) Prior works included either a commodity flow model~\cite{lei2019radiality} to ensure tree structure, or included multiple constraints to avoid the occurrence of cycles~\cite{manish2018OptimalDS}. Here, we have theoretically proved in Proposition~(\ref{prop-1}) that the power flow constraints are sufficient to ensure a tree structure. (ii) Prior papers on distribution network planning and expansion, which include a similar optimization framework assume that the number of feeders are known beforehand. In our framework, we do not make such assumptions; rather, the optimal set of feeders are identified as an output of the primary network creation problem. 
    \item  We construct an ensemble of networks which indeed is an important contribution of the paper. We do not consider the optimal network as the `sole' output; rather we consider it as `one' random realization of the actual distribution network and propose a framework to create multiple feasible and realistic (but not optimal) networks. Using our well-defined optimization framework in Step-1 and Step-2, we  are able to create an ensemble of feasible networks by solving a restricted version of the optimization problem. Prior works have not considered this aspect of an ensemble of synthetic power distribution networks.
\end{enumerate}

\begin{table}[tbhp]
\centering
\caption{Table showing comparison with other related works}
\label{tab:comp-relworks}
\begin{tabular}{llll}

\hline
\multicolumn{2}{l}{Previous Works}                                                   & \textbf{Includes}  & \textbf{Excludes}   \\ \hline
\multicolumn{2}{l}{D'Souza et.al~\cite{Souza2014}}                                   & I                  & II,III,IV,V,VI,VII  \\
\multicolumn{2}{l}{Zussman et.al~\cite{zussman_2016}}                                & I                  & II,III,IV,V,VI,VII  \\
\multicolumn{2}{l}{Birchfield et.al~\cite{overbye_101,overbye_102}}                  & I                  & II,III,IV,V,VI,VII  \\
\multicolumn{2}{l}{Overbye et.al~\cite{overbye_2019}}                                & I                  & II,III,IV,V,VI,VII  \\
\multicolumn{2}{l}{Trpovski et.al~\cite{trpovski_2018}}                              & I,IV,V             & II,III,VI,VII       \\
\multicolumn{2}{l}{Kadavil et.al~\cite{gm2016}}                                      & I                  & II,III,IV,V,VI,VII  \\
\multicolumn{2}{l}{Schweitzer et.al~\cite{schweitzer,anna_naps}}                     & I,II               & III,IV,V,VI,VII     \\
\multicolumn{2}{l}{Reference Network Model (RNM)~\cite{rnm_2011,rnm_2013,nrel_net}}  & I,II,V             & III,IV,VI,VII       \\
\multicolumn{2}{l}{Liang et.al~\cite{feeder_gan}}                                    & I,VII              & II,III,IV,V,VI      \\
\multicolumn{2}{l}{Bidel et.al~\cite{highres_net}}                                   & I,II               & III,IV,V,VI,VII     \\ \hline
I                                          & \multicolumn{3}{l}{Geographic information embedding}                               \\
II                                         & \multicolumn{3}{l}{High resolution residential hourly demand profile}              \\
III                                        & \multicolumn{3}{l}{Avoids usage of actual distribution networks}                   \\
IV                                         & \multicolumn{3}{l}{Well defined mathematical framework with a guaranteed solution} \\
V                                          & \multicolumn{3}{l}{Networks satisfies ANSI voltage constraints}                    \\
VI                                         & \multicolumn{3}{l}{Optimal choice of feeders}                                      \\
VII                                        & \multicolumn{3}{l}{Ensemble of networks}                                           \\ \hline
\end{tabular}
\end{table}
\FloatBarrier

\clearpage
\section*{Methods}\label{sec:methods}
\subsection*{Details of the dataset used}
We use open-source, publicly available information regarding several infrastructures to generate the synthetic distribution networks: (i) road network data from OpenStreetMap (OSM)~\cite{osm}, (ii) geographic locations of HV (greater than 33kV) substations from data sets published by EIA~\cite{eia_substations}, and (iii) residential electric power demand information developed in earlier work from our research group~\cite{swapna_2018}. The details of each dataset are provided in Table~\ref{tab:dataset-intro}. The rightmost column provides the size of each dataset for Montgomery County in southwest Virginia, US. 

\begin{table*}[tbhp]
\centering
\caption{Datasets and related attributes used to generate synthetic distribution network}
\label{tab:dataset-intro}
\begin{tabular}{clll}
\hline
\textbf{Dataset}                                        & \multicolumn{1}{c}{\textbf{Source}}                                                                                                             & \multicolumn{1}{c}{\textbf{Attributes}}                                                                                                             & \multicolumn{1}{c}{\textbf{\begin{tabular}[c]{@{}c@{}}Example for\\ Montgomery County\\ of Virginia, US\end{tabular}}} \\ \hline
Substation                                              & \begin{tabular}[c]{@{}l@{}}Electric substation data\\ published by the US \\ Department of \\ Homeland Security~\cite{eia_substations}\end{tabular} & \begin{tabular}[c]{@{}l@{}}\tabitem substation ID\\\tabitem longitude\\\tabitem latitude \end{tabular}                                              & 20 substations                                                                                \\ \hline
\begin{tabular}[c]{@{}c@{}}Road \\ network\end{tabular} & \begin{tabular}[c]{@{}l@{}}GIS and electronic \\ navigable maps \\ published by \\ OpenStreetMap~\cite{osm}\end{tabular}                     & \begin{tabular}[c]{@{}l@{}}\tabitem node ID\\\tabitem node longitude\\\tabitem node latitude\\\tabitem link ID\\\tabitem link geometry\end{tabular} & \begin{tabular}[c]{@{}l@{}}33882 nodes\\ 41261 edges\end{tabular}                             \\ \hline
Residences                                              & \begin{tabular}[c]{@{}l@{}}Synthetic population \\ and electric load \\ demand profiles\cite{swapna_2018}\end{tabular}          & \begin{tabular}[c]{@{}l@{}}\tabitem residence ID\\\tabitem longitude\\\tabitem latitude\\ \tabitem hourly load \\~~profile\end{tabular}             & 35629 homes                                                                                   \\ \hline
\end{tabular}
\end{table*}

\FloatBarrier
\subsection*{Step 1a: Map residences to the nearest road network link}\label{ssec:step1a}
\begin{figure}[tbhp]
\centering
	\includegraphics[width=0.24\textwidth]{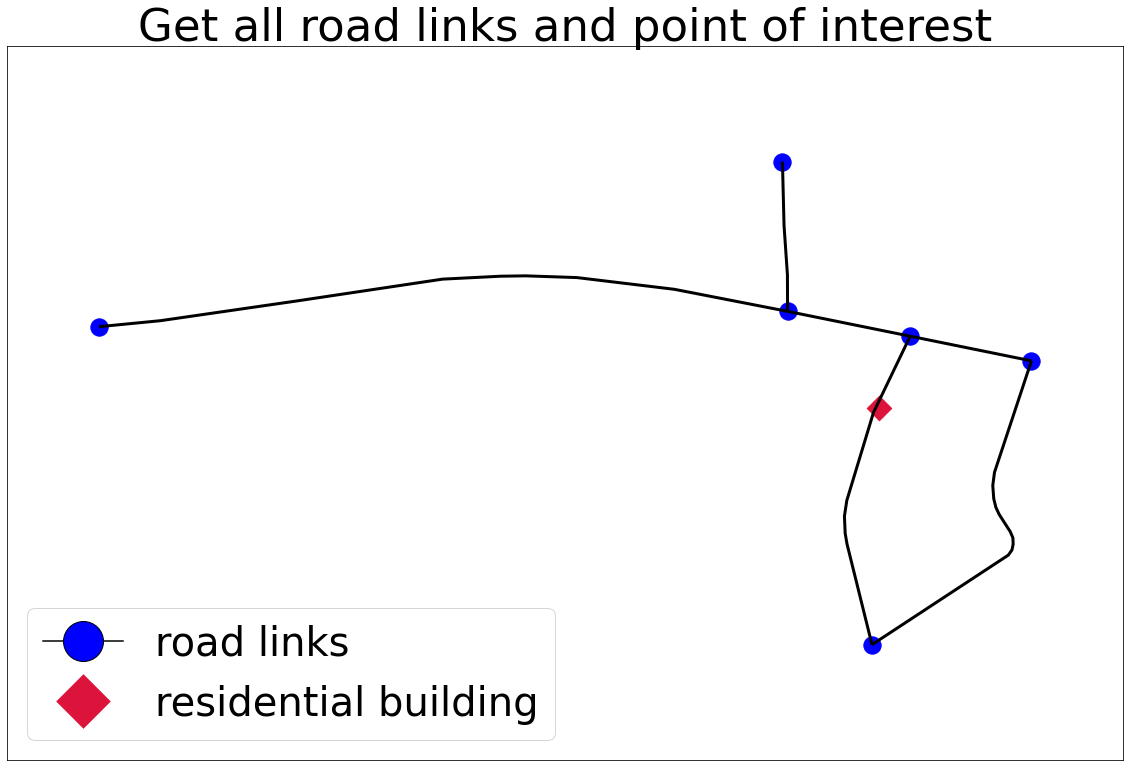}
	\includegraphics[width=0.24\textwidth]{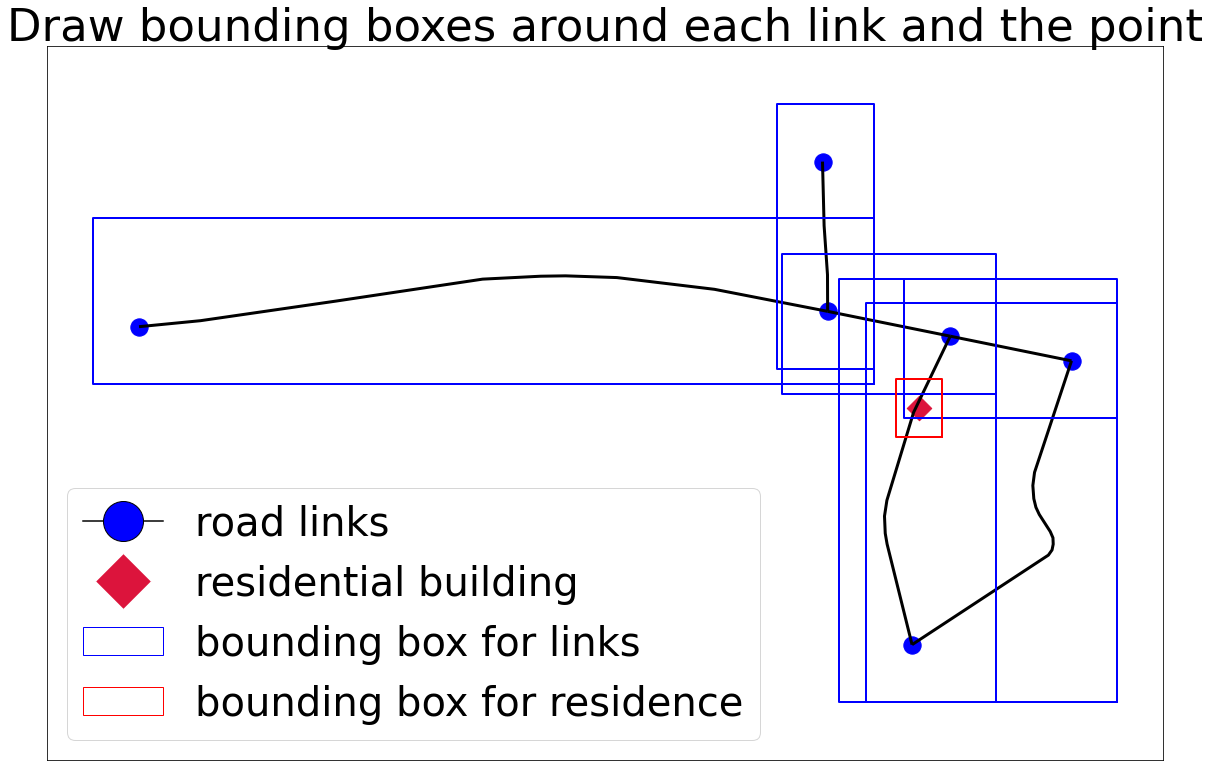}
	\includegraphics[width=0.24\textwidth]{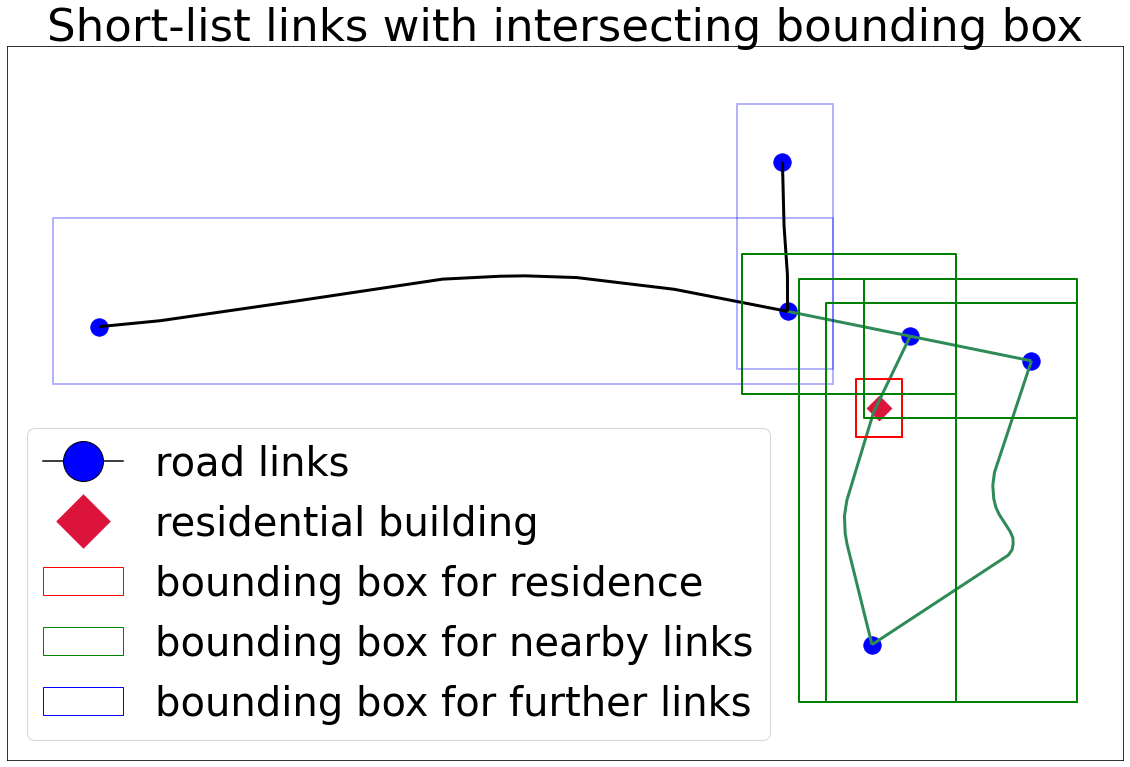}
	\includegraphics[width=0.24\textwidth]{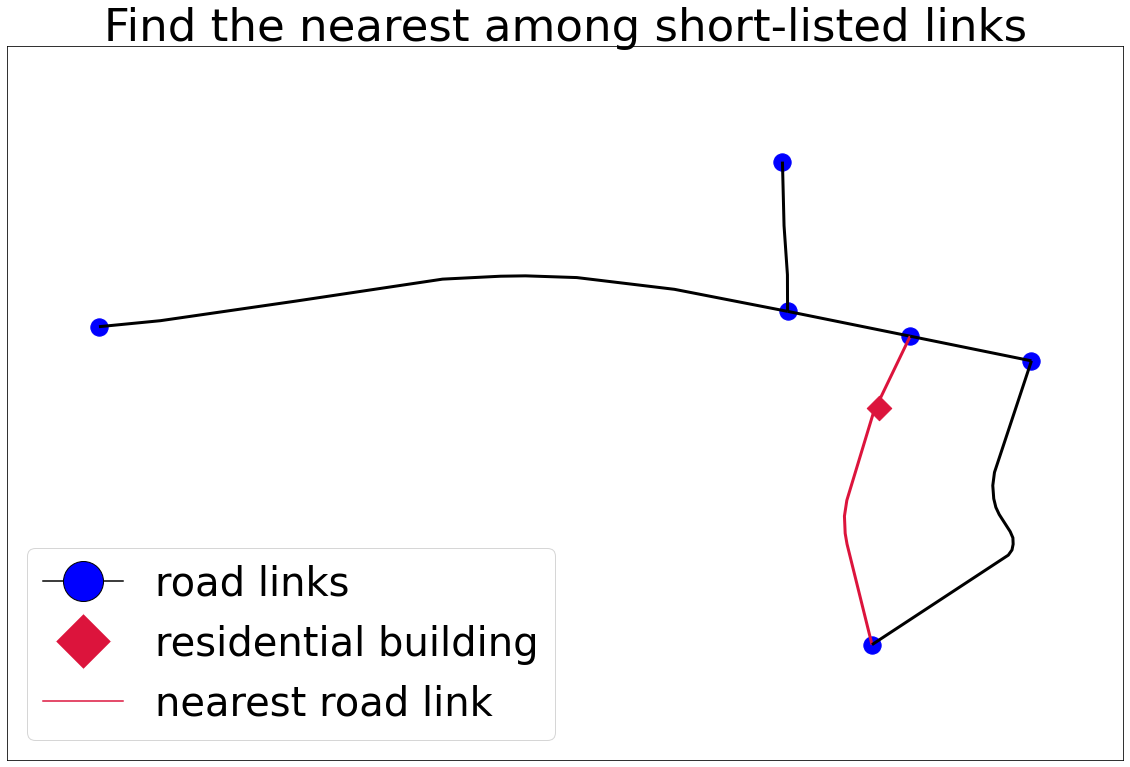}
	\caption{Plots showing the steps in mapping a residence to the nearest road network link. Bounding regions are drawn around each link (blue boxes) and the residence (red box). The intersections (green boxes) between these bounding regions are identified to shortlist nearby road links. The nearest road network link (red edge) is then selected from these shortlisted links.}
	\label{fig:mapping}
\end{figure}
\FloatBarrier
This section details the proposed mapping $\mathscr{F}_M \colon \mathscr{H} \rightarrow \mathscr{E}_R$ between the set of residences $\mathscr{H}$ and links $\mathscr{E}_R$ of road network $\mathscr{G}_R\left(\mathscr{V}_R,\mathscr{E}_R\right)$. We map each residence $h\in\mathscr{H}$ to the nearest road network link $e\in\mathscr{E}_R$. Thereafter, we compute the inverse mapping $\mathscr{F}^{-1}\left(e\right)$ to identify a set of residential buildings near each road network link $e\in\mathscr{E}_R$. This information would be used in the successive steps to generate the secondary distribution network. We denote the spatial embedding of residence $h\in\mathscr{H}$ as $\mathbf{p}_h\in\mathbb{R}^2$. A road network link $e\in\mathscr{E}_R$ is denoted by $e:\left(u,v\right)$, where the nodes $u,v\in\mathscr{V}_R$ have spatial embedding $\mathbf{p}_u,\mathbf{p}_v\in\mathbb{R}^2$ respectively. 

Algorithm~\ref{alg:dist} is used to compute the nearest road network link to a given residence. First, a bounding region of suitable size is evaluated for each road network link. This is done such that any point in the region is within a radius $r$ from any internal point of the road network link $e$. The spatial embedding of an interior point along link $e=(u,v)$ is the convex addition of the spatial embedding of the nodes $u$ and $v$. The bounding region for link $e=(u,v)\in\mathscr{E}_R$ denoted by $\mathcal{B}_e$ is,
\begin{equation}
    \mathcal{B}_e=\big\{\mathbf{p};\lvert\lvert\mathbf{p}-\mathbf{p}_e\rvert\rvert_2~\leq r,\forall \mathbf{p}_e=\theta\mathbf{p}_u+(1-\theta)\mathbf{p}_v,\theta\in[0,1]\big\}\label{eq:bound-link}
\end{equation}
Similarly, a bounding region is considered for a residential building $h\in\mathscr{H}$ and denoted it by $\mathcal{B}_h$,
\begin{equation}
    \mathcal{B}_h=\big\{\mathbf{p}\big|||\mathbf{p}-\mathbf{p}_h||_2\leq r\big\}\label{eq:bound-point}
\end{equation}
The intersections between the bounding region of the building and those for the links are stored and indexed in a \emph{quad-tree} data structure~\cite{quadtree1974}. We identify links $e_1,e_2,\cdots,e_k$ with bounding regions $\mathcal{B}_{e_1},\mathcal{B}_{e_2},\cdots,\mathcal{B}_{e_k}$ which intersect with $\mathcal{B}_h$. These $k$ links are comparably nearer to the residential building than the others. Thus, the algorithm reduces the computational burden of evaluating the distance between all road links and residential buildings. Finally, the nearest road network link is identified by computing the minimum perpendicular distance from $\mathbf{p_h}$ as,
\begin{equation}
    \mathscr{F}_M\left(h\right) = \textrm{arg}~\min_{e=\left(u,v\right)\in\{e_i\}_{i=1}^{k}}\frac{\lvert\lvert\left(\mathbf{p}_u-\mathbf{p}_h\right)\times\left(\mathbf{p}_u-\mathbf{p}_v\right)\rvert\rvert}{\lvert\lvert\mathbf{p}_u-\mathbf{p}_v\rvert\rvert}\label{eq:nearest}
\end{equation}
\begin{algorithm}
	\caption{Map residences to the nearest road network link.}
	\label{alg:dist}
	\textbf{Input} Set of residences $\mathscr{H}$, road network $\mathscr{G_R}\left(\mathscr{V}_R,\mathscr{E}_R\right)$, radius for bounding region $r$.
	\begin{algorithmic}[1]
		\For {each link $e\in\mathscr{E}_R$}
		\State Define bounding region $\mathcal{B}_e$ using (\ref{eq:bound-link}).
		\EndFor
		\For {each residence $h\in\mathscr{H}$}
		\State Define bounding region $\mathcal{B}_h$ using (\ref{eq:bound-point}).
		\State Find links $e_1,\cdots,e_k$ with bounding regions $\mathcal{B}_{e_1},\cdots,\mathcal{B}_{e_k}$ which intersect with $\mathcal{B}_h$.
		\State Compute the nearest link to residence $h$ using (\ref{eq:nearest})
		\EndFor
	\end{algorithmic}
	\textbf{Output} Mapping~$\mathscr{F}_M \colon \mathscr{H} \rightarrow \mathscr{E}_R$
\end{algorithm}

\noindent\textbf{Implementation examples.}~
The steps involved in mapping a given to the nearest road network link are shown in Fig.~\ref{fig:mapping}. The first figure shows the residence (magenta), which is required to be mapped to the nearest road link. We draw bounding regions around each road link as well as the residence. Thereafter, we identify the bounding regions which intersect with the bounding region of the residence. The nearest road link is identified from these short-listed links.

\subsection*{Step1b: Connect residences to local transformers}
\begin{problem}[$\mathcal{P}_{\textrm{sec}}$ \textbf{construction}]
Given a road link~$e\in\mathscr{E}_R$ with a set of residences~$\mathscr{F}_M^{-1}\left(e\right)$ assigned to it, construct an optimal forest of trees, $\mathscr{G}_S\left(e\right)$, rooted at points (local transformers) along the link and connecting the residences.
\end{problem}
Here we present the detailed optimization framework for the problem $\mathcal{P}_{\textrm{sec}}\left(e\right)$ to create the secondary distribution network $\mathscr{G}_S\left(e\right)(\mathscr{V}_S\left(e\right),\mathscr{E}_S\left(e\right))$ connecting local transformers $\mathscr{V}_T\left(e\right)$ along a given road network link $e\in\mathscr{E}_R$ to the set of residences $\mathscr{V}_H=\mathscr{F}_M^{-1}\left(e\right)$ mapped to it. It has already been presented in an earlier work~\cite{rounak2020}; yet included here for the sake of completion. We drop the dependency on $e$ from every notation since we are discussing the same problem for each road link $e$. 

Note that we do not have the set of local transformers $\mathscr{V}_T$ to start with. Hence, we begin with a set of probable local transformers interpolated along the link and denoted by $\mathscr{V}_{\textrm{prob}}$. We need to connect the set of residences $\mathscr{V}_H$ to actual transformer nodes $\mathscr{V}_T\subseteq\mathscr{V}_{\textrm{prob}}$ in a forest of \emph{star-like} trees with each tree rooted at a transformer node. A candidate set of edges $\mathscr{E}_{D}$ between the residences and probable transformer nodes is chosen. The secondary network edges are to be selected from this candidate set. We construct an undirected graph $\mathscr{G}_{D}:=(\mathscr{V}_{D},\mathscr{E}_{D})$ with node set $\mathscr{V}_{D}=\mathscr{V}_{\textrm{prob}}\cup\mathscr{V}_H$ and edge set $\mathscr{E}_{D}$. We provide the formal problem statement to construct the secondary distribution network.

\begin{formal}[Secondary network creation problem]
Given the undirected graph $\mathscr{G}_D(\mathscr{V}_D,\mathscr{E}_D)$, find $\mathscr{E}_S\subseteq\mathscr{E}_D$ such that the induced subgraph network $\mathscr{G}_S(\mathscr{V}_S,\mathscr{E}_S)$ with $\mathscr{V}_S=\mathscr{V}_T\bigcup\mathscr{V}_H$ is a forest of starlike trees, $\mathscr{V}_T\subseteq\mathscr{V}_\textrm{prob}$ is the set of root nodes, and the overall length of the network is minimized.
\end{formal}
Note that a complete graph composed of the residence and transformer nodes can always be considered as the candidate set $\mathscr{E}_D$. In this paper, a Delaunay triangulation~\cite{delaunay} of the residential nodes is considered to reduce the size of the problem.
\begin{table}[htb]
\centering
\caption{Sets of nodes and edges in secondary network creation problem for each road link}
\label{tab:sets-sec}
\begin{tabular}{ll}
\hline
\textbf{Notation} & \textbf{Description} \\ \hline
$\mathscr{E}_D$ & Set of all candidate edges for the network \\
$\mathscr{E}_S$ & Set of chosen edges in the network \\
$\mathscr{V}_D$ & Set of all possible nodes in the network \\
$\mathscr{V}_\textrm{prob}$ & Set of all probable transformer nodes along link\\
$\mathscr{V}_T$ & Set of actual transformer nodes along link\\
$\mathscr{V}_H$ & Set of residence nodes mapped to the link \\ \hline
\end{tabular}
\end{table}

\noindent\textbf{Edge weight assignment} An edge $l:(i,j)\in\mathscr{E}_D$ is assigned a weight $w(i,j)$
\begin{equation*}
w(i,j)=
\begin{cases}
\infty,\quad\quad\quad\quad\quad\quad\quad\quad\textrm{if }i,j\in\mathscr{V}_\textrm{prob}\\
\mathsf{dist}(i,j)+\lambda \mathsf{C}(i,j),\quad\textrm{otherwise}
\end{cases}
\label{eq:weight}
\end{equation*}
where $\mathsf{dist}:\mathscr{V}_D\times\mathscr{V}_D\rightarrow\mathbb{R}$ denotes the geodesic distance between the nodes $i,j$. The function $\mathsf{C}(i,j)$ penalizes the cost function if the edge $l$ connecting nodes $i,j$ crosses the road network link $e$ and is defined as
$$
\mathsf{C}(i,j)=
\begin{cases}
0,\quad \textrm{if }i,j\textrm{ are on the same side of the road network link } e\\
2,\quad \textrm{if }i,j\textrm{ are on the opposite side of the road network link } e\\
1,\quad \textrm{if }i\in\mathscr{V}_\textrm{prob}\textrm{ or }j\in\mathscr{V}_\textrm{prob}
\end{cases}
$$
$\lambda$ is a weight factor to penalize multiple crossing of edges over the road links. It also penalizes multiple edges emerging from the root node. The weights are stacked in $|\mathscr{E}_D|$-length vector $\mathbf{w}$. Note that an edge between two probable transformer nodes is assigned a weight of \emph{infinity} which is equivalent to not considering them as candidate edges.

\noindent\textbf{Edge variables} We introduce binary variables $x_l\in\{0,1\}$ for each $l\in\mathscr{E}_D$. Variable $x_l=1$ indicates that the edge $l$ is present in the optimal topology and $x_l=0$ denotes otherwise.
Each edge $l:=(u,v)$ is assigned a flow variable $f_l$ (arbitrarily) directed from node $u$ to node $v$. The binary variable and flows can be respectively stacked in $|\mathscr{E}_D|$-length vectors $\mathbf{x}$ and $\mathbf{f}$.

\noindent\textbf{Node variables} The average hourly load demand at the $i^{th}$ residence node is denoted by $p_i$ and is strictly positive. We stack these average hourly load demands at all residence nodes in a $|\mathscr{V}_H|$-length vector $\mathbf{p}$. 
We also consider the reactive power load at each household. Since the models in~\cite{swapna_2018} did not explicitly model the reactive power demand at each residence, we consider a reactive power consumption of $q_i=\gamma p_i$ for each residence $i$. Here, $\gamma=\tan\left(\phi\right)$ with $\cos\phi$ denoting the power factor. For all cases, we consider a power factor of $0.95$ which renders $\gamma=0.33$. This is the addition we have made to the initial models which were proposed in~\cite{rounak2020}.

\noindent\textbf{Degree constraint.}~Statistical surveys on distribution networks in~\cite{mv_2011,review2017} show that residences along the secondary network are mostly connected in series with at most two neighbors. This is ensured by (\ref{eq:sec-degree}) which limits the degree of residence nodes to $2$.
\begin{equation}
    \sum_{l:(h,j)}x_{l}\leq 2,\quad \forall h\in\mathscr{V}_{H}\label{eq:sec-degree}
\end{equation}

\noindent\textbf{Power flow constraints.}~For the connected graph $\mathscr{G}_D(\mathscr{V}_D,\mathscr{E}_D)$, we define the $|\mathscr{E}_D|\times|\mathscr{V}_D|$ branch-bus incidence matrix $\mathbf{A}_{\mathscr{G}_D}$ with the entry along $l^{\textrm{th}}$ row and $k^{\textrm{th}}$ column as
\begin{equation*}
    \begin{matrix}
    \mathbf{A}_{\mathscr{G}_D}(l,k):=
    \begin{cases}
    ~~1,\quad k=i\\-1,\quad k=j\\~~0,\quad\textrm{otherwise}
    \end{cases}
    &\forall{l=(i,j)\in\mathscr{E}_D}
    \end{matrix}
    \label{eq:bus-incidence}
\end{equation*}
Since the order of rows and columns in $\mathbf{A}_{\mathscr{G}_D}$ is arbitrary, we can partition the columns as $\mathbf{A}_{\mathscr{G}_D}=\begin{bmatrix}\mathbf{A_{T}}&\mathbf{A_{H}}\end{bmatrix}$, without loss of generality, where the partitions are the columns corresponding to transformer and residence nodes respectively. We call $\mathbf{A_{H}}$ the reduced branch-bus incidence matrix.

Assuming no network losses, (\ref{eq:pf-sec1}) represents the power balance equations at all residence nodes. Note that the optimal network is obtained from $\mathscr{G}_D$ after removing the edges for which $x_l=0$. Therefore, we need to enforce zero flows $f_l$ for non-existing edges. The constraint (\ref{eq:pf-sec2}) performs this task along with constraining the flows $f_l$ for existing edges to be within pre-specified capacities $\overline{f}$.
\begin{subequations}
    \begin{align}
    &\mathbf{A_H}^T\mathbf{f}=\mathbf{p}\label{eq:pf-sec1}\\
    -&\overline{f}\mathbf{x}\leq \mathbf{f}\leq \overline{f}\mathbf{x}\label{eq:pf-sec2}
    \end{align}
    \label{eq:pf-sec}
\end{subequations}

\noindent\textbf{Ensuring radial topology} The radiality requirement of the secondary network $\mathscr{G}_S$ can be enforced from a known graph theory property: \emph{a forest with $n$ nodes and $m$ root nodes has $n-m$ edges}. In our case, $|\mathscr{V}_H|+|\mathscr{V}_T|$ nodes need to be covered in a forest of trees with $|\mathscr{V}_T|$ root nodes, which leads us to the following constraint.
\begin{equation}
    \sum_{l\in\mathscr{E}_D}x_l=|\mathscr{V}_H|\label{eq:radial}
\end{equation}

However, we need to ensure that there are no disconnected cycles in the optimal network. This can be done by ensuring that the residence points are connected to a transformer node~\cite{manish2019,lei2019radiality}. In our case, this condition is satisfied by the node power flow condition in (\ref{eq:pf-sec1}) if all the residential nodes consume non-zero power. 

\begin{proposition}
The graph $\mathscr{G}_S(\mathscr{V}_S,\mathscr{E}_S)$ with reduced branch-bus incidence matrix $\mathbf{A_H}$ (corresponding to columns of $\mathscr{V}_H$) and node power demand vector $\mathbf{p}\in\mathbb{R}^{|\mathscr{V}_H|}$, with strictly positive entries, has exactly $|\mathscr{V}_T|$ connected components if and only if there exists $\mathbf{f}\in\mathbb{R}^{|\mathscr{E}_D|}$, such that (\ref{eq:pf-sec1}) is satisfied.
\label{prop-1}
\end{proposition}
\begin{proof}
Proving by contradiction, suppose $\mathscr{G}_S(\mathscr{V}_S,\mathscr{E}_S)$ has more than $|\mathscr{V}_T|$ connected components and there exists $\mathbf{f}\in\mathbb{R}^{|\mathscr{E}_D|}$ satisfying the proposed equality. Therefore, there exists a connected component $\mathscr{G}_{C}(\mathscr{V}_{C},\mathscr{E}_{C})$ which is a maximal connected subgraph with $\mathscr{V}_{C}\subset\mathscr{V}_H$ and $\mathscr{V}_{C}\bigcap\mathscr{V}_T=\emptyset$. Let $\mathbf{A}_C$ denote the bus incidence matrix of ${\mathscr{G}}_{C}$. By definition, it holds that $\mathbf{A}_C\mathbf{1}=\mathbf{0}$. 

Since graph $\mathscr{G}_{C}(\mathscr{V}_{C},\mathscr{E}_{C})$ is a maximal connected subgraph of $\mathscr{G}_S$, there exists no edge $(i,j)$ with $i\in\mathscr{V}_C$ and $j\in\mathscr{V}_{\overline{C}}$, where $\mathscr{V}_{\overline{C}}=\mathscr{V}_S\setminus\mathscr{V}_{C}$. Since the order of rows and columns of $\mathbf{A_H}$ are arbitrary, we can partition without loss of generality as
\begin{equation}
    \mathbf{A_H}=\begin{bmatrix}\mathbf{A}_{\overline{C}}&\mathbf{0}\\\mathbf{0}&\mathbf{A}_{C}\end{bmatrix}\notag
\end{equation}
We can partition vectors $\mathbf{f}$ and $\mathbf{p}$ conformably to $\mathbf{A_H}$ to get the following equality 
\begin{equation}
  \begin{bmatrix}\mathbf{A}_{\overline{C}}&\mathbf{0}\\\mathbf{0}&\mathbf{A}_{C}\end{bmatrix}^T\begin{bmatrix}\mathbf{f}_{\overline{C}}\\\mathbf{f}_{C}\end{bmatrix}=\begin{bmatrix}\mathbf{p}_{{\overline{C}}}\\\mathbf{p}_{C}\end{bmatrix}\notag
\end{equation}
where $\mathbf{p}_{C},\mathbf{p}_{{\overline{C}}}$ are respectively the vectorized power demands for node sets $\mathscr{V}_C$ and $i\in\mathscr{V}_{\overline{C}}$. From the second block, it implies that 
\begin{equation}
\mathbf{p}_{C} = \mathbf{A}_{C}^T\mathbf{f}_{C} \Rightarrow \mathbf{1}^T\mathbf{p}_{C}=\mathbf{1}^T\mathbf{A}_{C}^T\mathbf{f}_{C}=\mathbf{0}\label{eq:claim}
\end{equation}
Since all entries of $\mathbf{p}$ are strictly positive from the initial assumption, we have $\mathbf{1}^T\mathbf{p}_{C}\neq0$ which contradicts (\ref{eq:claim}) and completes the proof. 
\end{proof}

\noindent\textbf{Generating optimal network topology}
The optimal secondary network is obtained after solving the optimization problem.
\begin{equation}
\begin{aligned}
    \min_{\mathbf{x}}&~\mathbf{w}^T\mathbf{x}\\
    \textrm{s.to.}&~(\ref{eq:sec-degree}),(\ref{eq:pf-sec}),(\ref{eq:radial})
\end{aligned}
\label{eq:sec-opt}
\end{equation}
Though the mixed integer optimization problem (MILP) in (\ref{eq:sec-opt}) has been relaxed, the time complexity depends on the size of the edge set $\mathscr{E}_D$. The computation time can be significantly reduced by considering the edges of a Delaunay triangulation of nodes $\mathscr{V}_H$ as the edge set instead of choosing $\binom{|\mathscr{V}_H|}{2}$ combinatorial edges between the nodes. The MILP is solved individually for each road link $e\in\mathscr{E}_R$ to construct the secondary network $\mathscr{G}_S\left(e\right)$ corresponding to the road link. The overall secondary network is a combination of such generated networks.

\begin{figure}[tbhp]
\centering
	\includegraphics[width=0.98\textwidth]{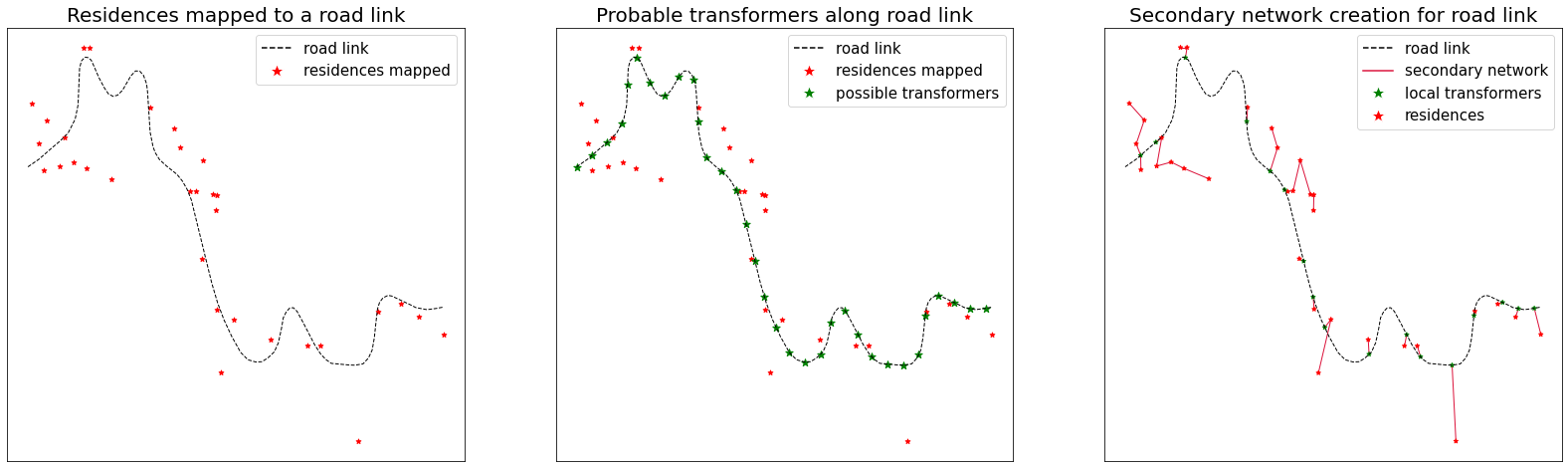}
	\caption{Plots showing steps in creating the secondary network for a road network link. First the probable transformer locations are identified along the link. Thereafter, the network is generated by solving the optimization problem (\ref{eq:sec-opt}). The network originates from the road link and connects residences mapped to it in a forest of star-like trees rooted at the transformers.}
	\label{fig:secondary}
\end{figure}

\noindent\textbf{Implementation examples.}~Here we present an example of the secondary distribution network created for the state of Virginia. The steps involved in constructing the synthetic secondary network connecting residences along a road network link are shown in Fig.~\ref{fig:secondary}. The first figure shows the residences (red) mapped to the road link and the second figure shows the probable transformers (green points) along the link. A Delaunay triangulation is considered to connect the points and obtain the set of possible edges. Thereafter, the MILP (\ref{eq:sec-opt}) is solved which identifies the optimal set of edges as shown in the third figure. We observe that all residences are connected in star-like trees with roots as the transformers along the road link.
\FloatBarrier

\subsection*{Step 2a: Map local transformers to the nearest substation}\label{ssec:voronoi}
The secondary network results in local transformer nodes $\mathscr{V}_T=\bigcup_{e\in\mathscr{E}_R}\mathscr{V}_T\left(e\right)$ along the road network links of the geographic region (in our case, throughout the state of Virginia). The task of primary network creation is to connect these transformer nodes to substation nodes in the region. In order to tackle the size of the problem, we partition the task into a number of subtasks. We perform this partition such that each substation is required to connect only the transformer nodes near to it. Therefore, we produce a map $\mathscr{F}_V$ from the set of local transformers $\mathscr{V}_T$ to the set of substations $\mathscr{S}$. Since we use the road network as a proxy for creating primary networks, we map each local transformer to the nearest substation along the road network. 
\begin{algorithm}[tbhp]
	\caption{Map local transformer to nearest substation.}
	\label{alg:voronoi}
	\textbf{Input} Set of substations $\mathscr{S}$, set of local transformers $\mathscr{V}_T$, road network $\mathscr{G_R}\left(\mathscr{V}_R,\mathscr{E}_R\right)$.
	\begin{algorithmic}[1]
		\For {each substation $s\in\mathscr{S}$}
		\State Find the nearest road network node $v_s\in\mathscr{V}_R$.
		\EndFor
		\For {each local transformer $t\in\mathscr{V}_T$}
		\State Compute the nearest substation using (\ref{eq:near-sub}).
		\EndFor
	\end{algorithmic}
	\textbf{Output} Mapping~$\mathscr{F}_V \colon \mathscr{V}_T \rightarrow \mathscr{S}$
\end{algorithm}

Algorithm~\ref{alg:voronoi} details the steps involved in creating this mapping. First, we find the geographically located nearest road network node $v_s\in\mathscr{V}_R$ for each substation $s\in\mathscr{S}$. Let $\mathsf{NetDist}(u,v)$ denotes the shortest path distance between nodes $u,v\in\mathscr{V}_R$ along road network $\mathscr{G}_R$. We identify the nearest substation to each local transformer using,
\begin{equation}
    \mathscr{F}_V\left(v\right) = \textrm{arg}\min_{s\in\mathscr{S}}\mathsf{NetDist}\left(v_s,v\right)\label{eq:near-sub}
\end{equation}

\begin{figure}[tbhp]
\centering
    \includegraphics[width=0.9\textwidth]{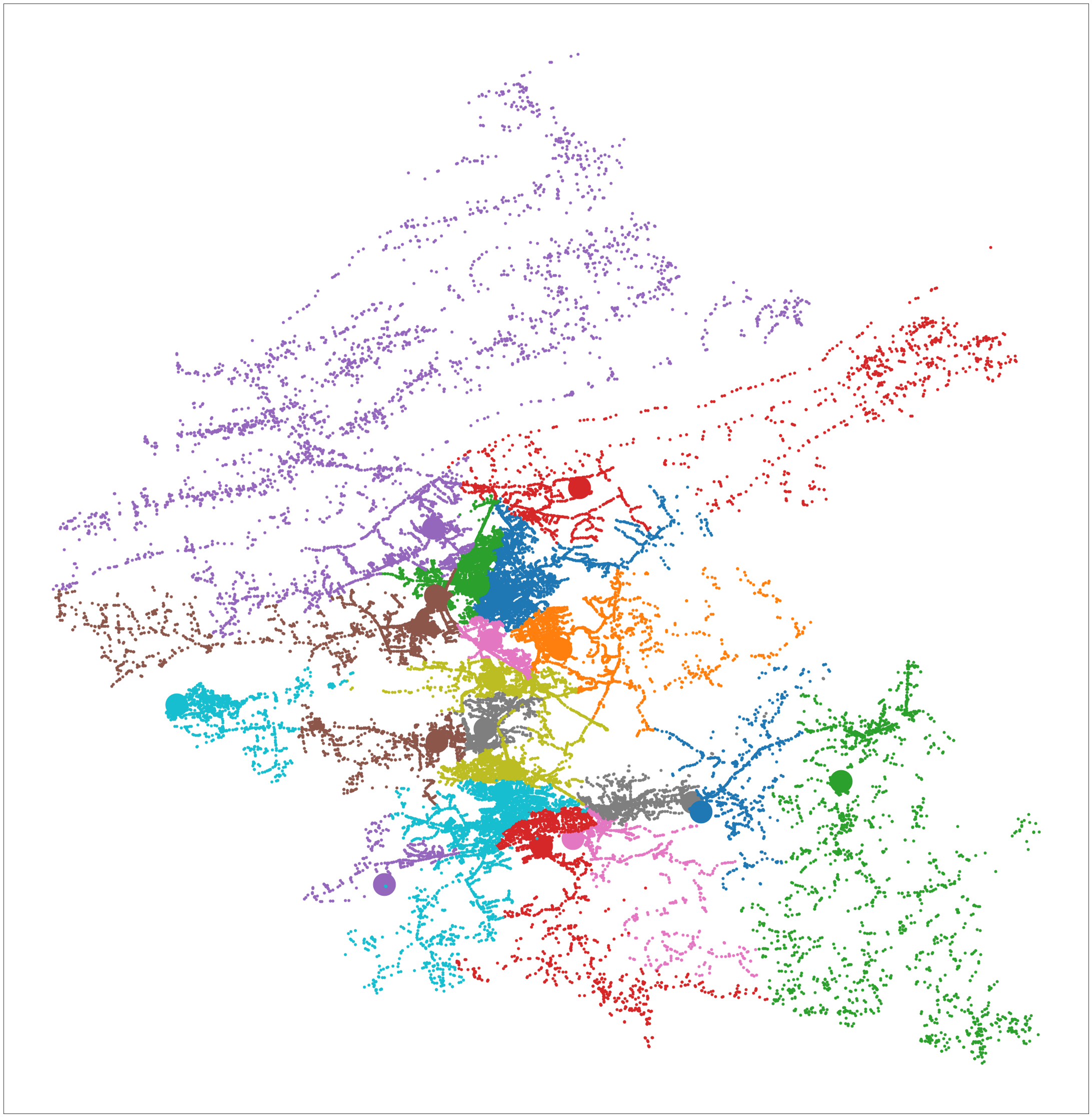}
	\caption{Plot showing the mapping between local transformer nodes and substations in Montgomery County of southwest Virginia. Each color depicts a particular set of transformer nodes mapped to a substation. Note that same color has been used to denote more than one mapping. The primary network creation problem (in the succeeding step) would generate a primary network connecting the substation to the mapped transformer nodes.}
	\label{fig:voronoi}
\end{figure}
\FloatBarrier
The inverse mapping $\mathscr{F}_V^{-1}\left(s\right)$ assigns a set of local transformer nodes to each substation $s\in\mathscr{S}$. These sets of local transformer nodes are mutually exclusive and are exhaustive. We perform the primary network creation for each of these sets individually.

\noindent\textbf{Implementation example.}~The secondary network creation step generates local transformer nodes all over the footprint of the region under consideration, which in our case, is the Montgomery County of southwest Virginia. Next, we perform the mapping of the transformer nodes to the nearest substation along the road network. We show the partitions in Fig.~\ref{fig:voronoi} where each color denotes a partition of the transformer nodes.

\subsection*{Step 2b: Connect local transformers to the substation}
\begin{problem}[$\mathcal{P}_{\textrm{prim}}$ construction]
Given a substation~$s\in\mathscr{S}$ with an assigned set of local transformer nodes~$\mathscr{F}_V^{-1}\left(s\right)$, construct a tree network $\mathscr{G}_P\left(s\right)$ using the road network $\mathscr{G}_R$ as a proxy which connects all local transformers while ensuring acceptable node voltages by power engineering standards.
\end{problem}
Here we present the detailed optimization framework for the problem $\mathcal{P}_{\textrm{prim}}\left(s\right)$ to create the primary distribution network $\mathscr{G}_P\left(s\right)(\mathscr{V}_P\left(s\right),\mathscr{E}_P\left(s\right))$ connecting substation $s$ to local transformers $\mathscr{V}_T=\mathscr{F}_V^{-1}\left(s\right)$ mapped to it. We drop the dependency on $s$ from every notation since we are discussing the same problem for each substation $s$. 

We start with an undirected graph $\mathscr{G}_R(\mathscr{V},\mathscr{E}_R)$ which is the road network subgraph induced from the mapped transformer nodes. Here $\mathscr{V}=\mathscr{V}_R\bigcup\mathscr{V}_T$ comprises of transformer nodes as well as road nodes. The goal of the primary network creation problem is to select edge set $\mathscr{E}_P\subseteq\mathscr{E}_R$ to generate the optimal primary network $\mathscr{G}_P(\mathscr{V}_P,\mathscr{E}_P)$. Note that all transformer nodes are required to be included. On the contrary, road nodes are dummy points with no load and are only used to connect the local transformers. For example, we cannot have a road node as a leaf node since it does not connect transformer nodes. The node set $\mathscr{V}_P=\mathscr{V}_R^\star\bigcup\mathscr{V}_T$ comprises road and transformer nodes respectively where $\mathscr{V}_R^\star\subseteq\mathscr{V}_R$ are the selected road nodes.

In all practical distribution networks, multiple feeder lines originate from a substation and connect the local transformers. To this end, we construct the primary network $\mathscr{G}_P$ as a forest of trees with the root of each tree connected to the substation through high voltage feeder lines. These root nodes are road network nodes with no loads connected to them directly. Note that some nodes in $\mathscr{V}_R^\star$ which are the root nodes in the constructed primary network. Here, we present the formal problem statement for constructing the primary distribution network.
\begin{formal}[Primary network creation problem]
Given a connected network $\mathscr{G}_R(\mathscr{V},\mathscr{E}_R)$ where $\mathscr{V}=\mathscr{V}_R\bigcup\mathscr{V}_T$, find $\mathscr{E}_P\subseteq\mathscr{E}_R$ such that the induced subgraph network $\mathscr{G}_P(\mathscr{V}_P,\mathscr{E}_P)$ is a forest of trees with each tree rooted at some $r\in\mathscr{V}_R^\star$ where $\mathscr{V}_P=\mathscr{V}_R^\star\bigcup\mathscr{V}_T$ and $\mathscr{V}_R^\star\subseteq\mathscr{V}_R$.
\end{formal}

\begin{table}[tbhp]
\centering
\caption{Sets of nodes and edges in primary network creation problem}
\label{tab:sets}
\begin{tabular}{ll}
\hline
\textbf{Notation} & \textbf{Description} \\ \hline
$\mathscr{E}_R$ & Set of all candidate edges in the network \\
$\mathscr{V}$ & Set of all nodes in the network \\
$\mathscr{V}_T$ & Set of all transformer nodes in the network \\
$\mathscr{V}_R$ & Set of all road nodes in the network \\ 
$\mathscr{E}_P$ & Set of all chosen edges in the optimal primary network \\
$\mathscr{V}_P$ & Set of all chosen nodes in the optimal primary network \\
$\mathscr{V}_R^\star$ & Set of all chosen road nodes in the optimal primary network\\ \hline
\end{tabular}
\end{table}

\begin{table}[tbhp]
\centering
\caption{Binary variables in primary network creation problem}
\label{tab:bin-var}
\begin{tabular}{ll}
\hline
\textbf{Notation} & \textbf{Description} \\ \hline
$x_e$ & 1 if edge $e\in\mathscr{E}_R$ is included in optimal network \\
$y_r$ & 1 if road node $r\in\mathscr{V}_R$ is included in optimal network \\
$z_r$ & 0 if road node $r\in\mathscr{V}_R$ is a root node \\ \hline
\end{tabular}
\end{table}

\noindent\textbf{Binary variables.}~We assign a binary variable $x_e\in\{0,1\}$ for each edge $e\in\mathscr{E}_R$. Variable $x_e=1$ indicates that the edge $e$ is included in the network, and $x_e=0$ otherwise. Further, we introduce binary variables $y_r,z_r\in\{0,1\}$ for each road network node $r\in\mathscr{V}_R$. Variable $y_r=1$ indicates that road node $r$ is part of the primary network and vice versa. A selected road node ($y_r=1$) may be chosen to be a root node or otherwise. Binary variable $z_r=0$ indicates that road node $r$ is a root node, and $z_r=1$ implies that $r$ is not a root node. Note that an unselected road node ($y_r=0$) needs to be treated as a non-root node ($z_r=1$) and hence 
\begin{equation}
    1-z_r\leq y_r,\quad \forall r\in\mathscr{V}_{R}\label{eq:sup-non-root}
\end{equation}

\noindent\textbf{Topology constraints.}~We need to ensure the following: (i) a selected non-root road node should not be a terminal node, (ii) the network does not consist of any cycles, and (iii) the network covers all transformer nodes.

Let $e=(r,j)\in\mathscr{E}_R$ denote an edge which is incident on the road node $r\in\mathscr{V}_{R}$. The degree of a road node $r$ in graph $\mathscr{G}_R$ is denoted by $\sum_{e:(r,j)}x_{e}$. For the first condition to hold true, we need to ensure that the degree of a selected non-root road node (with $y_r=1,z_r=1$) is at least $2$, and the degree of an unselected road node (with $y_r=0$) is $0$. Finally, the degree of a selected road node which is also a root node (with $y_r=1,z_r=0$) is positive. This is ensured through the following inequalities.
\begin{subequations}
\begin{align}
    \sum_{e=(r,j)}&x_{e}\leq |\mathscr{E}_R|y_r, &\forall r\in\mathscr{V}_{R}\label{eq:sup-non-chosen}\\
    \sum_{e:(r,j)}&x_{e}\geq2(y_r+z_r-1), &\forall r\in\mathscr{V}_{R}\label{eq:sup-transfer}\\
    \sum_{e=(r,j)}&x_{e}\geq y_r, &\forall r\in\mathscr{V}_{R}\label{eq:sup-chosen}
    \end{align}
    \label{eq:sup-prim-connectivity}
\end{subequations}

To enforce the second condition (no cycles in the network) or `radiality' condition, we use results from graph theory. We know that a forest with $n$ nodes and $m$ components has $n-m$ edges. In our case, the total number of nodes is $|\mathscr{V}_T|+\sum_{r\in\mathscr{V}_R}y_r$, while the number of components is the number of root nodes, i.e, $\sum_{r\in\mathscr{V}_R}(1-z_r)$. Therefore, the radiality constraint is given below 
\begin{equation}
    \sum_{e\in\mathscr{E}_R}x_e=|\mathscr{V}_T|+\sum_{r\in\mathscr{V}_{R}}y_r-\sum_{r\in\mathscr{V}_{R}}(1-z_r)\label{eq:sup_prim_radial}
\end{equation}
However, this is not a sufficient condition for radiality since it does not avoid the formation of disconnected components with cycles. We can extend Proposition 1 in~\cite{rounak2020} and ensure radiality with the power balance equations at each node in the network.

\begin{table}[htb]
\centering
\caption{Power flow variables and parameters in primary network creation problem}
\label{tab:pf-var}
\begin{tabular}{ll}
\hline
\textbf{Notation} & \textbf{Description} \\ \hline
$V_j=v_je^{\mathsf{j}\theta_j}$ & complex bus voltage phasor at bus $j\in\mathscr{V}$; magnitude: $v_j$, angle: $\theta_j$ \\
$I_e$ & complex current flowing through edge $e\in\mathscr{E}_R$ \\
$S_{ij}=P_{ij}+\mathsf{j}Q_{ij}$ & complex power flowing through edge $e=(i,j)\in\mathscr{E}_R$ from node $i$ to node $j$; real power flow: $P_{ij}$, reactive power flow: $Q_{ij}$ \\
$s_j=p_j+\mathsf{j}q_j$ & complex power injection at bus $j\in\mathscr{V}$; real power injection: $p_j$, reactive power injection: $q_j$ \\
$Z_e=R_e+\mathsf{j}X_e$ & complex impedance of edge $e\in\mathscr{E}_R$; resistance: $R_e$, reactance: $X_e$ \\
$f_e$ & power flow limit in edge $e\in\mathscr{E}_R$ \\
$\overline{s}_j$ & substation feeder rating \\
$\underline{v},\overline{v}$ & lower and upper limits of bus voltage\\ \hline
\end{tabular}
\end{table}
\noindent\textbf{Power balance constraints.}~The complex power injection at bus $j\in\mathscr{V}$ is denoted by $s_j=p_j+\mathsf{j}q_j$. Note that the power injections are the difference between generation and load demand at each bus. For transformer nodes, the power injection is the negative load demand; for non-root road nodes, the power injection is zero; and for root nodes, the power injection is bounded by a defined limit. It is assumed that the root nodes are connected to the substation through high voltage feeders. Therefore, the feeder capacity can be considered as the defined limit.  
\begin{subequations}
\begin{align}
    \sum_{e:(i,j)}x_eS_{ij}-\sum_{e:(j,k)}x_eS_{jk}=-s_j,\quad&\forall j\in\mathscr{V}_T\label{seq:sup-trans-bal}\\
    \sum_{e:(i,j)}x_eS_{ij}-\sum_{e:(j,k)}x_eS_{jk}=0,\quad&\forall j\in\mathscr{V}_R,z_j=1\label{seq:sup-road-bal}\\
    \bigg|\sum_{e:(i,j)}x_eS_{ij}-\sum_{e:(j,k)}x_eS_{jk}\bigg|\leq \overline{s}_j,\quad&\forall j\in\mathscr{V}_R,z_j=0\label{seq:sup-root-bal}
\end{align}
\label{eq:sup-prim-pbal}
\end{subequations}
In the secondary network creation method, we discuss about the residential reactive power model. We assumed a constant power factor of $0.95$ for all residences. This leads to a linear relationship between the real and reactive power throughout the network with $Q_{ij}=\gamma P_{ij}$ for all edges. We can drop either of the real or reactive power terms in (\ref{eq:sup-prim-pbal}), since they would lead to the same equality constraints. Such assumptions are standard for problems related to network planning~\cite{mv_2011,manish2019,lei2019radiality}. Therefore, we can rewrite (\ref{eq:sup-prim-pbal}) by dropping the reactive power terms (imaginary parts) and combining (\ref{seq:sup-road-bal}) and (\ref{seq:sup-root-bal}) to get:
\begin{subequations}
\begin{align}
    &\sum_{e:(i,j)}P_{ij}-\sum_{e:(j,k)}P_{jk}=-p_j,\quad\forall j\in\mathscr{V}_T\label{seq:sup-trans-bal-relax}\\
    &-\overline{p}_j(1-z_j)\leq\sum_{e:(i,j)}P_{ij}-\sum_{e:(j,k)}P_{jk}\leq \overline{p}_j(1-z_j),\forall j\in\mathscr{V}_R\label{seq:sup-road-pbal-relax}
\end{align}
\label{eq:sup-prim-pbal-relax}
\end{subequations}

\noindent\textbf{Power flow constraints.}~We define complex node voltage $V_j=v_je^{\mathsf{j}\theta_j}$ (magnitude $v_j$ and angle $\theta_j$) for each node~$j\in\mathscr{V}$ and complex power $S_{i,j}=P_{i,j}+\mathsf{j}Q_{i,j}$ for each edge $e=(i,j)\in\mathscr{E}_R$ flowing from node $i$ to $j$, where $P_{i,j}$ and $Q_{i,j}$ respectively denote real and reactive power flowing along edge $e\in\mathscr{E}_R$. The current flowing through the edge is defined by $I_e$ and the complex impedance of the edge is denoted by $Z_e=R_e+\mathsf{j}X_e$. 
The following equations/inequalities describe the relationship between the variables and the associated constraints. Note that the constraints are only activated for those edges which are selected in the optimal network. Therefore, we have introduced the binary variable~$x_e$ in~(\ref{seq:sup-ohm}). We assume that HV feeder lines from the substation to the root nodes in the optimal primary distribution network end in voltage regulators, which ensure that the root nodes have a voltage of $1.0$pu.
\begin{subequations}
\begin{align}
    &x_e(V_i-V_j-Z_eI_e)=0,\quad&\forall e\in\mathscr{E}_R\label{seq:sup-ohm}\\
    &S_{ij}=V_jI_e^{\star},\quad&\forall e\in\mathscr{E}_R\label{seq:sup-complex}\\
    &|S_{ij}|\leq\overline{f}_e,\quad&\forall e\in\mathscr{E}_R\label{seq:sup-slim}\\
    &\underline{v}\leq v_j\leq\overline{v},\quad&\forall j\in\mathscr{V}\label{seq:sup-voltlim}\\
    &v_j=1,\quad&\forall j\in\mathscr{V},z_j=0\label{seq:sup-subvolt}
\end{align}
\label{eq:sup-prim-pf}
\end{subequations}

\noindent\textbf{Generating optimal primary network.} Each edge $e=(i,j)\in\mathscr{E}_R$ is assigned a weight $w_e=w(i,j)=\mathsf{dist}(i,j)$ which is the geodesic distance between the nodes. Additionally, for every road node $j\in\mathscr{V}_{R}$, we compute its geodesic distance from the substation $s$, denoted by $\mathsf{d}_j$. The optimal primary network topology is obtained by solving the optimization problem:
\begin{equation}
\begin{aligned}
    \min_{\mathbf{x},\mathbf{y},\mathbf{z}}&~\sum_{e\in\mathscr{E}_R}x_ew_e+\sum_{j\in\mathscr{V}_{R}}(1-z_j)\mathsf{d}_j\\
    \textrm{s.to.}&~(\ref{eq:sup-non-root}),(\ref{eq:sup-prim-connectivity}),(\ref{eq:sup_prim_radial}),(\ref{eq:sup-prim-pbal-relax}),(\ref{eq:sup-prim-pf})
\end{aligned}
\label{eq:sup-prim-prob}
\end{equation}

\noindent\textbf{Relaxing non-linear power flow constraints.}~Note that power flow constraints in (\ref{eq:sup-prim-pf}) are non-convex because of the quadratic equality constraint in (\ref{seq:sup-complex}) and the bilinear terms in (\ref{seq:sup-ohm}). First, we consider relaxing the quadratic equality constraint. Multiplying by the complex conjugates in (\ref{seq:sup-ohm}) and (\ref{seq:sup-complex}), we get the following result:
\begin{equation}
    v_j^2=v_i^2-2R_eP_{ij}-2X_eQ_{ij}+(R_e^2+X_e^2)|I_e|^2\label{eq:sup-pf-relax}
\end{equation}
Assuming small $R_e,X_e$, the current magnitudes can be eliminated. Note that (\ref{eq:sup-pf-relax}) is still non-linear in $v_i,v_j$; however, it is linear in squared magnitude. This relaxed model is known in the literature as the Linearized Distribution Flow (LDF) model. 
The squared voltage is often approximated as $|v_i|^2\approx 2v_i-1$ which leads to the relation $v_j=v_i-R_eP_{i,j}-X_eQ_{i,j}$~\cite{ldf}. 
Notice that the constraint in (\ref{eq:sup-pf-relax}) is required to be enforced on only those edges $e\in\mathscr{E}_R$ for which $x_e=1$, which leads us to the relaxed version of (\ref{seq:sup-ohm}) as
\begin{equation}
    x_e(v_i-v_j-R_eP_{ij}-X_eQ_{ij})=0 \;.\label{eq:sup-relax-ohm}
\end{equation}
Here (\ref{eq:sup-relax-ohm}) is non-convex due to the bi-linear terms in the equality. In order to deal with this non-convexity, McCormick relaxation has been used widely in several previous works~\cite{manish2018OptimalDS,manish2019}. In general, McCormick relaxation replaces the non-convex equality constraint with its convex envelope~\cite{mccormick}. However, in the case of bilinear variables with at least one binary variable, this relaxation becomes exact. The convex relaxed version of (\ref{eq:sup-prim-pf}) is:
\begin{subequations}
\begin{align}
    -&(1-x_e)M\leq v_i-v_j-R_eP_{ij}-X_eQ_{ij}\leq(1-x_e)M,~\forall e\in\mathscr{E}_R\label{eq:sup-volt-edge}\\
    -&\overline{f}_e{x_e}\leq {P_{ij}}\leq \overline{f}_e\mathbf{x_e},~\quad\forall e\in\mathscr{E}_R\label{eq:sup-pf-flow}\\
    &(1-v_r)\leq z_r,~\quad\forall r\in\mathscr{V}_{R}\label{eq:sup-volt-root}\\
    &\underline{v}\leq {v_j}\leq \overline{v},~\quad\forall j\in\mathscr{V}\label{eq:sup-volt-limit}
    \end{align}
    \label{eq:sup-prim-pf-relax}
\end{subequations}

The overall relaxed optimization problem for creating the primary network topology is:
\begin{equation}
\begin{aligned}
    \min_{\mathbf{x},\mathbf{y},\mathbf{z}}&~\sum_{e\in\mathscr{E}_R}x_ew_e+\sum_{j\in\mathscr{V}_{R}}(1-z_j)d_j\\
    \textrm{s.to.}&~(\ref{eq:sup-non-root}),(\ref{eq:sup-prim-connectivity}),(\ref{eq:sup_prim_radial}),(\ref{eq:sup-prim-pf-relax}),(\ref{eq:sup-prim-pbal-relax})
\end{aligned}
\label{eq:sup-prim-relax}
\end{equation}

\begin{figure}[tbhp]
\centering
    \includegraphics[width=0.32\textwidth]{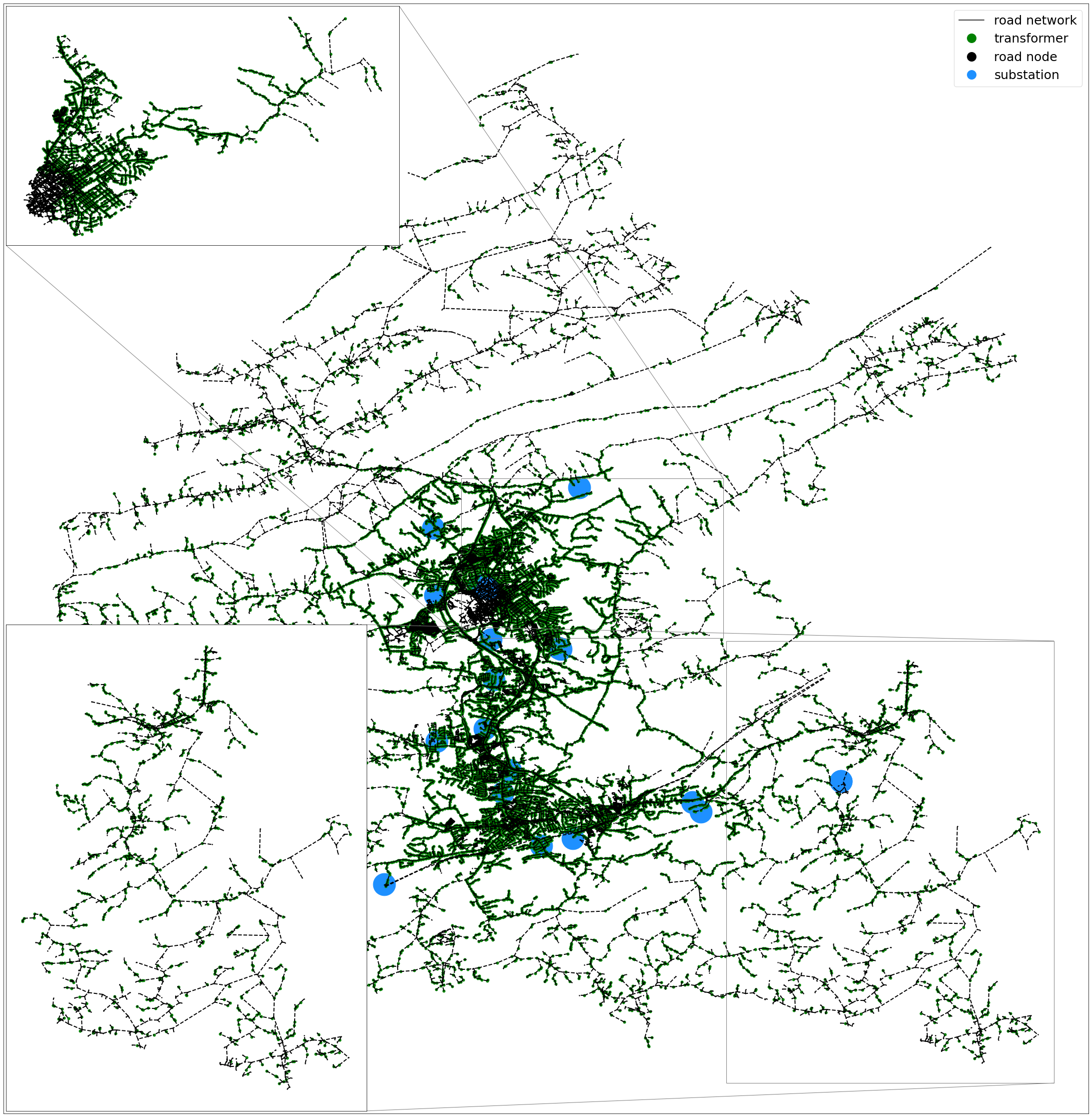}
	\includegraphics[width=0.25\textwidth]{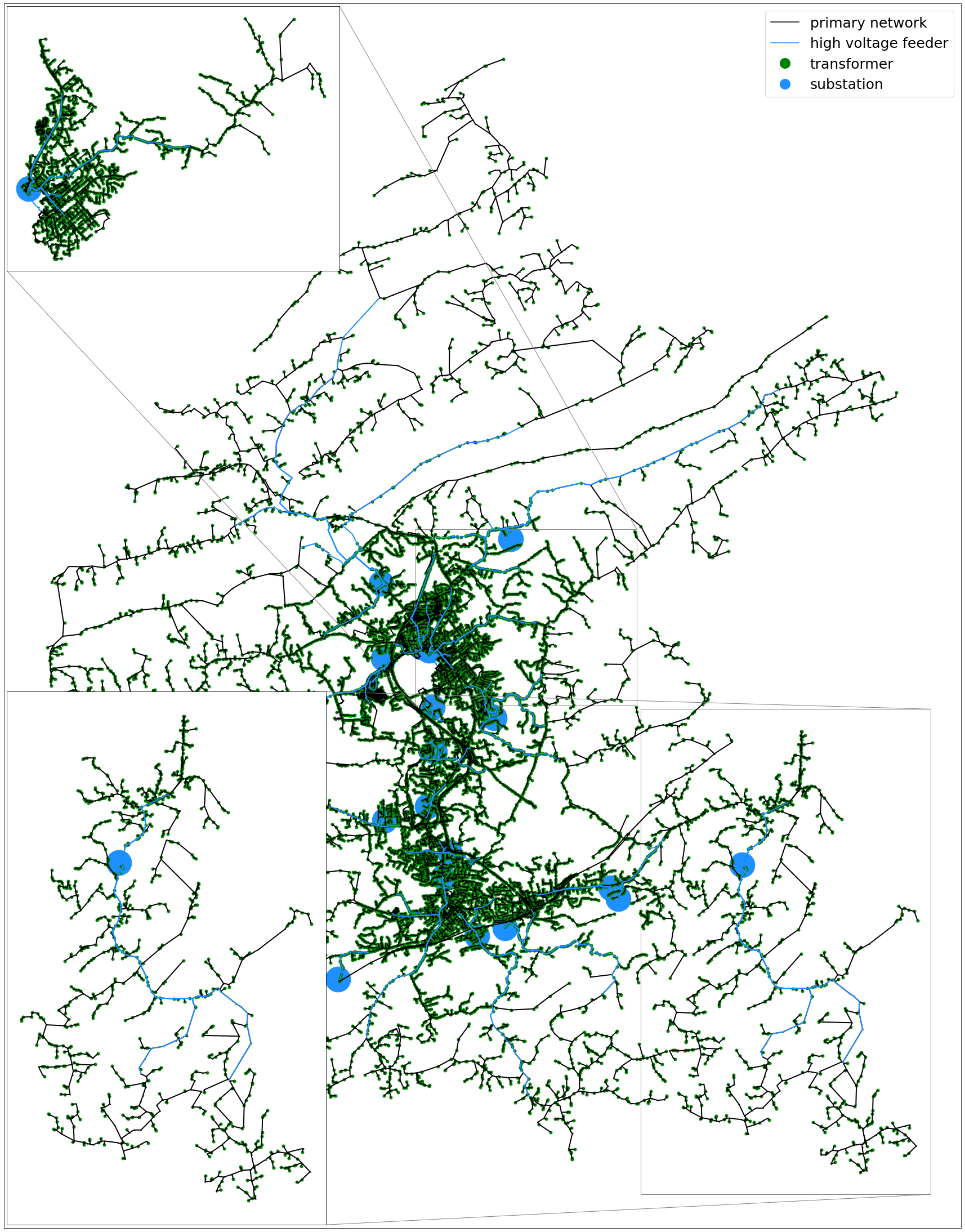}
	\includegraphics[width=0.25\textwidth]{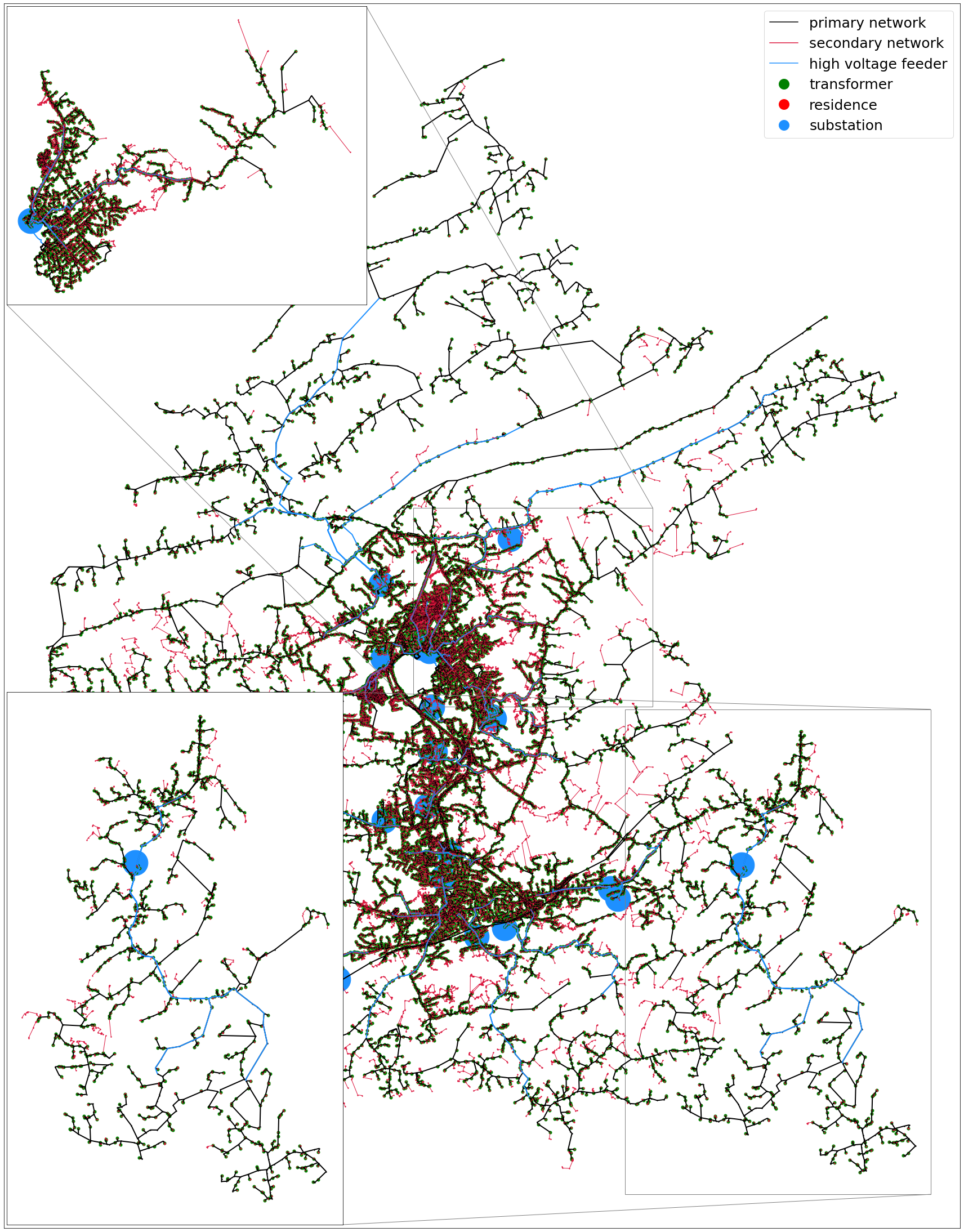}
	\caption{Plots showing the steps involved in creating the primary network of Montgomery County in southwest Virginia. The edges of the primary network (middle figure) are chosen from the road network (left figure) by solving the optimization problem in (\ref{eq:sup-prim-relax}). The network originates from the substations and follows the road links to connect the local transformers in a tree structure with no loops. The secondary network is appended to the local transformers to obtain the final synthetic distribution network (right figure).}
	\label{fig:primary}
\end{figure}

\noindent\textbf{Implementation example.}~Next, we show the steps involved in creating the primary distribution network within the boundary of Montgomery County of Virginia, US in Fig.~\ref{fig:primary}. We use the road network obtained from Open Street Maps as a proxy for the primary network, i.e., the edges in the primary network are chosen from the road network edges by solving the optimization problem in (\ref{eq:sup-prim-relax}). The first figure shows the road network which is used to construct the primary network. The zoomed-in inset figures show two examples of road network subgraphs assigned to two substations. Note that this assignment is computed from the inverse mapping $\mathscr{F}_V^{-1}$ in the preceding step. The second figure shows the solution of the optimization problem where the primary network is created from the set of edges in the road network. The primary network is a tree originating from a substation and connects the local transformers through multiple feeders. Note that the road network in the first figure had multiple cycles, which are no longer present in the primary network in the second figure. Finally, the secondary network is appended to the local transformers to obtain the final distribution network, which is shown in the third figure.
\FloatBarrier

\subsection*{Step 3: Constructing an ensemble of networks}
\begin{figure}[tbhp]
\centering
	\includegraphics[width=0.28\textwidth]{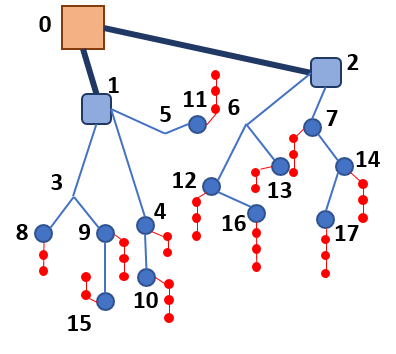}
	\includegraphics[width=0.28\textwidth]{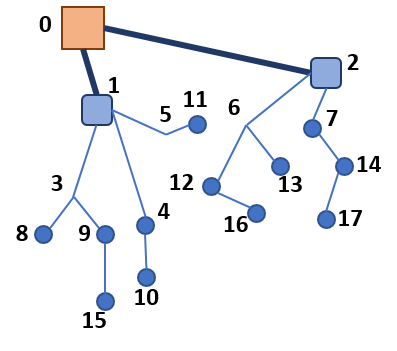}
	\includegraphics[width=0.28\textwidth]{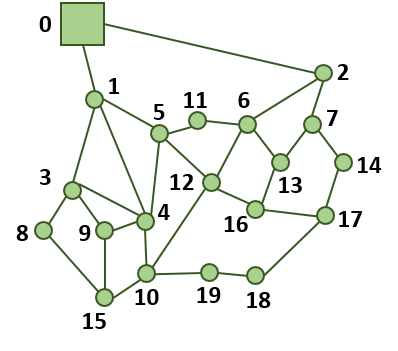}
	\caption{Plots summarizing the optimal distribution network constructed using Steps 1 and 2. The overall network (left figure) consists of the primary network (blue) and secondary network (red). The optimal primary network (middle figure) is constructed from the underlying road network (right figure). The edges in the primary network are a subset of the road network graph edges.}
	\label{fig:dist}
\end{figure}
\FloatBarrier
Using the optimization frameworks in (\ref{eq:sec-opt}) and (\ref{eq:sup-prim-relax}), we create the optimal (sometimes near-optimal) distribution network as shown in the left figure of Fig.~\ref{fig:dist}. 
It consists of the primary network which connects the substation to the local transformers in a tree structure, and these transformers are connected to the residences in chains, forming the secondary network. The middle figure of Fig.~\ref{fig:dist} shows the near-optimal primary network denoted by $\mathscr{G}_P^{0}(\mathscr{V}_P^{0},\mathscr{E}_P^{0})$. Note that the MILP (\ref{eq:sup-prim-relax}) identified $\mathscr{G}_P^{0}(\mathscr{V}_P^{0},\mathscr{E}_P^{0})$ as the solution by choosing edges from the underlying road network $\mathscr{G}_R(\mathscr{V}_R,\mathscr{E}_R)$. The right figure of Fig.~\ref{fig:dist} shows the underlying road network corresponding to the primary network. We note that all primary network edges are selected from the road network graph. The resulting primary network is a tree, and it connects all the transformer nodes. The road network nodes $\{3,5,6\}$ are used to construct the network, whereas some other road network nodes such as~$\{18,19\}$ are not included in the generated primary network.

We model the variant network creation process as a Markov chain $\mathcal{M}$ where each state denotes a valid realization of the primary network. 
The transitions $\mathscr{G}_P^{t}(\mathscr{V}_P^{t},\mathscr{E}_P^{t})\rightarrow \mathscr{G}_P^{t+1}(\mathscr{V}_P^{t+1},\mathscr{E}_P^{t+1})$ of $\mathcal{M}$ are specified in the following manner. Let $s$ denote the substation node in the network $\mathscr{G}_P^{t}(\mathscr{V}_P^{t},\mathscr{E}_P^{t})$ and $\mathscr{N}(s)$ denote the neighbors of $s$. 
We choose a random set of edges $\mathscr{E}_D=\{e_d\in\mathscr{E}_P^{t}|e_d=(u,v),u\neq v\neq s\}$ to be deleted and thereafter solve a restricted version of (\ref{eq:sup-prim-relax}) with the following additional constraints.
\begin{subequations}
\begin{align}
    & x_e=0,\textrm{ for }e\in\mathscr{E}_D;\quad x_e=1,\forall e\in\mathscr{E}_P^{t}\setminus\mathscr{E}_D\label{seq:restrict-edge}\\
    & y_r=1,\forall r\in\mathscr{V}_R\cap\mathscr{V}_P^{t}\label{seq:restrict-road}\\
    & z_r=0,\forall r\in\mathscr{N}(s);\quad z_r=1,\forall r\in\mathscr{V}_R\setminus\mathscr{N}(s)\label{seq:restrict-root}
\end{align}
\label{eq:restrict}
\end{subequations}
\begin{algorithm}
	\caption{Constructing ensemble of networks.}
	\label{alg:markov}
	\textbf{Input} Near-optimal primary network $\mathscr{G}_P^{0}(\mathscr{V}_P^{0},\mathscr{E}_P^{0})$, optimal secondary network $\mathscr{G}_S$, underlying road network $\mathscr{G}_R(\mathscr{V}_R,\mathscr{E}_R)$, size of ensemble $N$.
	\begin{algorithmic}[1]
	\State Initialize network count, $t\leftarrow 0$
	\While {$t<N$}
	\State Choose a random set of edges $\mathscr{E}_D\subset \mathscr{E}_P^{t}$ to be deleted.
	\State Formulate optimization problem $\mathcal{P}_t$ by combining (\ref{eq:sup-prim-relax}) with constraints in (\ref{eq:restrict}).
	\If {$\mathcal{P}_t$ is feasible}
	\State Solve $\mathcal{P}_t$ to get the variant primary network $\mathscr{G}_P^{t+1}(\mathscr{V}_P^{t+1},\mathscr{E}_P^{t+1})$
	\State Increment network count $t\leftarrow t+1$.
	\EndIf
	\State Augment the secondary network to the variant primary network; $\mathscr{G}^{t}\leftarrow \mathscr{G}_S\bigcup\mathscr{G}_P^t(\mathscr{V}_P^{t},\mathscr{E}_P^{t})$
	\EndWhile
	\end{algorithmic}
	\textbf{Output} Ensemble of networks $\mathscr{G}^{1},\mathscr{G}^{2},\cdots,\mathscr{G}^{N}$
\end{algorithm}

\begin{figure*}[tbhp]
    \centering
    \includegraphics[width=0.90\textwidth]{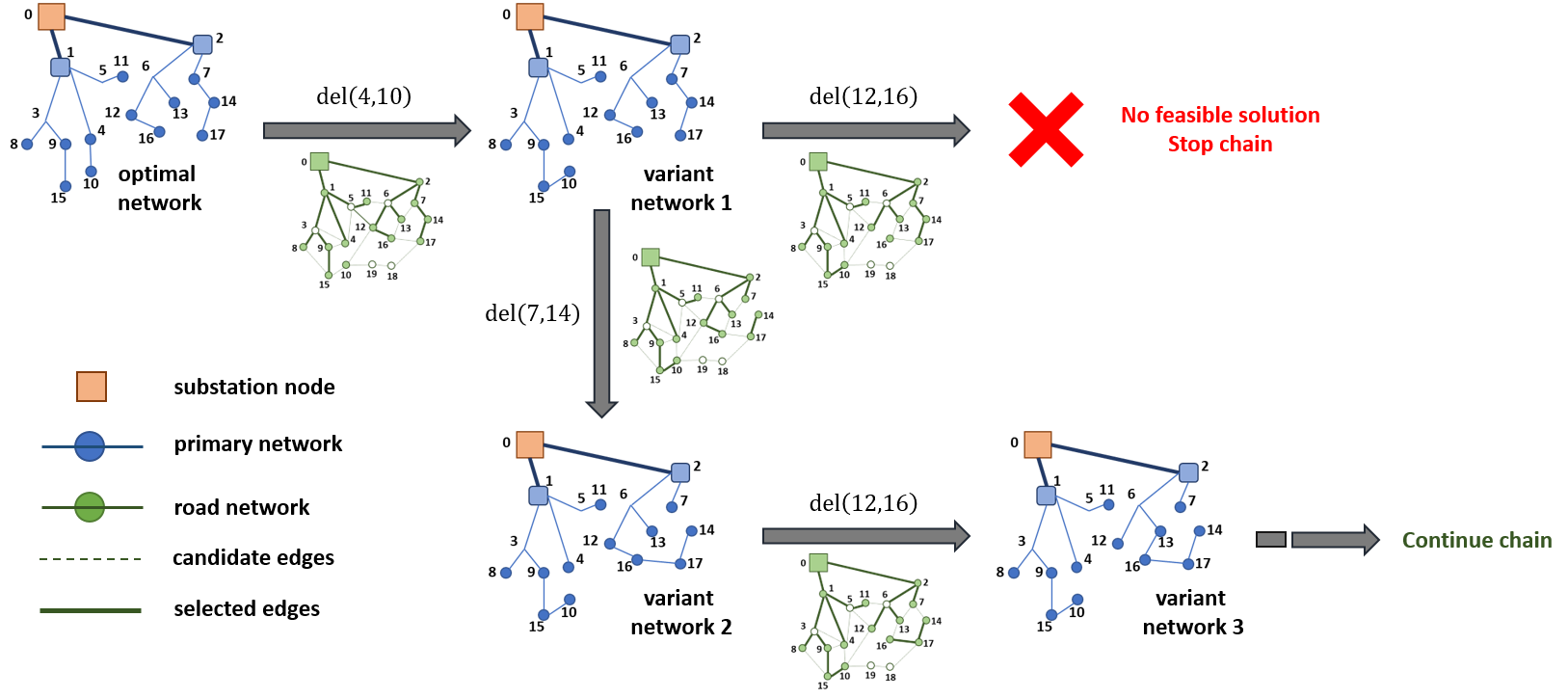}
    \caption{Schematic of the Markov chain process to generate an ensemble of primary networks from the optimal primary distribution network. Every transition in the Markov chain involves deletion of a random edge from the primary network followed by solving a restricted version of the primary network creation problem. If no feasible solution exists for the restricted version of the problem, the chain transitions to itself.}
    \label{fig:mc-ensemble}
\end{figure*}
\FloatBarrier

The restrictions reduce the order of the optimization problem (\ref{eq:sup-prim-relax}) to a significant extent and thereby requires less time to reach an optimal solution. (\ref{seq:restrict-edge}) forces all edges in the set $\mathscr{E}_P^{t}\setminus\mathscr{E}_D$ to be selected in the new network $\mathscr{G}_P^{t+1}(\mathscr{V}_P^{t+1},\mathscr{E}_P^{t+1})$. The edge set $\mathscr{E}_D$ is not selected and the optimization problem chooses edges from $\mathscr{E}_R\setminus\mathscr{E}_P^{t}$. Further, (\ref{seq:restrict-road}) forces all road network nodes in $\mathscr{G}_P^{t}$ to be selected in $\mathscr{G}_P^{t+1}$; the other road nodes are left free to be selected. Finally (\ref{seq:restrict-root}) keeps the feeder nodes (which are connected to substation node through HV feeders) similar in $\mathscr{G}_P^{t}$ and $\mathscr{G}_P^{t+1}$. Therefore, the number of binary variables reduces from $|\mathscr{E}_R|+2|\mathscr{V}_R|$  to $\left|\mathscr{E}_R\setminus\mathscr{E}_t\right|+\left|\mathscr{V}_R\setminus\mathscr{V}_t\right|$, which is a significant improvement. Here $|\cdot|$ denotes the cardinality of a set. Algorithm~\ref{alg:markov} lists the steps involved in constructing the ensemble of $N$ networks. We start with the near-optimal primary network and create the variant primary networks. Finally, we augment the variant primary networks to the optimal secondary network to create the networks in the ensemble.


The schematic of the process is shown in Fig.~\ref{fig:mc-ensemble} where each transition involves deletion of a random edge followed by solving the restricted version of the optimization problem. For example, in case of the first transition, the edge $(4,10)$ is chosen at random to be deleted. The restricted version of the optimization problem forces all the edges in the optimal primary network except $(4,10)$ to be selected. The remaining edges are selected from the candidate set of edges (shown in dotted green lines) so that the output result is a feasible network. 

\subsection*{Post-processing step}
The final step of the synthetic network creation process is to assign \emph{line types} to the different distribution lines in the network. We use the catalog of distribution lines~\cite{lines} to assign the line types and parameters to secondary and primary distribution lines. A simultaneity factor of $0.8$ has been used to rate the transformers and distribution lines similar to the steps used in~\cite{rnm_2011}. This means that when all residences are consuming their respective peak hourly load, the distribution lines are loaded to at most $80\%$ of their rating. Table~\ref{tab:node-edge-details} lists the node and edge attributes in the generated synthetic power distribution networks.
\begin{figure}[tbhp]
\centering
	\includegraphics[width=0.32\textwidth]{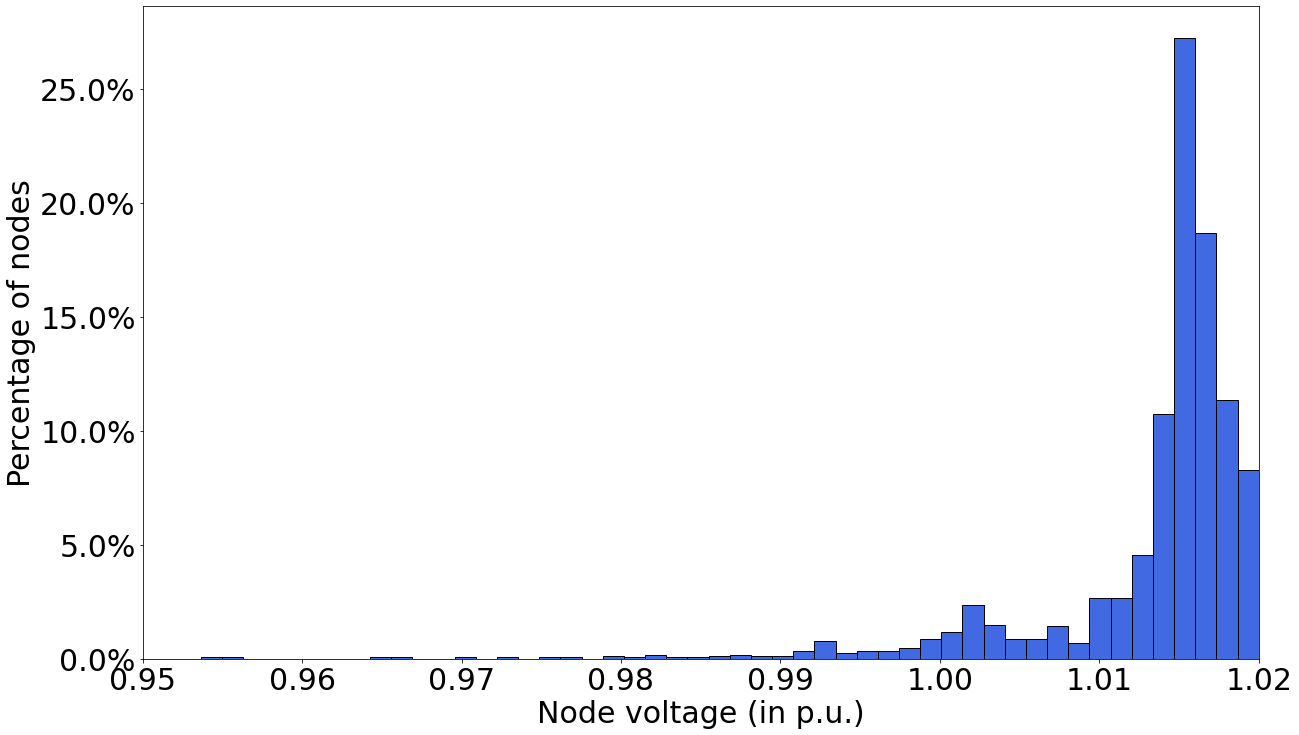}
	\includegraphics[width=0.32\textwidth]{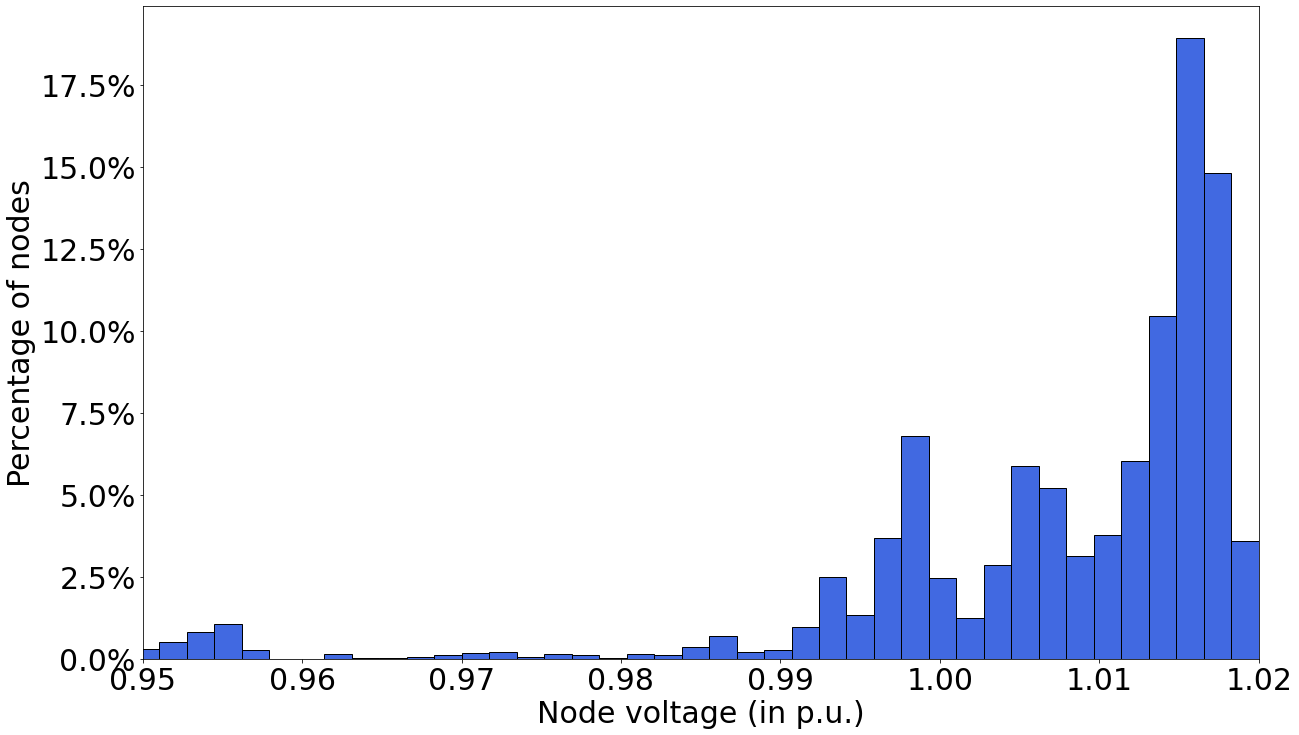}
	\includegraphics[width=0.32\textwidth]{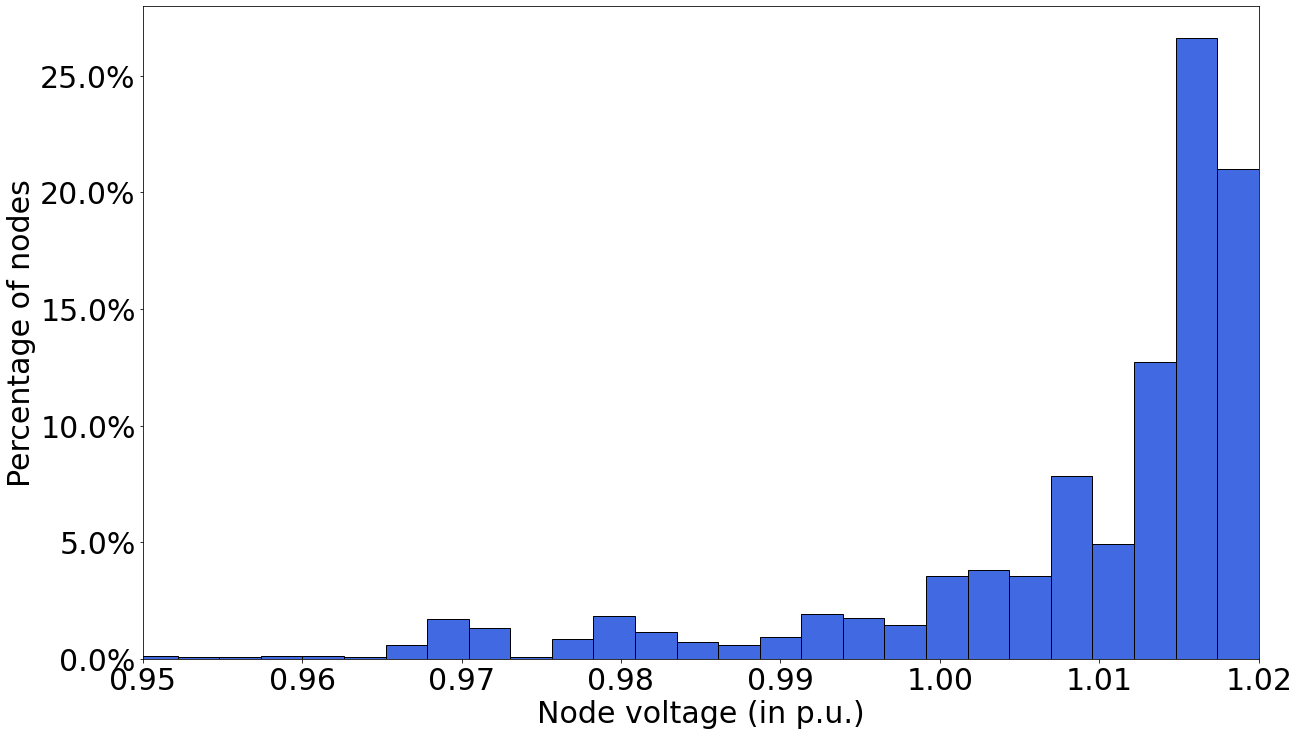}
	\caption{Plots showing histogram of node voltages for three synthetic distribution networks operating at peak load. We observe that majority of the nodes are within acceptable voltage limits set by the ANSI C.84 A standard. The substation voltage is set at 1.02 p.u. to avoid low voltages at the leaf residence nodes.}
	\label{fig:voltage-hist}
\end{figure}
\FloatBarrier
\begin{figure}[tbhp]
\centering
	\includegraphics[width=0.32\textwidth]{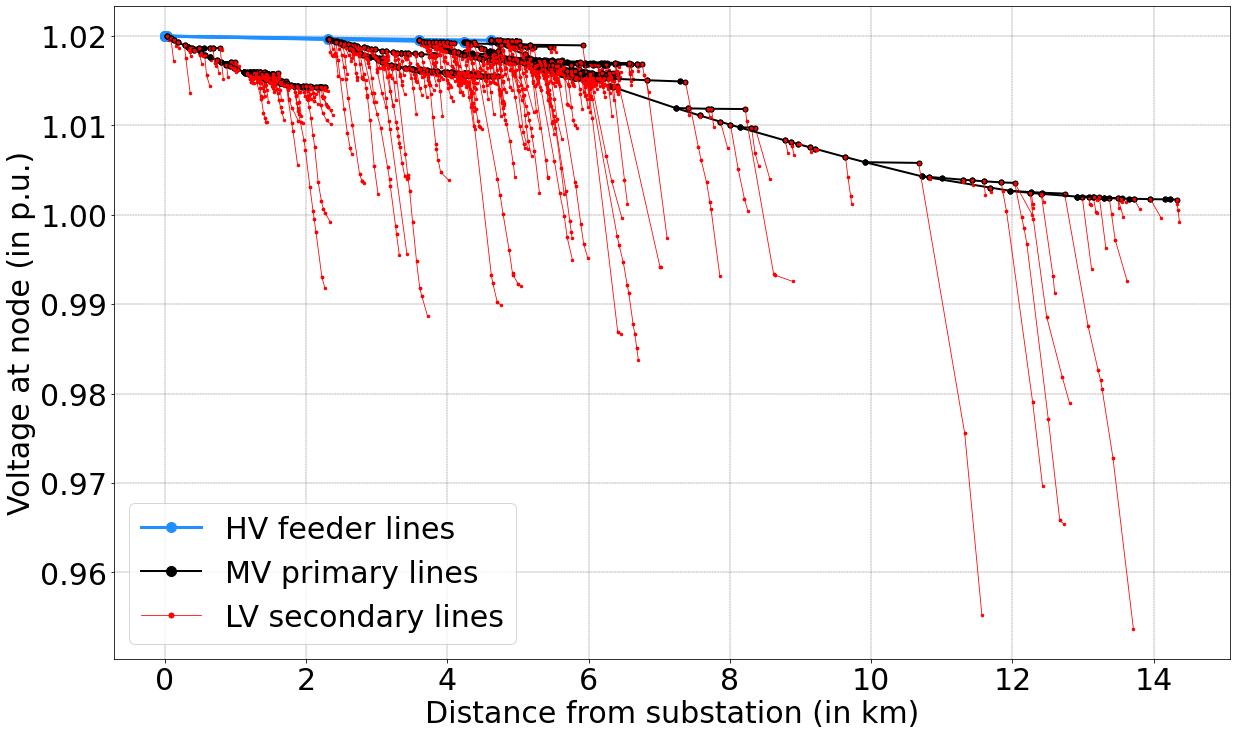}
	\includegraphics[width=0.32\textwidth]{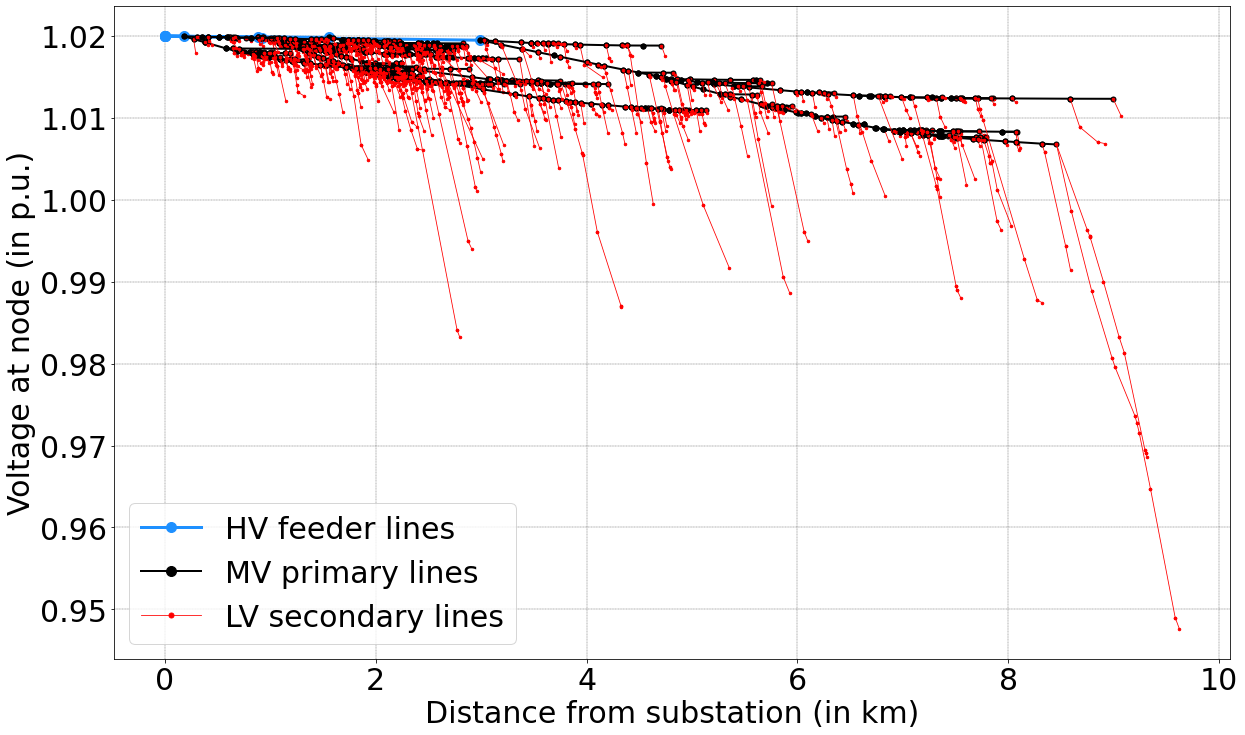}
	\includegraphics[width=0.32\textwidth]{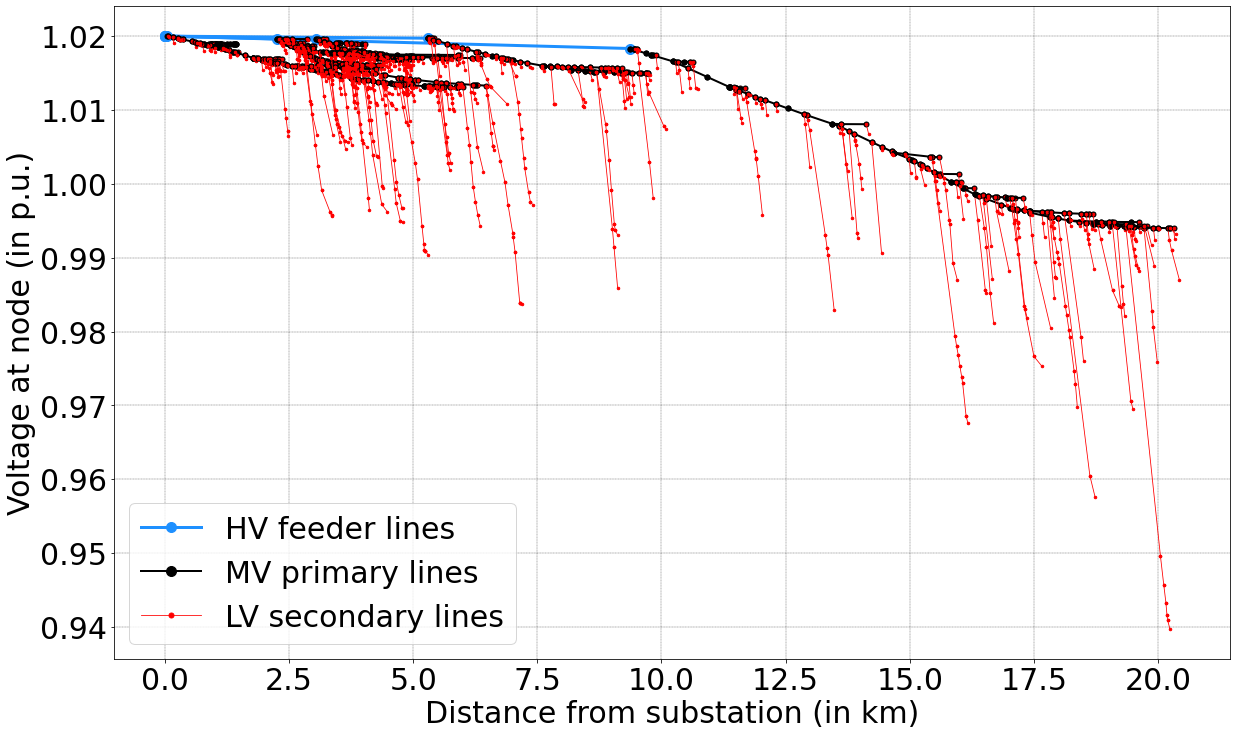}
	\caption{Plots showing variation of voltage with distance from substation (bottom three figures) for three synthetic distribution networks operating at peak load. The HV feeders are used to connect distant residences while maintaining a healthy voltage profile. The primary network maintains the voltage level within acceptable engineering standard for distribution; while significant voltage drop occurs at in the LV secondary network.}
	\label{fig:voltage-tree}
\end{figure}
\FloatBarrier

A key aspect of creating the synthetic networks is regulating the voltage at different nodes in the network. 
ANSI C.84 Range A limits the acceptable node voltage within $0.95-1.05$ p.u.; however, in certain sections, we observe undervoltage. Since such occurrences of undervoltage are common in real distribution networks as well, we relax the acceptable limits as follows: accept networks where voltage at $>99.5\%$ of residences are within the limits set by ANSI C.84 Range A ($0.95 - 1.05$ p.u.). In the rejected cases, we perform either of the following post-processing steps: (i) increase the load tap changer (LTC) setting at the distribution setting to 1.02 p.u. (ii) add line regulators at the end of HV feeders, which are used to connect remote loads. The three figures in Fig.~\ref{fig:voltage-hist} show the histogram of node voltages after performing these post-processing steps. The three plots in Fig.~\ref{fig:voltage-tree} show the variation of node voltage with distance from the substation. The blue lines show the HV feeder lines, which results in the minimum voltage drop. The black lines denote the MV primary network lines, which causes voltage drop as we tend to move away from the substation. Finally, the red lines are the LV secondary lines resulting in a major voltage drop at the residences. However, in all cases, we notice that $>99.5\%$ of customer node voltages are within the acceptable ANSI~C.84 A limits. The details of the output synthetic power distribution networks are provided in Table~\ref{tab:node-edge-details}. We have uploaded our data to the GitHub repository~\cite{git}. Table~\ref{tab:line-type} provides a list of lines used in the synthetic secondary and primary distribution networks. A summary of the percentage of these lines in the entire network of Virginia is also appended in the table. 

\begin{table}[tbhp]
\scriptsize
\centering
\caption{Node and edge attributes in created synthetic power distribution networks}
\label{tab:node-edge-details}
\begin{tabular}{llll}
\hline
 & \textbf{Attribute Name} & \textbf{Attribute Details} & \textbf{\begin{tabular}[c]{@{}l@{}}Size of Output Network for\\ Montgomery County, Virginia (US)\end{tabular}} \\ \hline
\textbf{} & Node ID & integer ID of node & 63220 nodes \\
\textbf{} & Node geometry & $(x,y)$: longitude and latitude of node & 20 substation nodes \\
\textbf{Nodes} & Node label & `S': substation, `T': local transformer, `H': residence, `R': auxiliary node & 18759 transformer nodes \\
\textbf{} & Node average load & Average hourly load demand in Watts (only for residences) & 35629 residence nodes \\
\textbf{} & Node peak load & Peak hourly demand in Watts (only for residences) & 8812 auxiliary nodes \\ 
& Node phase & Phase assigned to node (A, B or C) &  \\ \hline
\textbf{} & Edge ID & integer IDs of connecting nodes &  \\
 & Edge geometry & shapely LineString geometry of edge & 63200 edges \\
 & Edge label & `E': HV feeder line, `P': MV primary network, `S': LV secondary network & 156 HV feeder lines \\
\textbf{Edges} & Edge name & Name of conductor type used for the edge & 27415 primary edges \\
 & Length & Length of the edge in meters & 35629 secondary edges \\
 & Resistance & Resistance of the edge conductor in p.u. &  \\
 & Reactance & Reactance of the edge conductor in p.u. &  \\
 & Edge phase & Phase assigned to edge (A, B, C or mixed) &  \\ \hline
\end{tabular}
\end{table}
\FloatBarrier

\begin{table*}[tbhp]
\centering
\caption{Catalog of LV and MV distribution network lines}
\label{tab:line-type}
\begin{tabular}{llccccc}
\hline
\textbf{Line Name} & \textbf{Line Type} & \textbf{\begin{tabular}[c]{@{}c@{}}Resistance \\ (ohms/$1000$ ft)\end{tabular}} & \textbf{\begin{tabular}[c]{@{}c@{}}Reactance\\ (ohms/$1000$ ft)\end{tabular}} & \textbf{\begin{tabular}[c]{@{}c@{}}Current\\ (A)\end{tabular}} & \textbf{\begin{tabular}[c]{@{}c@{}}Voltage\\ (kV)\end{tabular}} & \textbf{\begin{tabular}[c]{@{}c@{}}Percentage \\ of lines \\ in Virginia (\%)\end{tabular}} \\ \hline
OH Voluta & TRPLX \#6 & 0.661 & 0.033 & 95 & 0.24 & 73.40 \\
OH Periwinkle & TRPLX \#4 & 0.416 & 0.031 & 125 & 0.24 & 8.70 \\
OH Conch & TRPLX \#2 & 0.261 & 0.030 & 165 & 0.24 & 7.00 \\
OH Neritina & TRPLX 1/0 & 0.164 & 0.030 & 220 & 0.24 & 5.00 \\
OH Runcina & TRPLX 2/0 & 0.130 & 0.029 & 265 & 0.24 & 2.20 \\
OH Zuzara & TRPLX 4/0 & 0.082 & 0.027 & 350 & 0.24 & 3.70 \\ \hline
OH Swanate & ACSR \#4 & 0.407 & 0.113 & 140 & 12.47 & 90.80 \\
OH Sparrow & ACSR \#2 & 0.259 & 0.110 & 185 & 12.47 & 3.90 \\
OH Raven & ACSR 1/0 & 0.163 & 0.104 & 240 & 12.47 & 2.70 \\
OH Pigeon & ACSR 3/0 & 0.103 & 0.099 & 315 & 12.47 & 1.70 \\
OH Penguin & ACSR 4/0 & 0.082 & 0.096 & 365 & 12.47 & 0.80 \\ \hline
\end{tabular}
\end{table*}
\FloatBarrier


\subsection*{Three phase networks}
We also created three phase synthetic distribution networks as a post-processing step. The synthetic networks created thus far are positive sequence networks, which are suitable for planning studies. We create networks comprising of three phases (A, B, and C) which would be useful to run studies on the operation of the network involving phase unbalances.
We aim to create a balanced three phase network where the power delivered at each substation feeder is balanced in the three phases. For the sake of simplicity, we consider branching of three different phases only in secondary networks. This implies that each residence is assigned one of the three phases, and the branching occurs at the local pole top transformers.  Each branch originating from a transformer feeds residences with the same phase. Thus, we have a problem of three way partitioning the set of residences, such that the total load supplied to residences in each partition is balanced. 

\begin{problem}[Three phase network creation problem]
Given a set of residences $\mathscr{H}$ connected to a substation feeder with respective load demand $p_h$ for every $h\in\mathscr{H}$, find a partition into three sets $\mathscr{H}_A,\mathscr{H}_B,\mathscr{H}_C$ such that $$\sum_{h\in\mathscr{H}_A}p_h=\sum_{h\in\mathscr{H}_B}p_h=\sum_{h\in\mathscr{H}_C}p_h$$
\end{problem}
We define an optimization problem similar to the one proposed in~\cite{anna_naps}, which assigns a phase to every residence connected to a substation feeder and minimizes the deviation between the total load fed by each pair of phases. We provide the details of the optimization framework, which we solve to assign one of the three phases to each residence connected to a substation feeder. We assign binary variables $u^A_h,u^B_h,u^C_h\in\{0,1\}$ to every residence $h\in\mathscr{H}$ in order to denote the phase assigned to the residence. For example, $u^A_h=1$ denotes that residence $h$ is assigned phase A. We stack the binary variables to construct vectors $\mathbf{u}^A,\mathbf{u}^B,\mathbf{u}^C\in\{0,1\}^{|\mathscr{H}|}$ respectively. Since we assign exactly one of the phases (A,B,C) to each feeder, we enforce the following constraint.
\begin{equation}
    \mathbf{u}^A+\mathbf{u}^B+\mathbf{u}^C = \mathbf{1} \label{eq:3phase-sum}
\end{equation}
Let $p_n$ denote the average hourly power demand at residence $n$. We stack the power demand to obtain the vector $\mathbf{p}\in\mathbb{R}^{|\mathscr{H}|}$. Hence, we can compute the total power fed by three phases at the substation feeder as follows.
\begin{subequations}
\begin{align}
    &p^A = \mathbf{p}^T\mathbf{u}^A \label{seq:a-phase}\\
    &p^B = \mathbf{p}^T\mathbf{u}^B \label{seq:b-phase}\\
    &p^C = \mathbf{p}^T\mathbf{u}^C \label{seq:c-phase}
\end{align}
\label{eq:phase-load}
\end{subequations}
The phases are assigned to each residence by solving the optimization problem in (\ref{eq:3phase-create}).
\begin{equation}
\begin{aligned}
    \min_{\mathbf{u}^A,\mathbf{u}^B,\mathbf{u}^C}&~|p^A-p^B|+|p^B-p^C|+|p^C-p^A|\\
    \textrm{s.to.}&~(\ref{eq:3phase-sum}),(\ref{eq:phase-load})
\end{aligned}
\label{eq:3phase-create}
\end{equation}


\begin{figure}[tbhp]
\centering
	\includegraphics[width=0.45\textwidth]{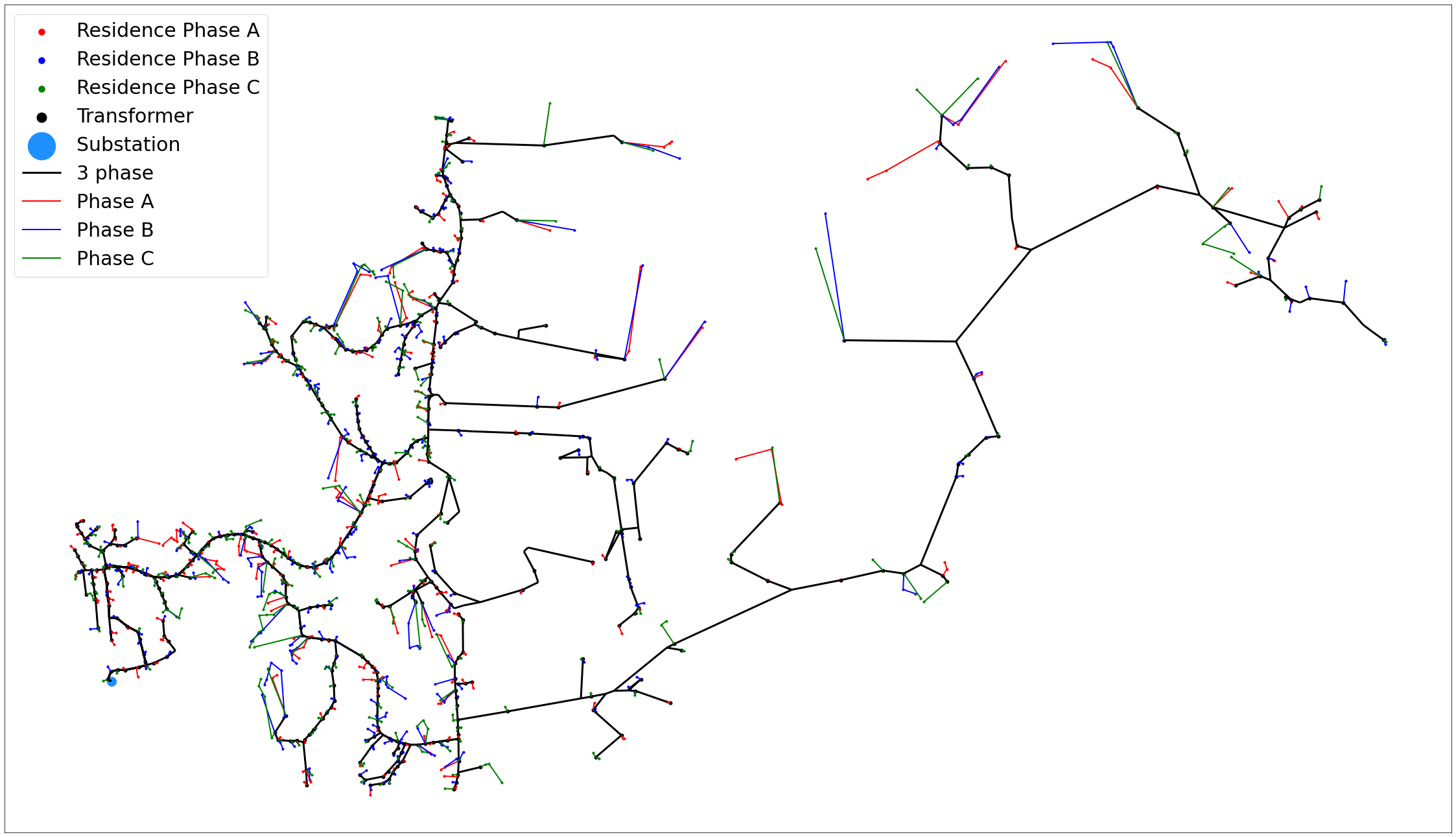}
	\includegraphics[width=0.45\textwidth]{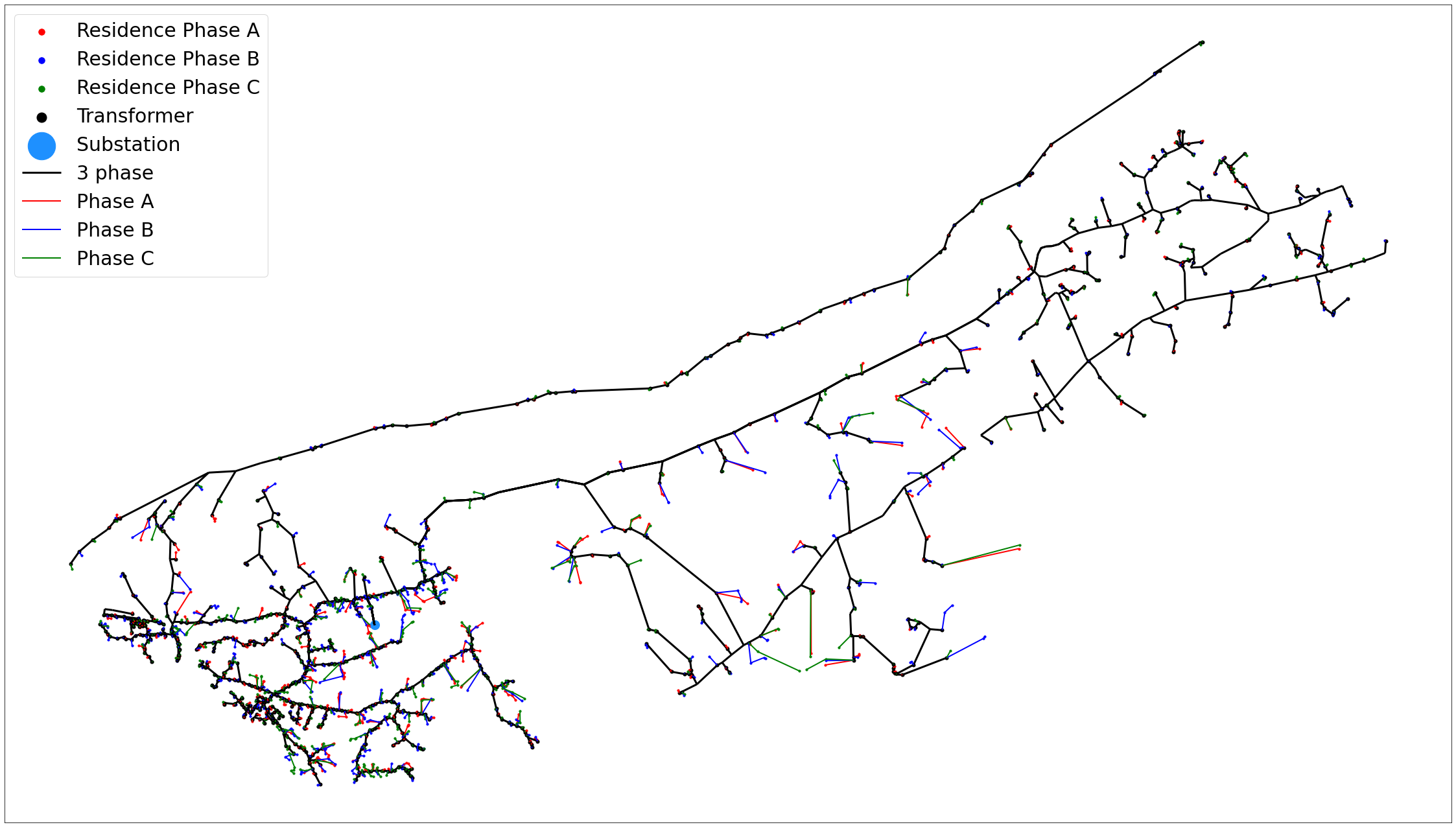}
	\caption{Plots showing two 3-phase networks generated from the optimal positive sequence networks.}
	\label{fig:3-phase-net}
\end{figure}
\FloatBarrier

\section*{Results}
\subsection*{Motifs in the network}
\begin{figure}[tbhp]
    \centering
    \includegraphics[width=0.24\textwidth]{figs/4path-motif-comp.png}
    \includegraphics[width=0.24\textwidth]{figs/4star-motif-comp.png}
    \includegraphics[width=0.24\textwidth]{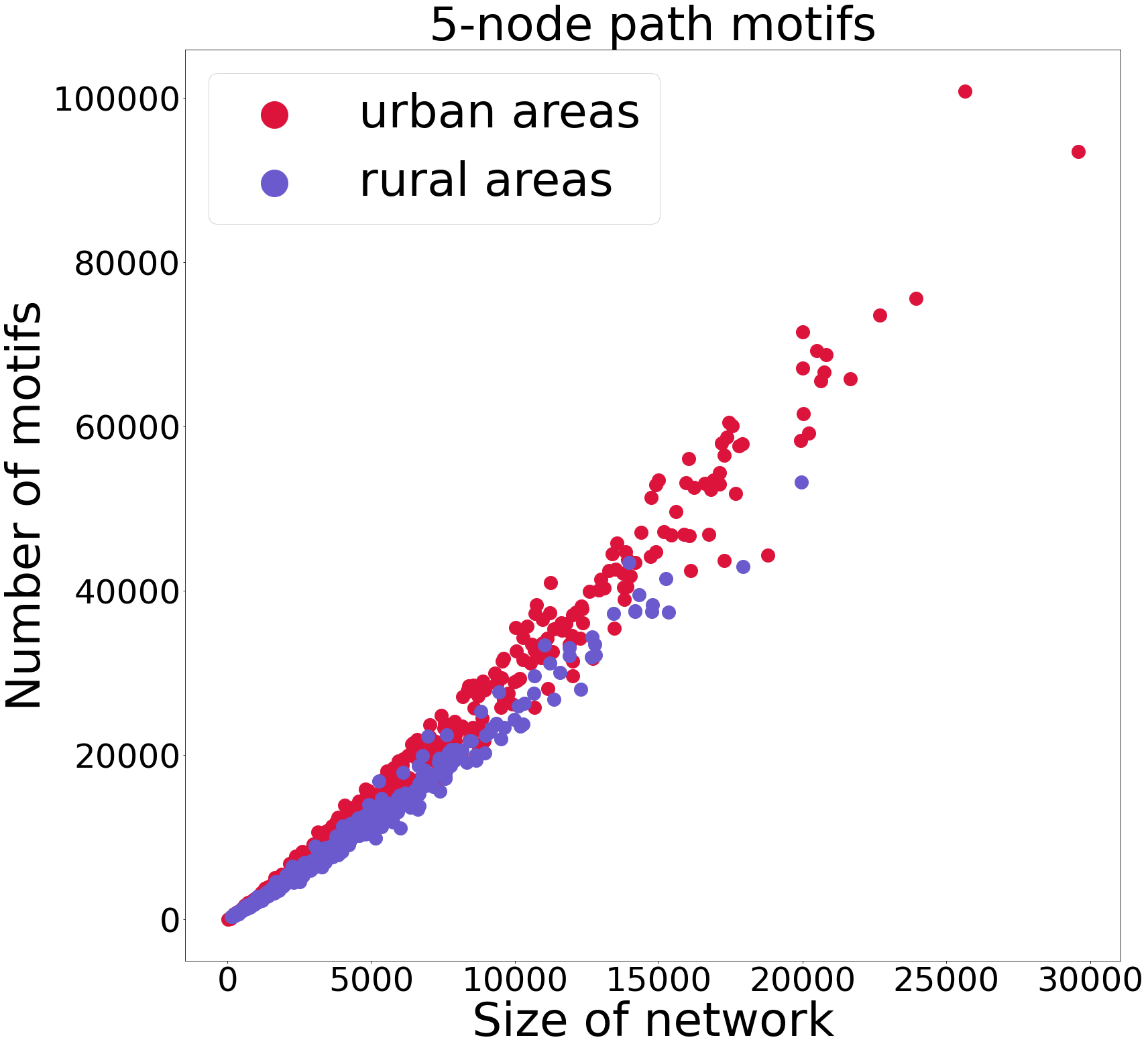}
    \includegraphics[width=0.24\textwidth]{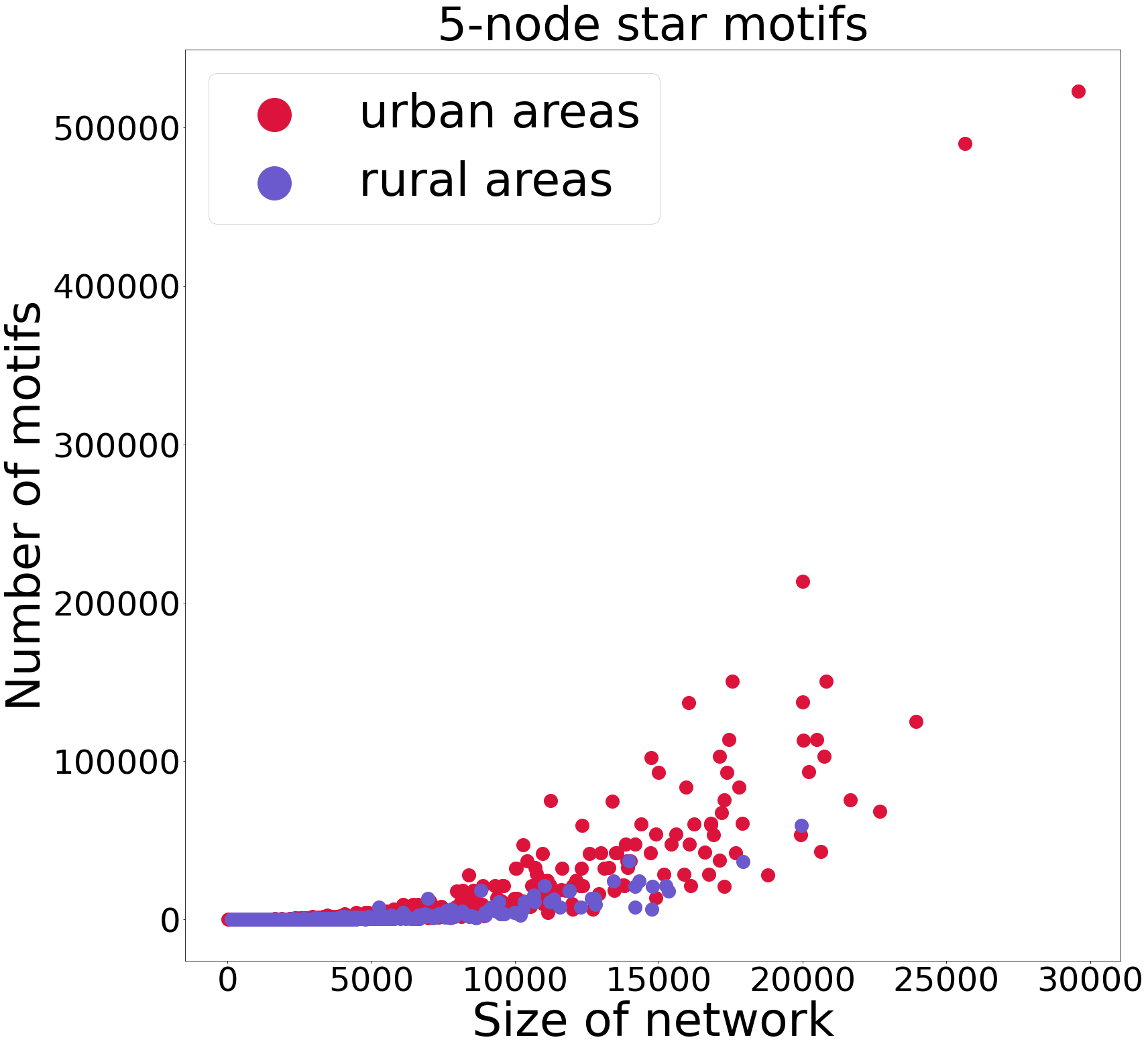}
    \caption{Plots showing number of 4-node paths (top left), 4-node star motifs (top right), 5-node paths (bottom left), and 5-node star motifs (bottom right) as a function of network size (measured as number of nodes in the network). Colors depict motif numbers in urban versus rural areas. Urban distribution networks have a larger number of star motifs than rural networks. In contrast, the path motif count does not differ significantly across rural and urban areas.  Urban networks are often larger than rural networks as measured by number of nodes due to larger population size.}
    \label{fig:motif-supplement}
\end{figure}
\FloatBarrier
We consider subgraphs of sizes $4$ and $5$ in the generated synthetic power distribution networks. Since the distribution networks are tree graphs, the relevant subgraphs are paths and stars. We define a $k$-path motif as a subgraph of $k$ nodes that form a path. A $k$-star motif is a subgraph of $k$ nodes which form a star, i.e., it consists of a single central node with degree $k-1$, and the remaining $k-1$ nodes are connected to the central node. We compare the path and star motifs in the created synthetic distribution networks in Fig.~\ref{fig:motif-supplement}. The two colors denote the motif counts for rural and urban networks. 

\subsection*{Structural Comparison of Network Ensembles}
\begin{figure}[tbhp]
    \centering
    \includegraphics[width=0.32\textwidth]{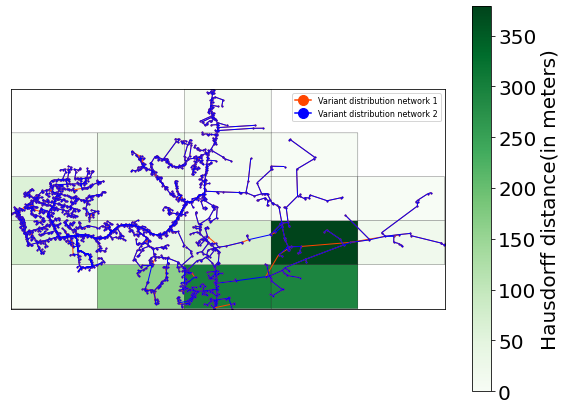}
    \includegraphics[width=0.32\textwidth]{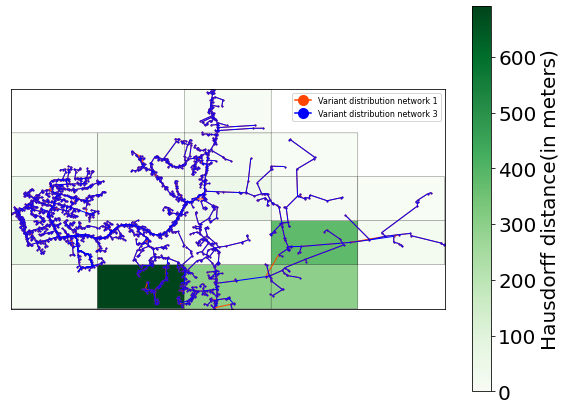}
    \includegraphics[width=0.32\textwidth]{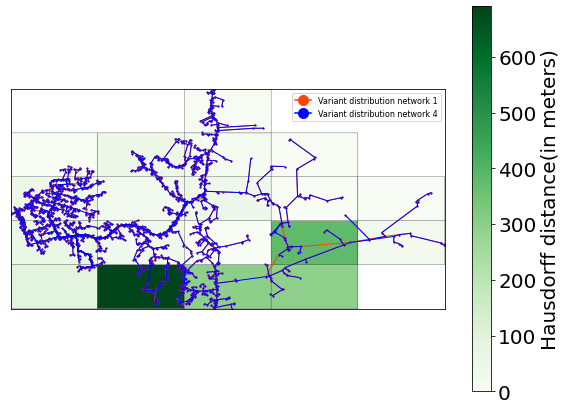}
    \includegraphics[width=0.32\textwidth]{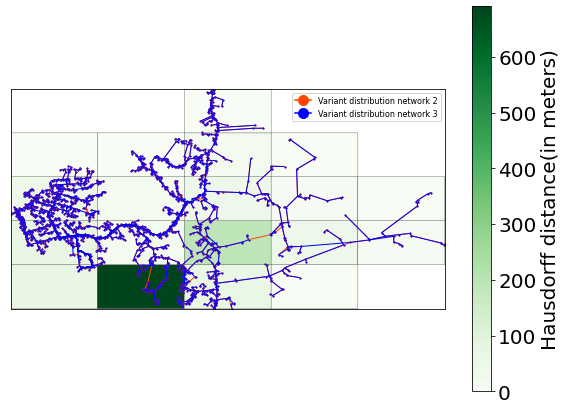}
    \includegraphics[width=0.32\textwidth]{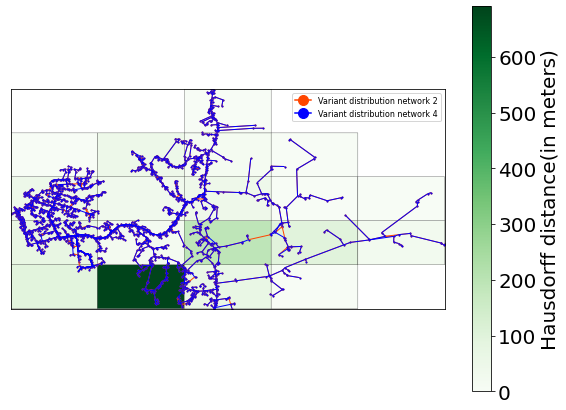}
    \includegraphics[width=0.32\textwidth]{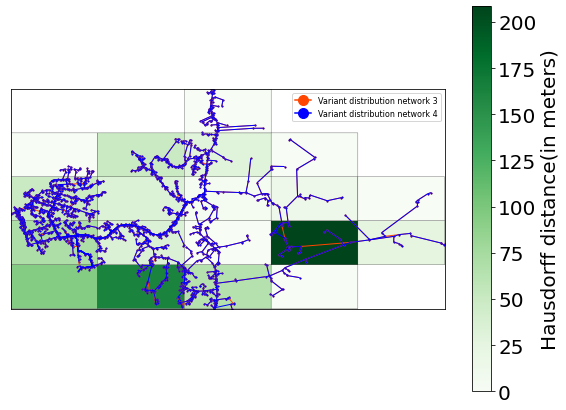}
    \caption{Plots comparing the Hausdorff distance between pairs of networks in the ensemble. The comparison is performed for each rectangular grid cell and color depicts the magnitude of deviation in meters.}
    \label{fig:ens-haus}
\end{figure}

We consider the structural comparison of the ensemble of networks here. We create an ensemble of $20$ networks for each distribution network created for Montgomery County in southwest Virginia. Fig.~\ref{fig:ens-haus} shows the results of structural comparison among $4$ of these networks. Note that the networks consist of the same residences, transformers and substations and hence the number of nodes and edges are similar to each other. However, the geometries of each network differ from the others. We do not see many structural differences between the networks at a visual level. However, a comparison of the geometry of the networks in the rectangular grid cells shows some differences, but these are not too large, as that would create networks where the power engineering constraints do not hold true.

\FloatBarrier

\begin{figure*}[htbp]
\centering
	\includegraphics[width=0.8\textwidth]{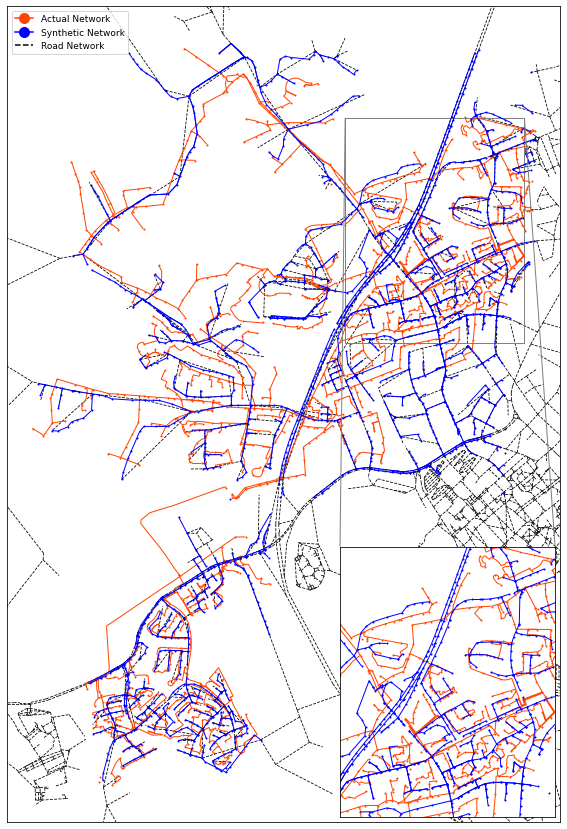}
	\caption{Plot comparing the structure of the actual distribution network (red) with the synthetic distribution network (blue) for the town of Blacksburg in southwest Virginia. The black dotted network denotes the underlying road network, which is used as a proxy to create the synthetic network.}
	\label{fig:validate-structure}
\end{figure*}
\FloatBarrier
\section*{Validation}
\subsection*{Variation in network structure}
In this section, we perform a visual comparison of the generated synthetic network to the actual network covering the same geographical region. Fig.~\ref{fig:validate-structure} shows the actual distribution network of the town of Blacksburg in southwest Virginia along with the synthetic network generated for the same. At first sight, the two networks are adjacent to and almost overlap each other. The inset figure confirms this hypothesis.  A zoomed-in view of the figure shows that the two networks are similar to each other. While the synthetic network retraces the road network, as expected based on our assumption, the actual network is also adjacent to it. This validates our primary assumption that the primary distribution network follows the road network. 

\begin{figure*}[htbp]
\centering
	\includegraphics[width=0.8\textwidth]{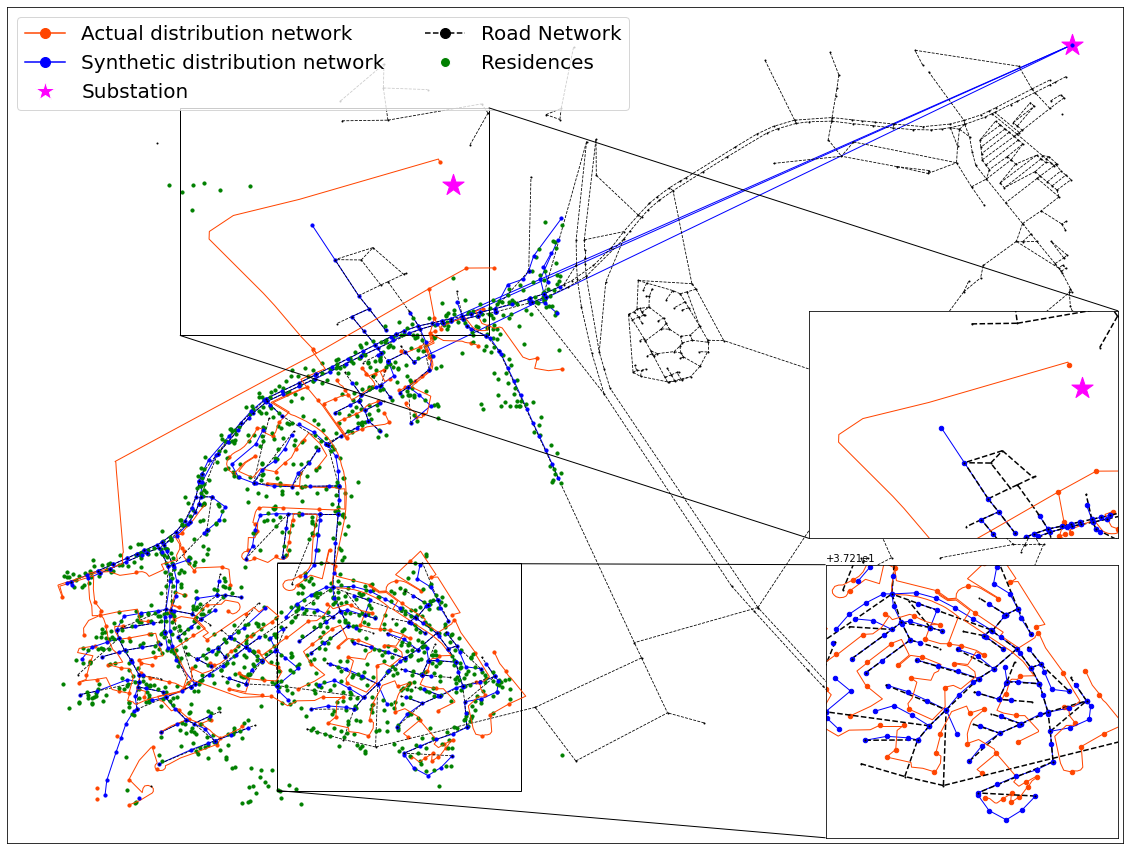}
	\caption{Plot explaining the visual structural difference between actual distribution network (red) and synthetic distribution network (blue) generated using first principles from the knowledge of road network (black).}
	\label{fig:hethwood-structure-compare}
\end{figure*}

\noindent\textbf{Feeder selection.} The significant structural difference between the two networks is the substation feeder to which it is connected. Fig~\ref{fig:hethwood-structure-compare} depicts the structural differences between the two networks. The actual network is connected to the nearest geographically located substation, where as the synthetic network is connected to a distant feeder. It is to be noted that the synthetic distribution network is generated using first principles with the assumption that it follows the road network to the maximum extent (it is easier to place distribution poles along the road network). It is seen that though the nearest substation (from which the actual network is fed) has no road links connecting itself to the neighborhood. On the contrary, the other substation, though located at a further distance, has road links connecting itself to the neighborhood under consideration. Therefore, the synthetic network generation algorithm prefers the latter substation over the former.
\FloatBarrier

\subsection*{Statistical Validation}
This section contains additional results related to the statistical validation of created synthetic networks. The created networks of the town of Blacksburg are compared to the actual physical counterpart obtained from a power company.

\noindent\textbf{Degree Distribution.}~The degree $k_i$ of a node $i$ in a graph with $n$ nodes and internode adjacency matrix $\mathbf{A}$ with elements $a_{ij}$ is defined by $k_i=\sum_{j=1}^{n}a_{ij}$. In general, the node degree computes the number of lines connected to it in the network. We consider three separate sections of the two networks and compare the degree distribution for each section. We observe that the two distributions match each other very closely. We observe that the degree of the nodes ranges between $1$ and $4$ for either network and the degree distribution shows that majority of nodes have a degree of $2$. The networks are created in a manner so that when operating with average hourly load demand, the network maintains an acceptable voltage profile. Every branch originating from a node leads to a small voltage drop caused by the load demand of the children nodes. This, in turn, causes a voltage drop in the parent nodes. Hence, for a feasible network to operate within acceptable voltage limits, it is expected that multiple branching from a single node is avoided as much as possible. This is the same observation we notice in the actual and synthetic networks, where the node degree hardly exceeds a degree of $3$.
\begin{figure}[tbhp]
    \centering
    \includegraphics[width=0.32\textwidth]{figs/degree-distribution-194.png}
    \includegraphics[width=0.32\textwidth]{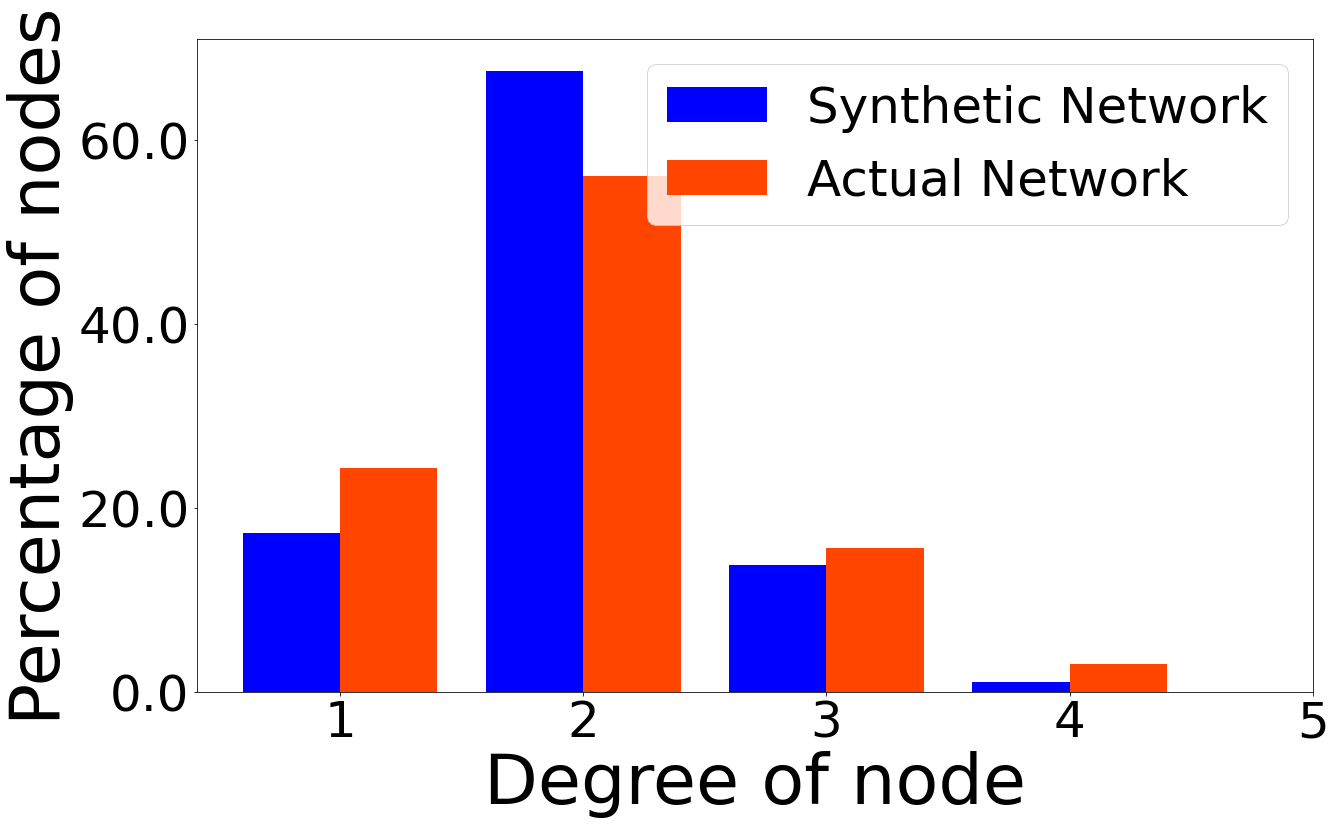}
    \includegraphics[width=0.32\textwidth]{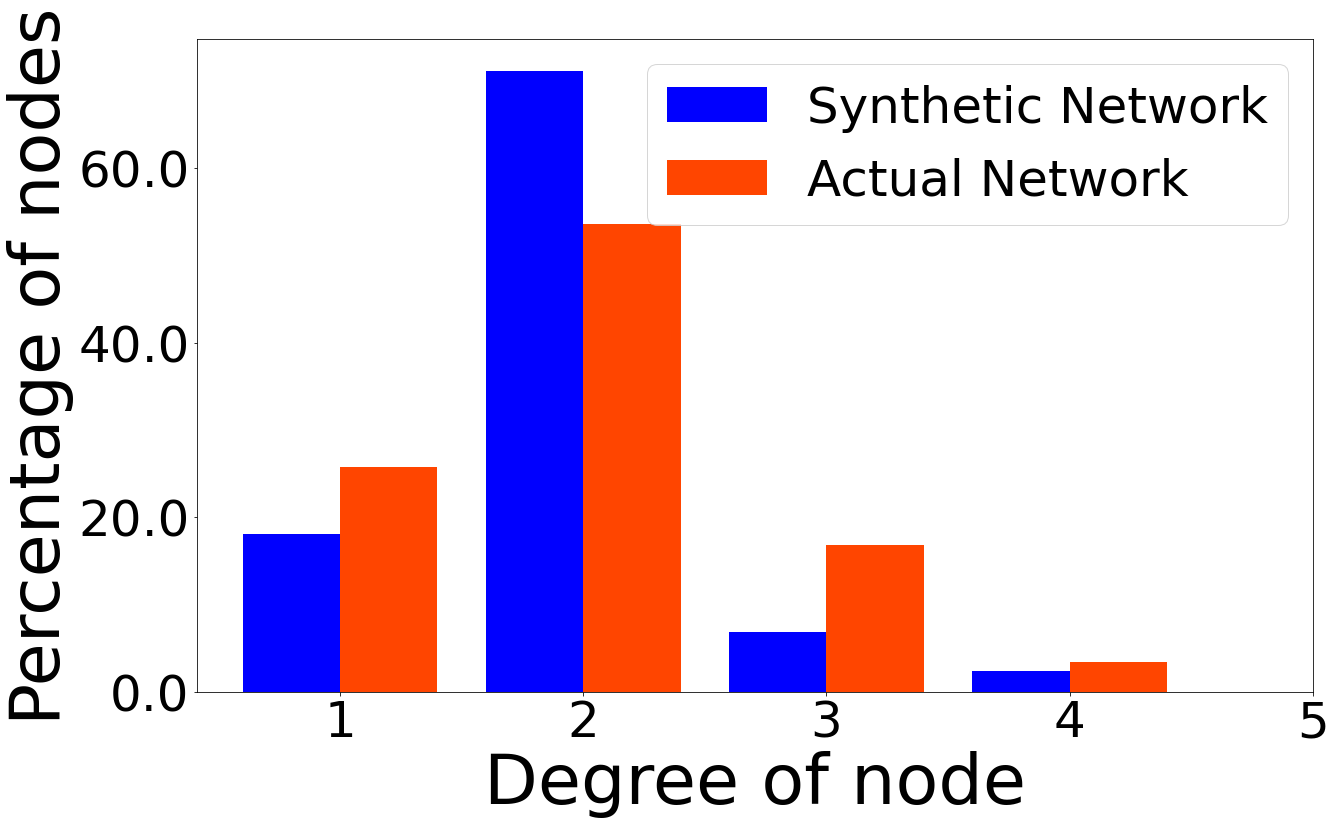}
    \caption{Plots comparing degree distribution of the actual and synthetic networks for three sections of region.}
    \label{fig:degree}
\end{figure}
\FloatBarrier

\noindent\textbf{Hop Distribution.}~
A \emph{path} $\mathscr{P}_{ij}$ in a graph between two nodes $i$ and $j$ is defined as a sequence of adjacent edges starting from node $i$ and terminating at node $j$. A \emph{tree} network (devoid of any cycles) has a unique path between any pair of nodes. The hops between nodes $i$ and $j$ in a tree network is defined as $|\mathscr{P}_{ij}|$ where $|\cdot|$ denotes the cardinality of a set. Essentially, it denotes the number of edges between the nodes $i$ and $j$. Here we are interested in the number of edges between any node $i$ and the substation node, which is the `root' node in the synthetic networks. Therefore, we define the number of hops between any node $i$ and substation (root) node by $h_i=|\mathscr{P}_{ri}|$. We consider the empirical distribution for the hops $h_i$'s of all nodes and term it as the \emph{hop distribution} of the networks. In this analysis, we study how the nodes in each network are distributed around the root node. We observe that the hop distributions for the synthetic and actual networks are significantly different. This is primarily because of the dissimilarity in the spatial distribution of nodes in each network which is studied later on in the later section.
\begin{figure}[tbhp]
    \centering
    \includegraphics[width=0.32\textwidth]{figs/hop-distribution-194.png}
    \includegraphics[width=0.32\textwidth]{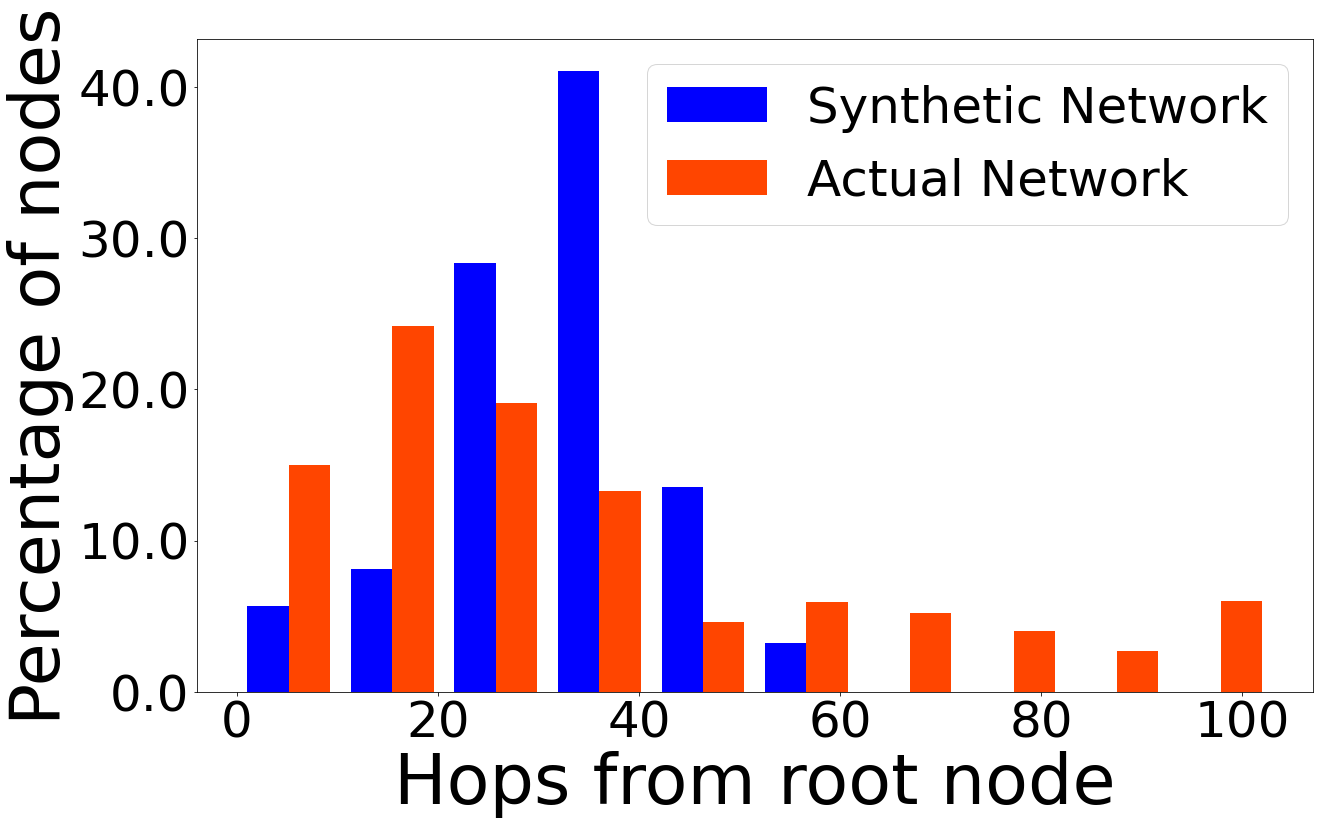}
    \includegraphics[width=0.32\textwidth]{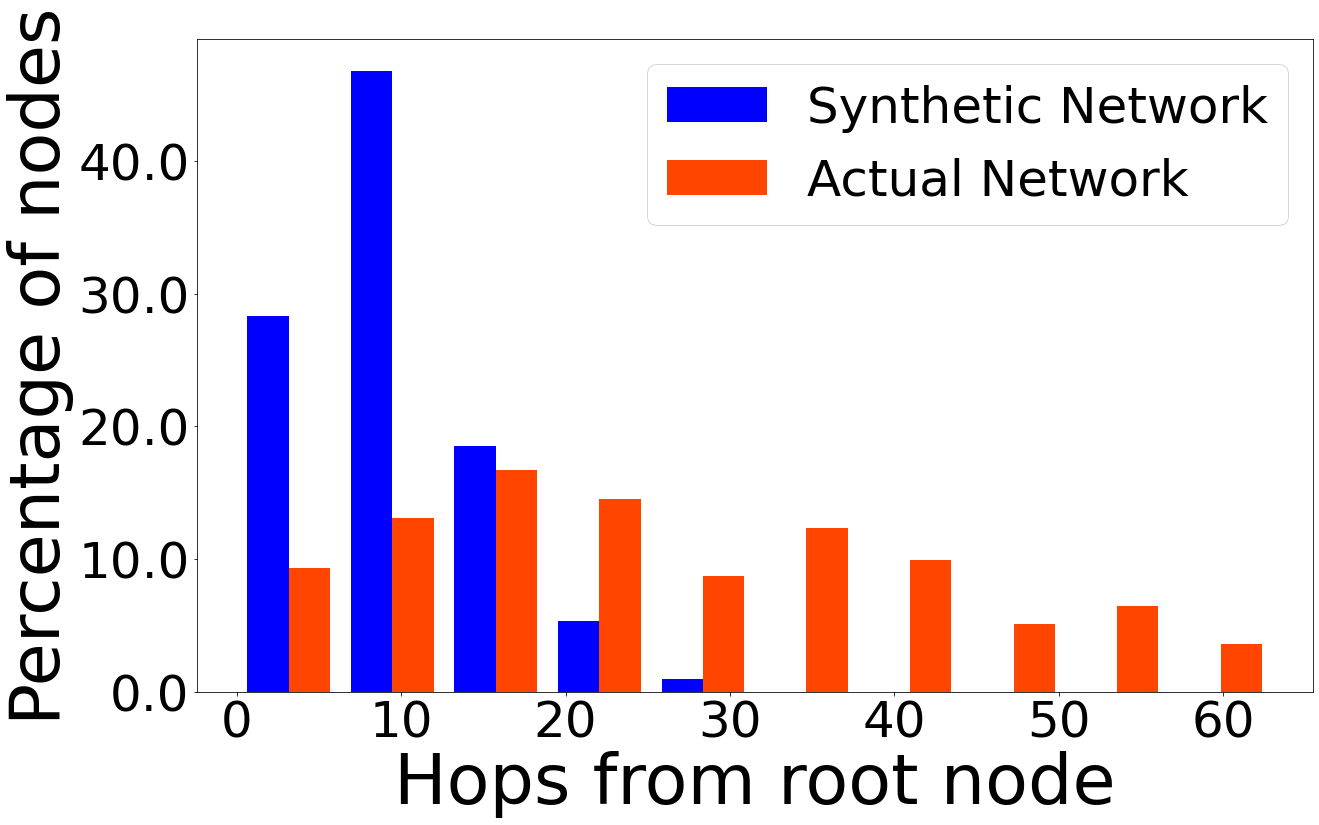}
    \caption{Plots comparing hop distribution of the actual and synthetic networks for three sections of the geographical region.}
    \label{fig:hop}
\end{figure}
\FloatBarrier

\begin{figure}[tbhp]
    \centering
    \includegraphics[width=0.30\textwidth]{figs/dist-distribution-194.png}
    \includegraphics[width=0.30\textwidth]{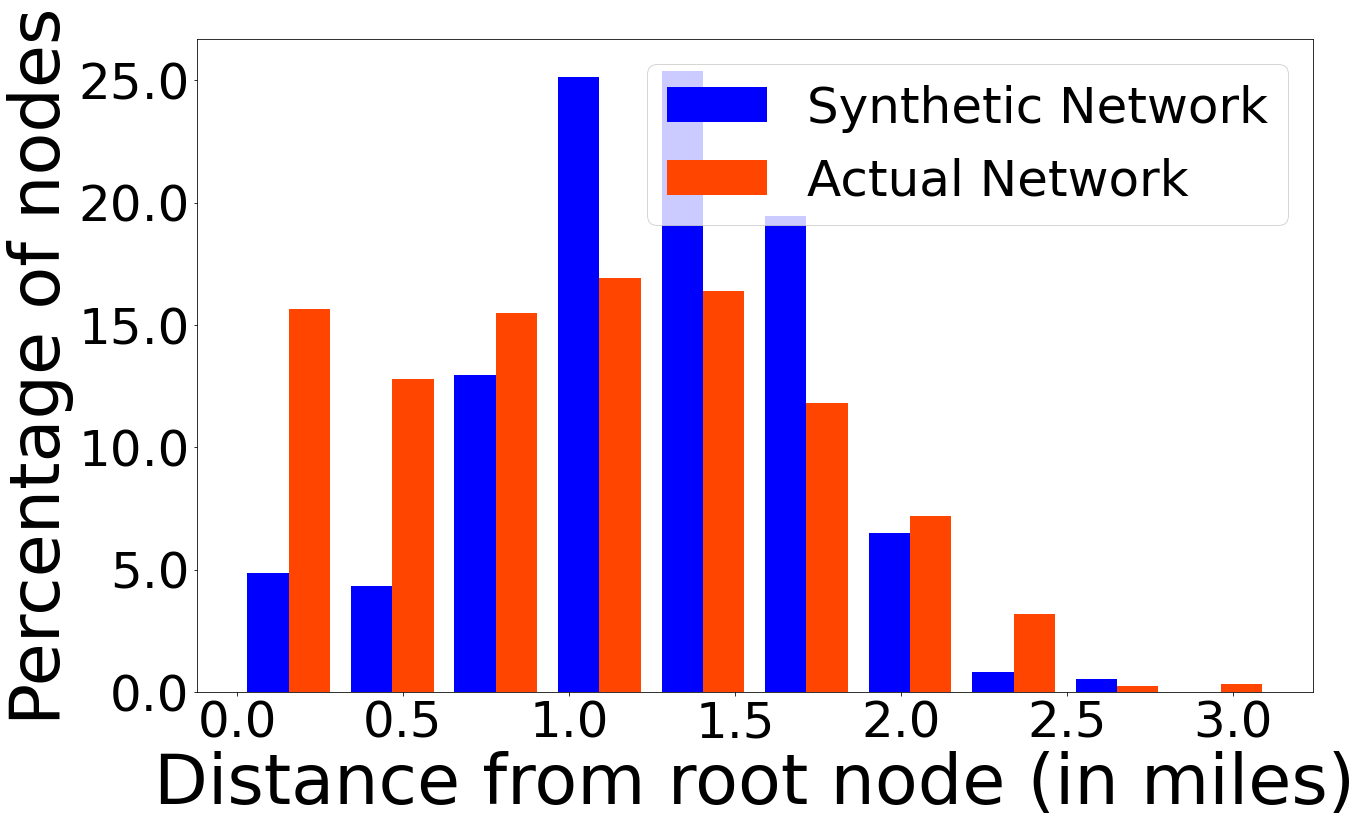}
    \includegraphics[width=0.30\textwidth]{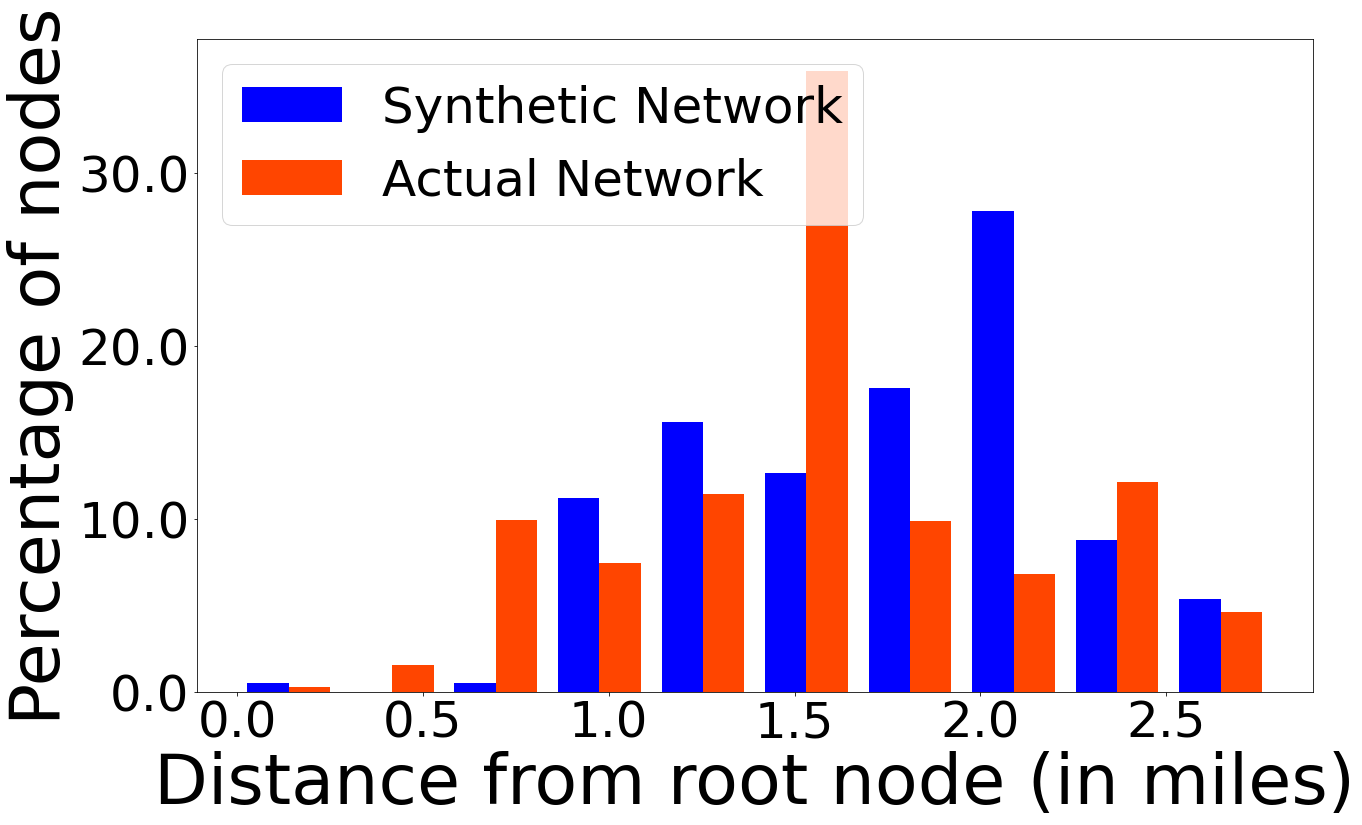}
    \caption{Plots comparing distribution of distance of each node to root in the actual and synthetic networks for three sections.}
    \label{fig:reach}
\end{figure}
\FloatBarrier
\noindent\textbf{Reach Distribution.}~
Since the synthetic networks have a geographic attribute associated with them, we can use the geographic distance of each node from the substation node to study how the nodes in the network are physically present around the root node. Each edge $e$ has an attribute $w_e$ which denotes the length of the edge in meters. We define \emph{reach} of a node $i$ in the synthetic network as $l_i=\sum_{e\in\mathscr{P}_{ri}}w_e$. Essentially, the reach of a node denotes the physical distance between the node and the substation (root) node. We consider the empirical distribution for the reach $l_i$'s of all nodes and term it as the \emph{reach distribution} of the networks. Fig.~\ref{fig:reach} compares the reach distribution of synthetic and actual network for three different sections. An interesting observation is that the actual network consists of a large number of long edges. This is primarily because of the fact that the network has been built over multiple years as the population grew in the geographical location. However, the synthetic networks are generated using first principles and as an output of an optimization problem where the generated network has the minimum length (or requires minimal installation and maintenance cost).

\begin{figure}[tbhp]
    \centering
    \includegraphics[width=0.24\textwidth]{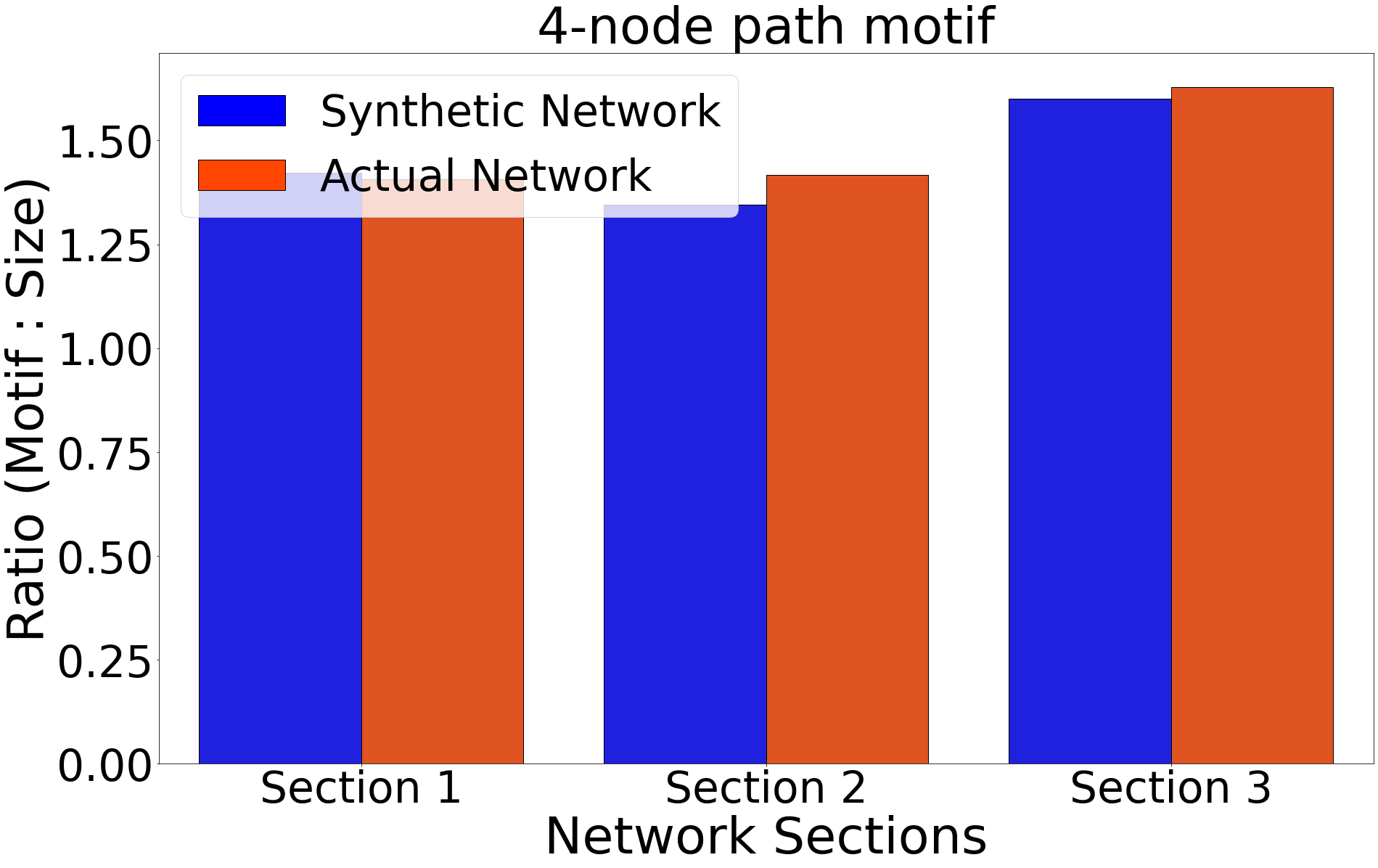}
    \includegraphics[width=0.24\textwidth]{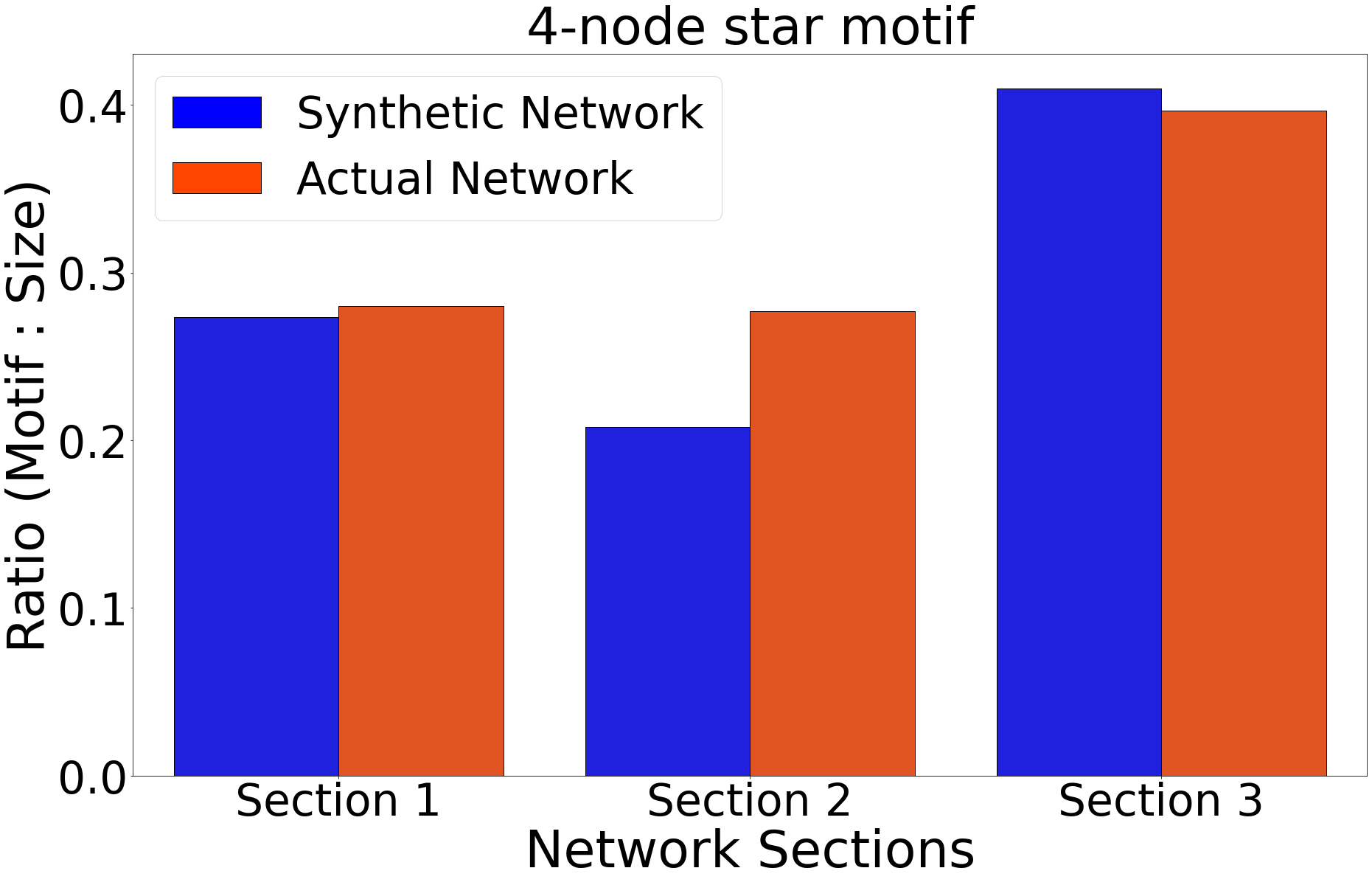}
    \includegraphics[width=0.24\textwidth]{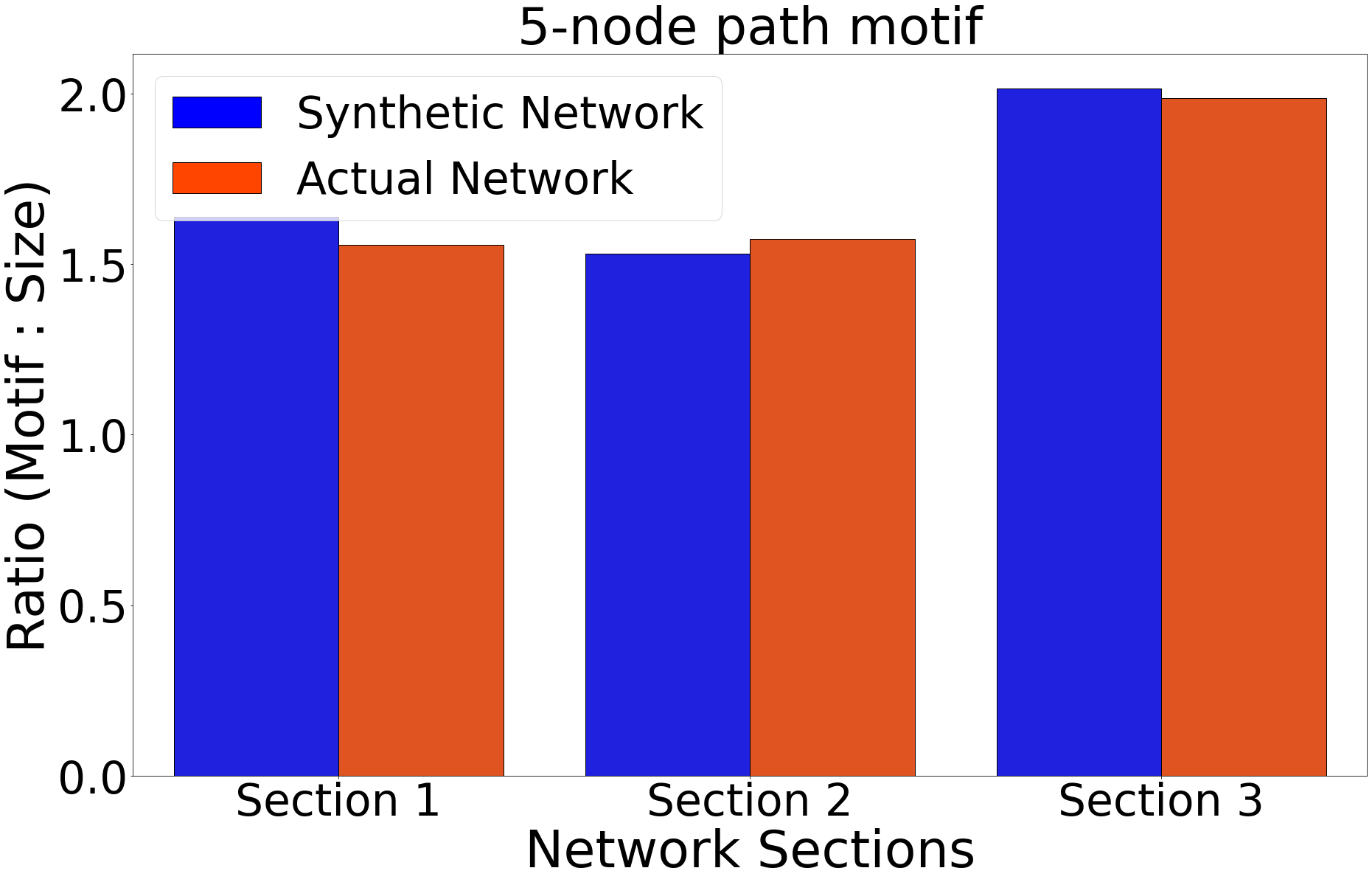}
    \includegraphics[width=0.24\textwidth]{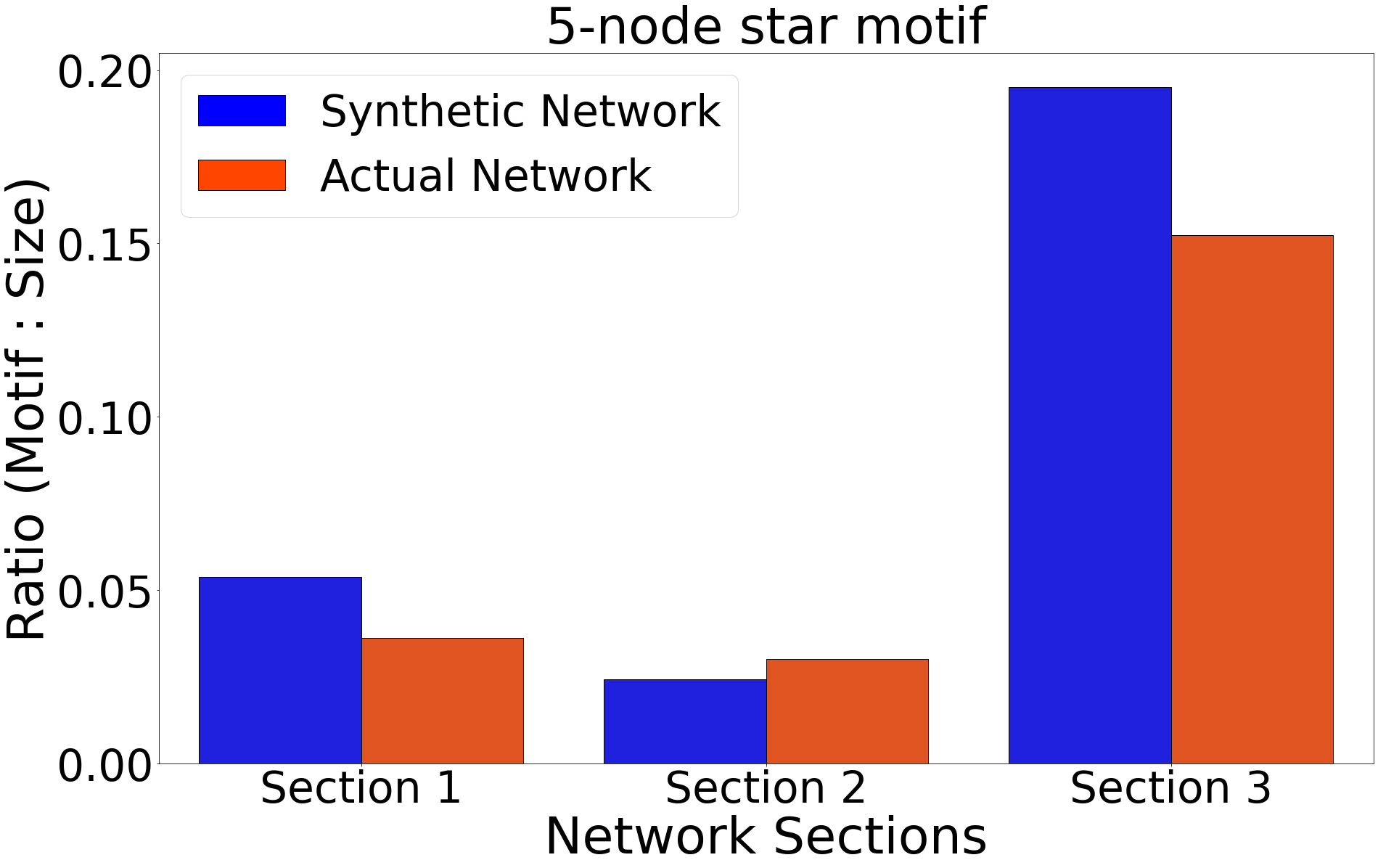}
    \caption{Plots comparing motif count to network size ratio of actual and synthetic distribution networks over three different sections.}
    \label{fig:motif-comp}
\end{figure}
\FloatBarrier
\noindent\textbf{Motif counts.}~We also compare the 4-node and 5-node path and star motifs in actual and synthetic distribution networks. We consider three different sections in the network. Since the size of the actual and synthetic networks are very different, we compare the ratio of motif counts to the network size in Fig.~\ref{fig:motif-comp}. We observe that the ratios are very similar for all three sections of the networks. This validates the structural resemblance of the networks through statistical attributes.

\subsection*{Structural Validation}
This section contains additional results related to the structural validation of created synthetic networks. The created networks of the town of Blacksburg are compared to the actual physical counterpart.

\begin{figure}[tbhp]
    \centering
    \includegraphics[width=0.30\textwidth]{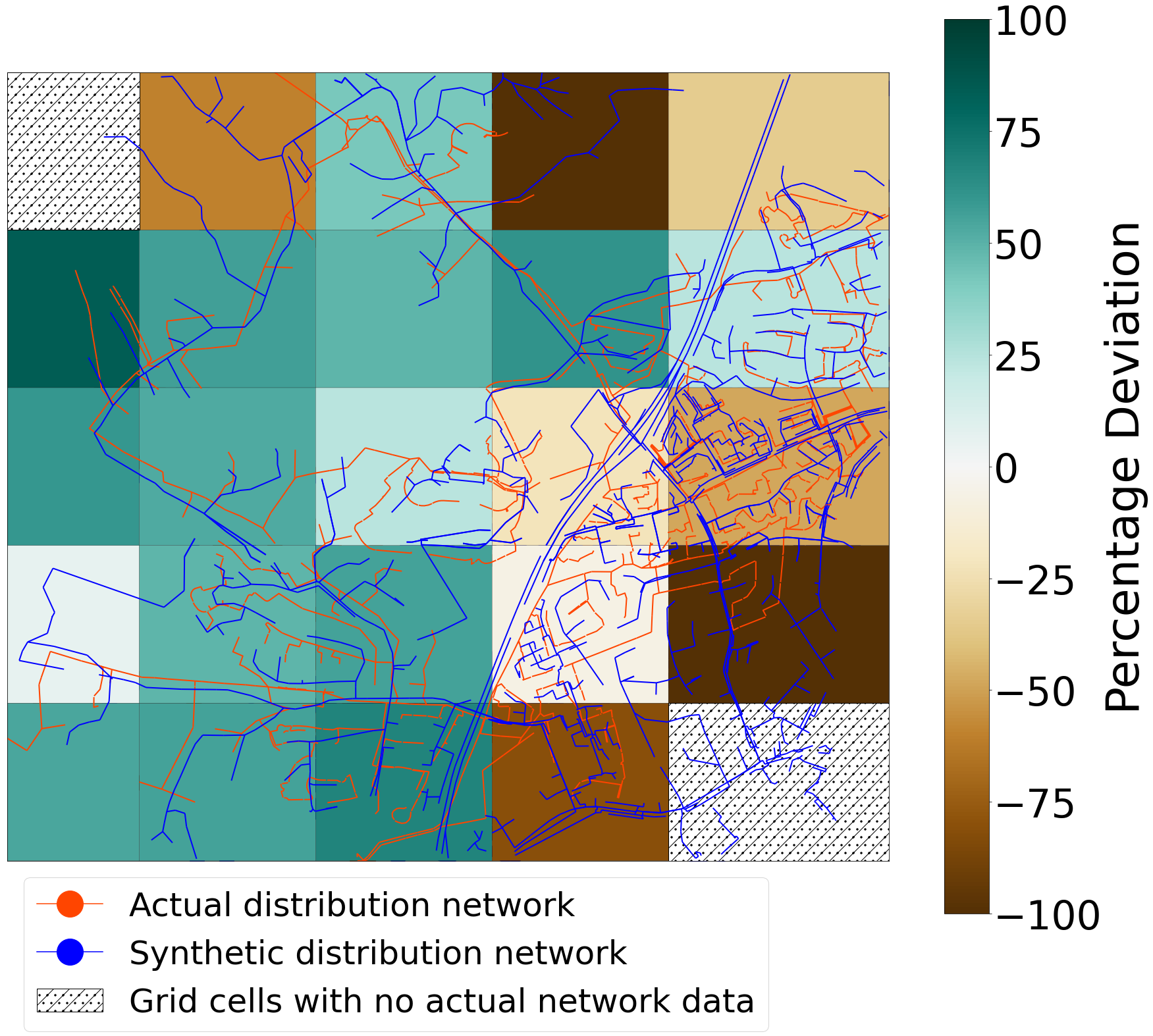}
    \includegraphics[width=0.30\textwidth]{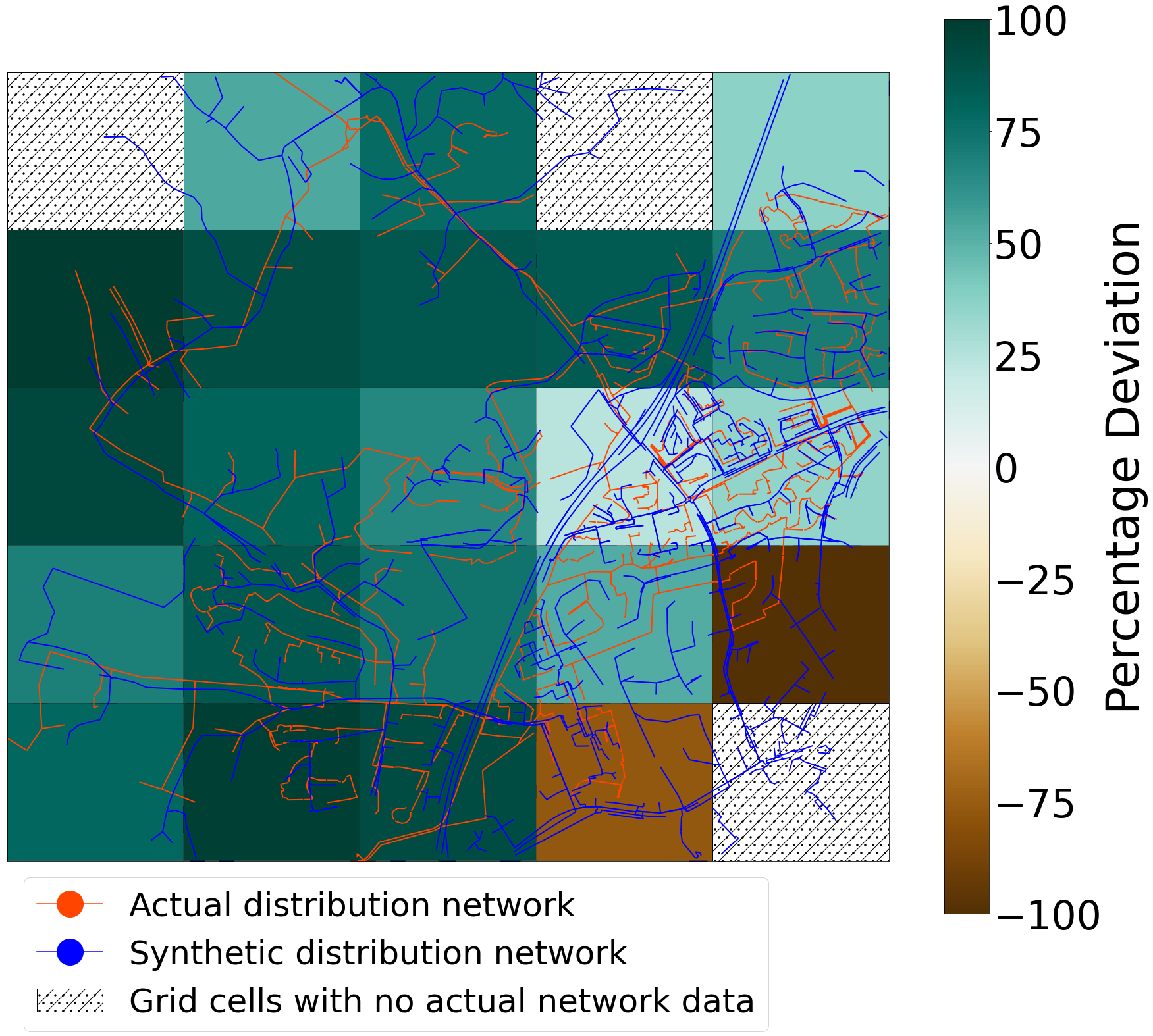}
    \includegraphics[width=0.30\textwidth]{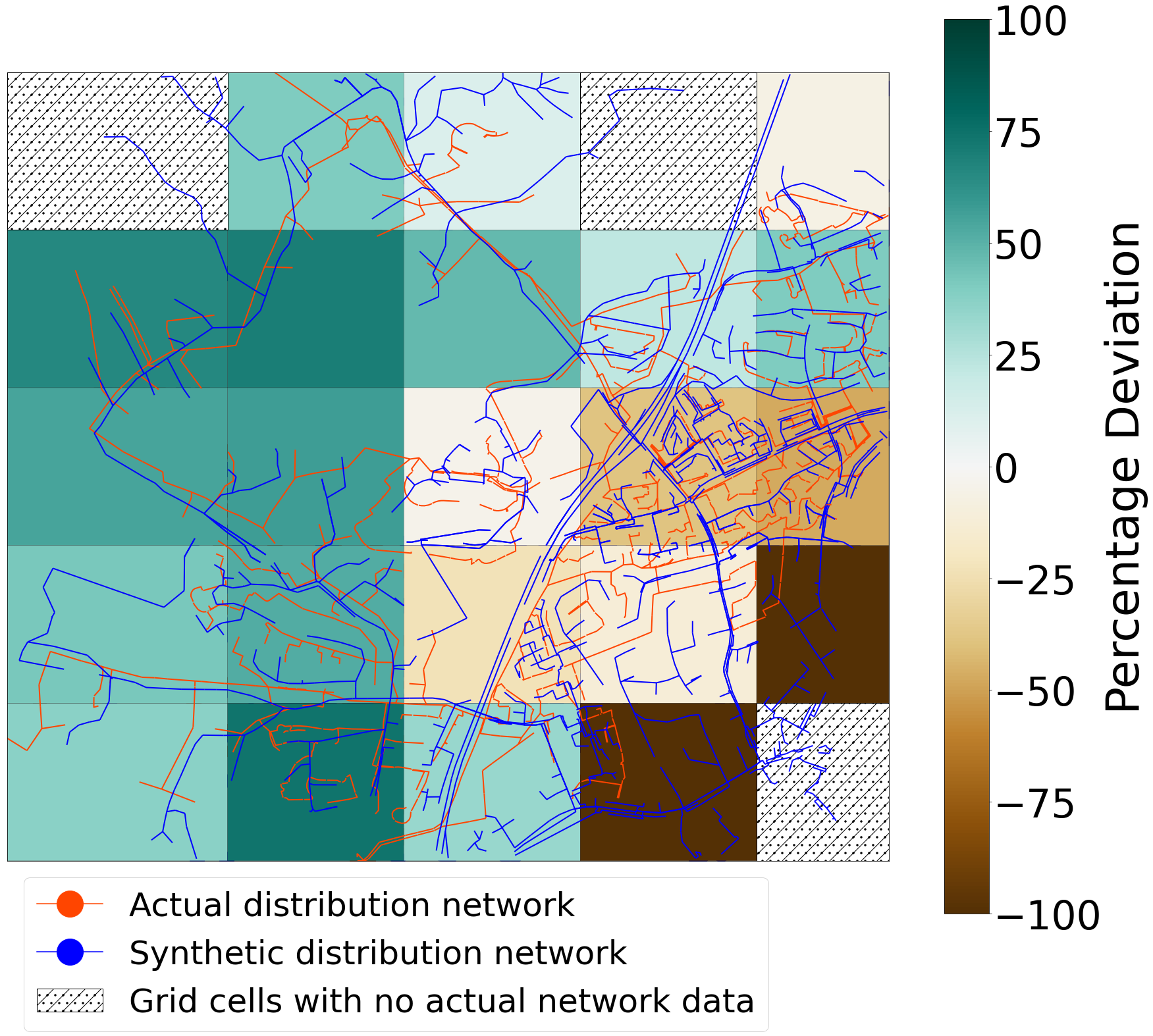}
    \includegraphics[width=0.30\textwidth]{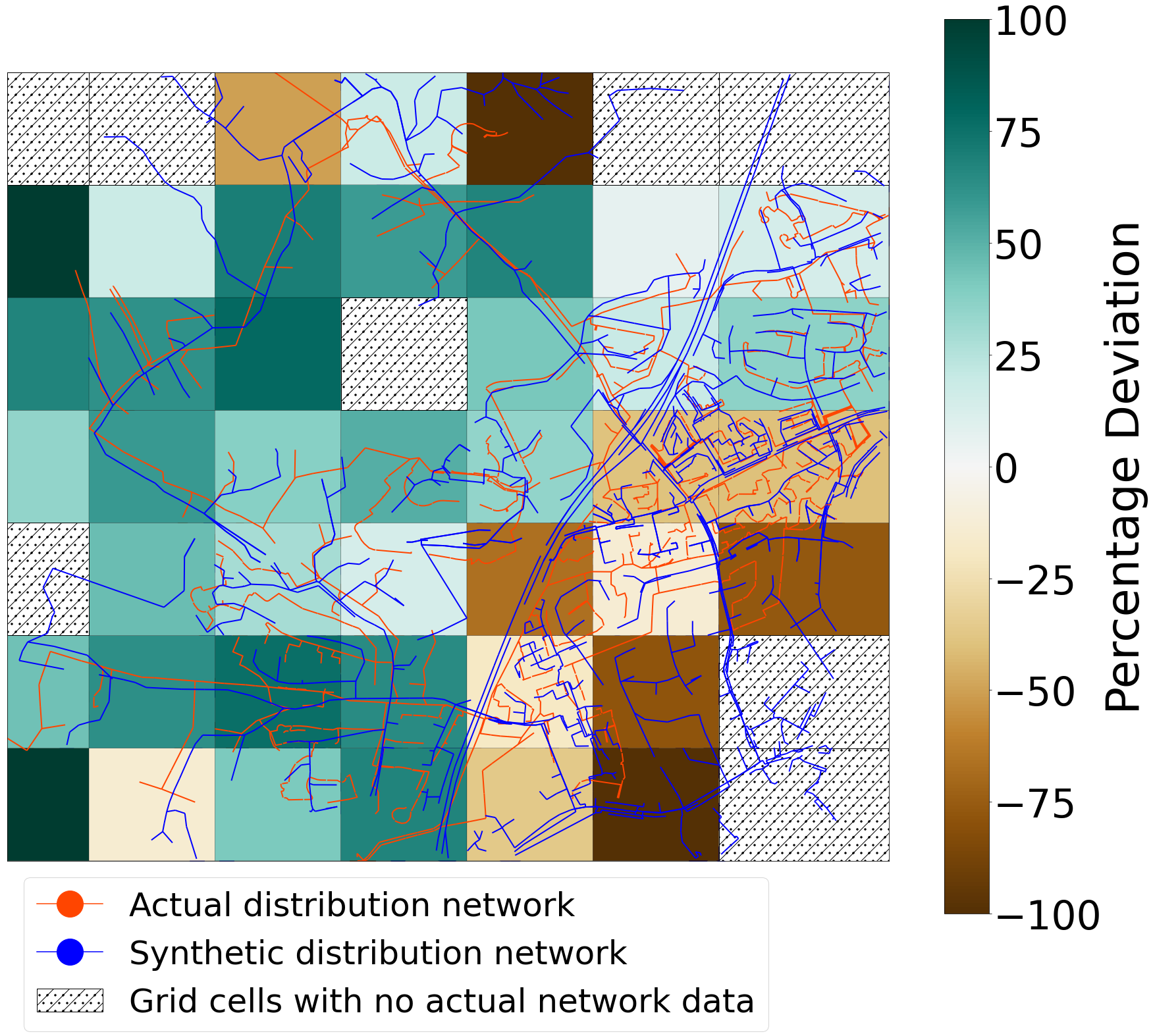}
    \includegraphics[width=0.30\textwidth]{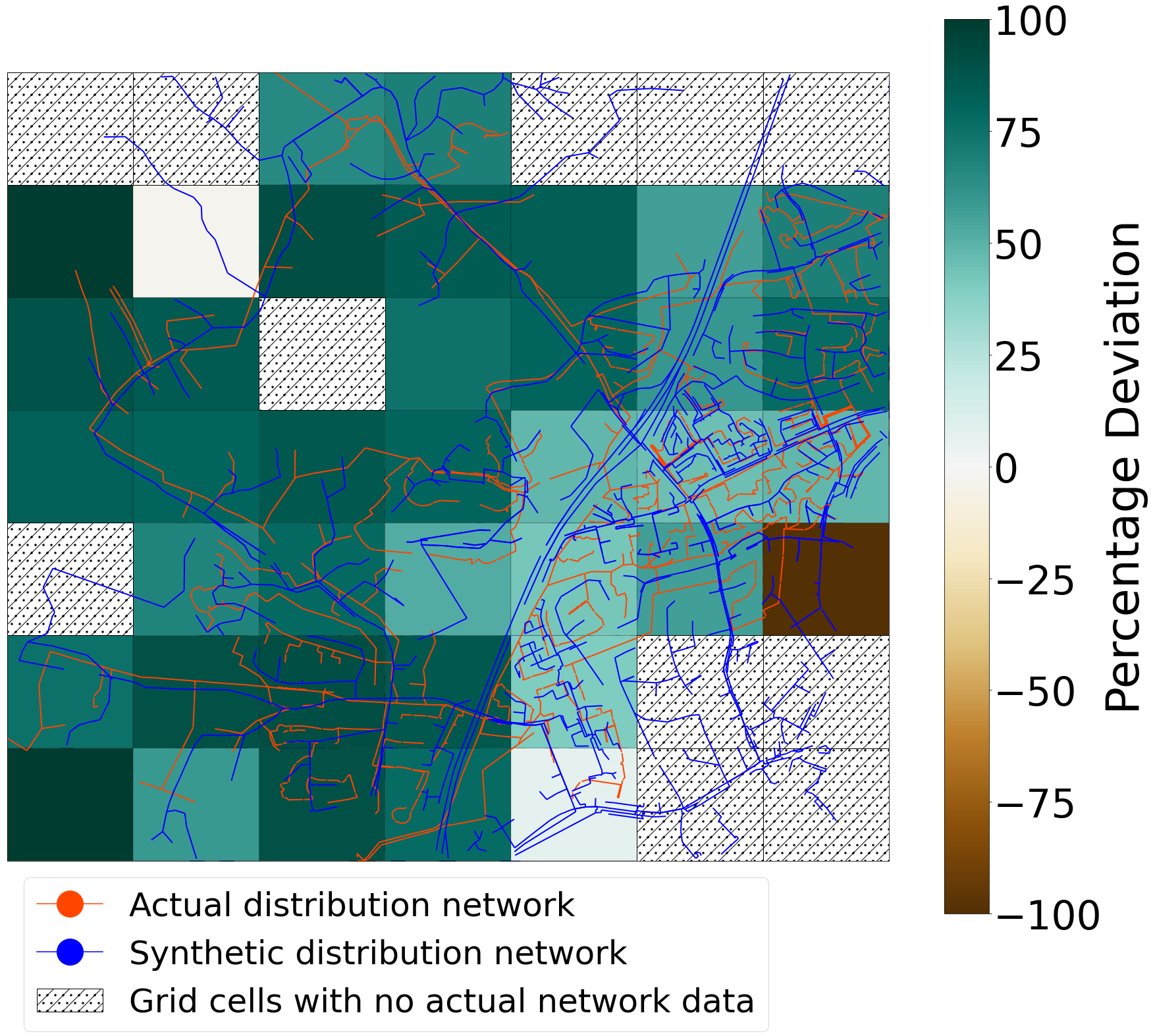}
    \includegraphics[width=0.30\textwidth]{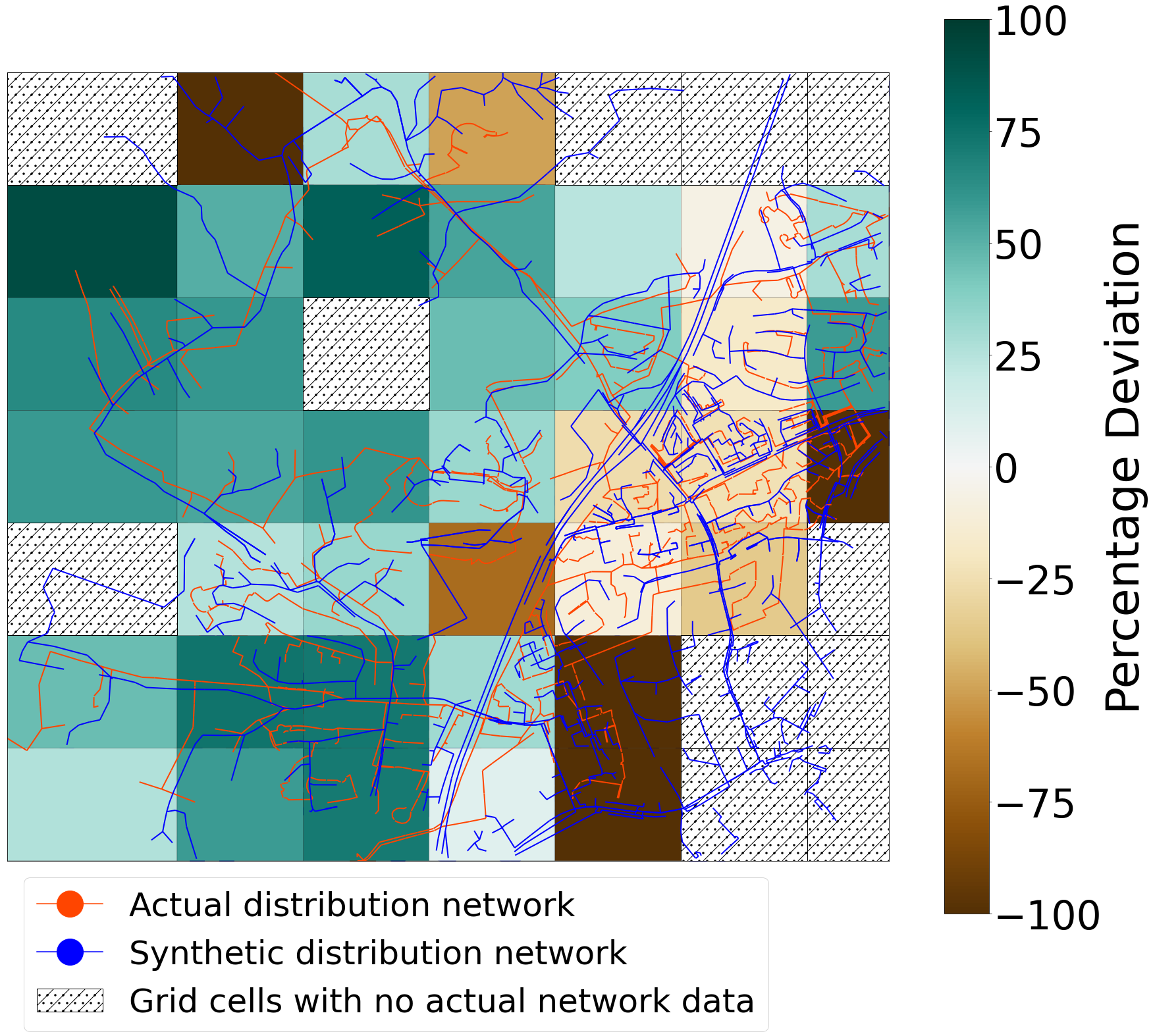}
    \caption{Plots comparing spatial distribution of nodes in the actual and synthetic networks. The comparison is made for rectangular grid cells where data regarding actual network is available. Two different grid resolutions are considered for the spatial distribution comparison. Also we compare the node distribution by moving the horizontal grid demarcations and observe varied results.}
    \label{fig:node-stat-spatial}
\end{figure}
\FloatBarrier
\noindent\textbf{Problems related to structural validation.}~The actual distribution network has been obtained from a power distribution company as images. The associated geographic topology is obtained by overlaying these images on maps from OpenStreetMap. The distribution network in the considered region is owned by multiple power companies and we have obtained the network for only one of them. For this reason, we focus our comparison on only those portions where we have data pertaining to both networks. Further, the above-mentioned overlaying process has been performed manually and thereby has introduced some errors in assigning the geographical attributes to the actual network. To this end, the effective way to compare the structural attributes of the networks is to divide the entire geographic region into multiple rectangular grid cells and perform a comparison in each cell separately.

\noindent\textbf{Spatial distributions of nodes.}~We compare the spatial distribution of nodes in the actual and synthetic networks as follows. Let the actual and synthetic networks for a region be denoted by $\mathscr{G}_{\textrm{act}}(\mathscr{V}_{\textrm{act}},\mathscr{E}_{\textrm{act}})$ and $\mathscr{G}_{\textrm{syn}}(\mathscr{V}_{\textrm{syn}},\mathscr{E}_{\textrm{syn}})$, 
and let $\mathscr{V}^{\textrm{CELL}}_{\textrm{act}}\subseteq\mathscr{V}_{\textrm{act}}$ and $\mathscr{V}^{\textrm{CELL}}_{\textrm{syn}}\subseteq\mathscr{V}_{\textrm{syn}}$ denote the set of nodes in the actual and synthetic networks lying within a rectangular grid cell. We use the following metric to compare the percentage deviation in the node distribution for grid cells where $\lvert\mathscr{V}_{\textrm{act}^{\textrm{CELL}}}\rvert\neq0$: 
\begin{equation*}
    \mathsf{D}_{\textrm{N}}^{\textrm{CELL}}=\dfrac{\dfrac{\lvert\mathscr{V}^{\textrm{CELL}}_{\textrm{act}}\rvert}{\lvert\mathscr{V}_{\textrm{act}}\rvert}-\dfrac{\lvert\mathscr{V}^{\textrm{CELL}}_{\textrm{syn}}\rvert}{\lvert\mathscr{V}_{\textrm{syn}}\rvert}}{\dfrac{\lvert\mathscr{V}^{\textrm{CELL}}_{\textrm{act}}\rvert}{\lvert\mathscr{V}_{\textrm{act}}\rvert}}\times 100
\end{equation*}
Fig.~\ref{fig:node-stat-spatial} shows the spatial comparison of node distribution for uniform rectangular cell partitions of two different resolutions followed by shifting the demarcations horizontally.. The color code denotes the intensity of the percentage deviation in the distribution. Note that some grid cells are shaded with black dots to denote the unavailability of network data. It is easy to note that the spatial distribution of nodes is a function of the size of the rectangular grid cells as well as the location of the demarcations between cells. To address this issue, one might consider comparing networks through multiple grid sizes (varying the resolution) or by considering a single-sized rectangular grid cells followed by shifting the horizontal and vertical demarcations of the grid to account for different sized cells.

\noindent\textbf{Geometry comparison}
We consider a metric space $(\mathscr{M},d)$ where $d(x,y)$ is the measure between two elements $x,y\in\mathscr{M}$. Given two non-empty subsets $\mathscr{X},\mathscr{Y}\subseteq\mathscr{M}$ of the metric space, the Hausdorff distance $d_{\textrm{H}}(\mathscr{X},\mathscr{Y})$ between the sets is defined as
\begin{equation}
    d_{\textrm{H}}(\mathscr{X},\mathscr{Y}) = \sup_{x\in\mathsf{X}}~\inf_{y\in\mathsf{Y}}d(x,y)
\end{equation}
\begin{figure}[tbhp]
    \centering
    \includegraphics[width=0.32\textwidth]{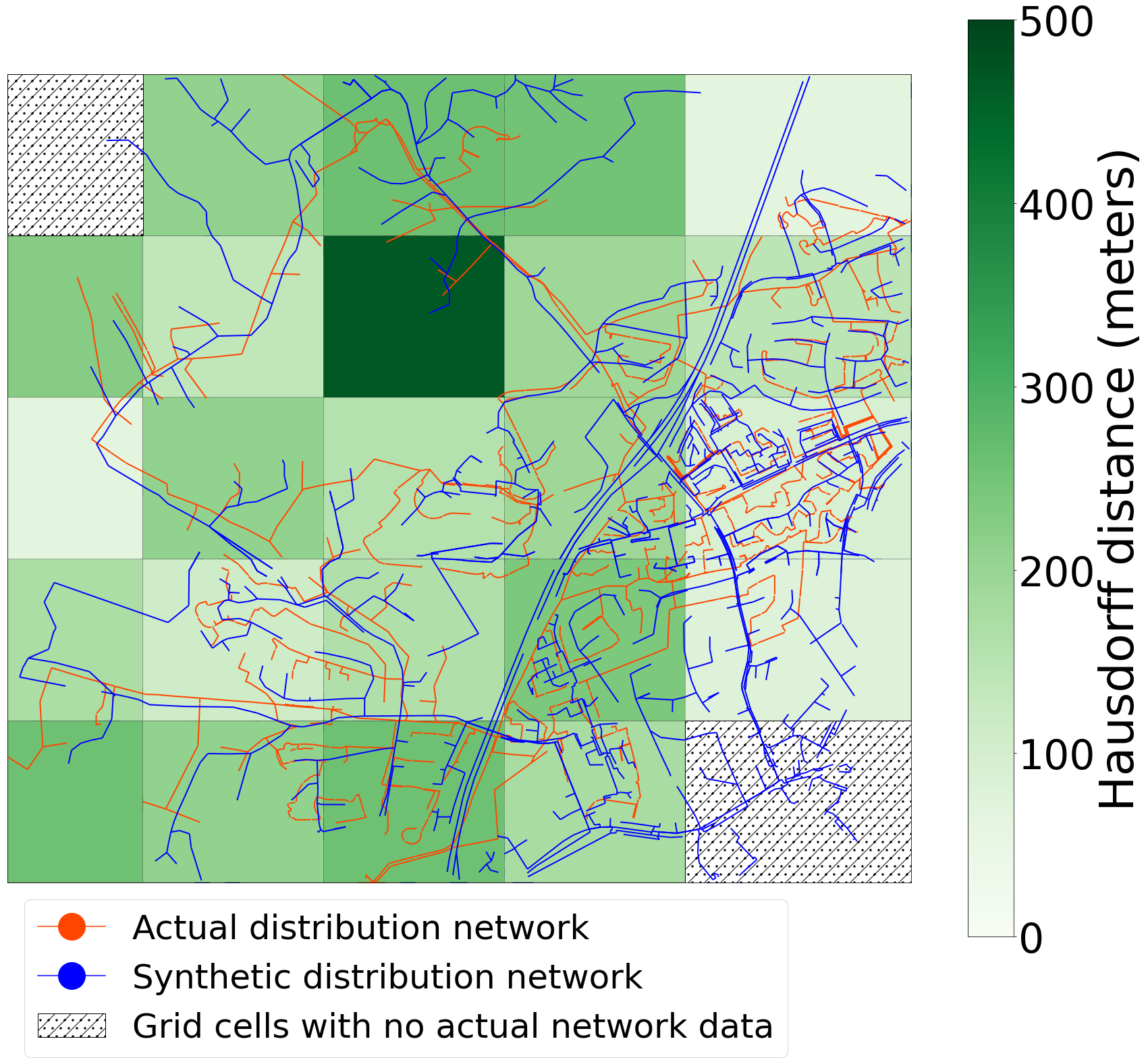}
    \includegraphics[width=0.32\textwidth]{figs/hauss-comparison-5-5.png}
    \includegraphics[width=0.32\textwidth]{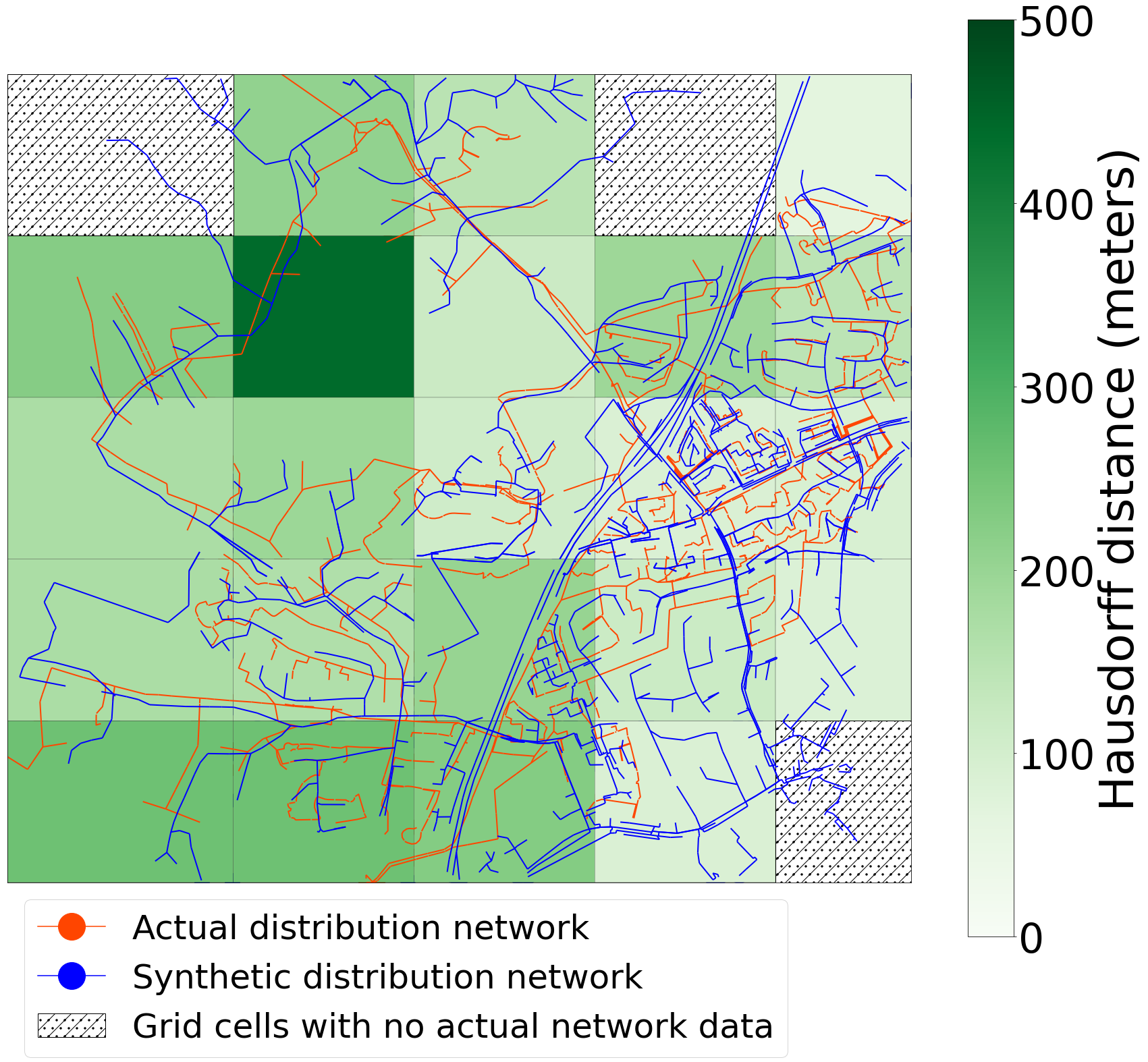}
    \includegraphics[width=0.32\textwidth]{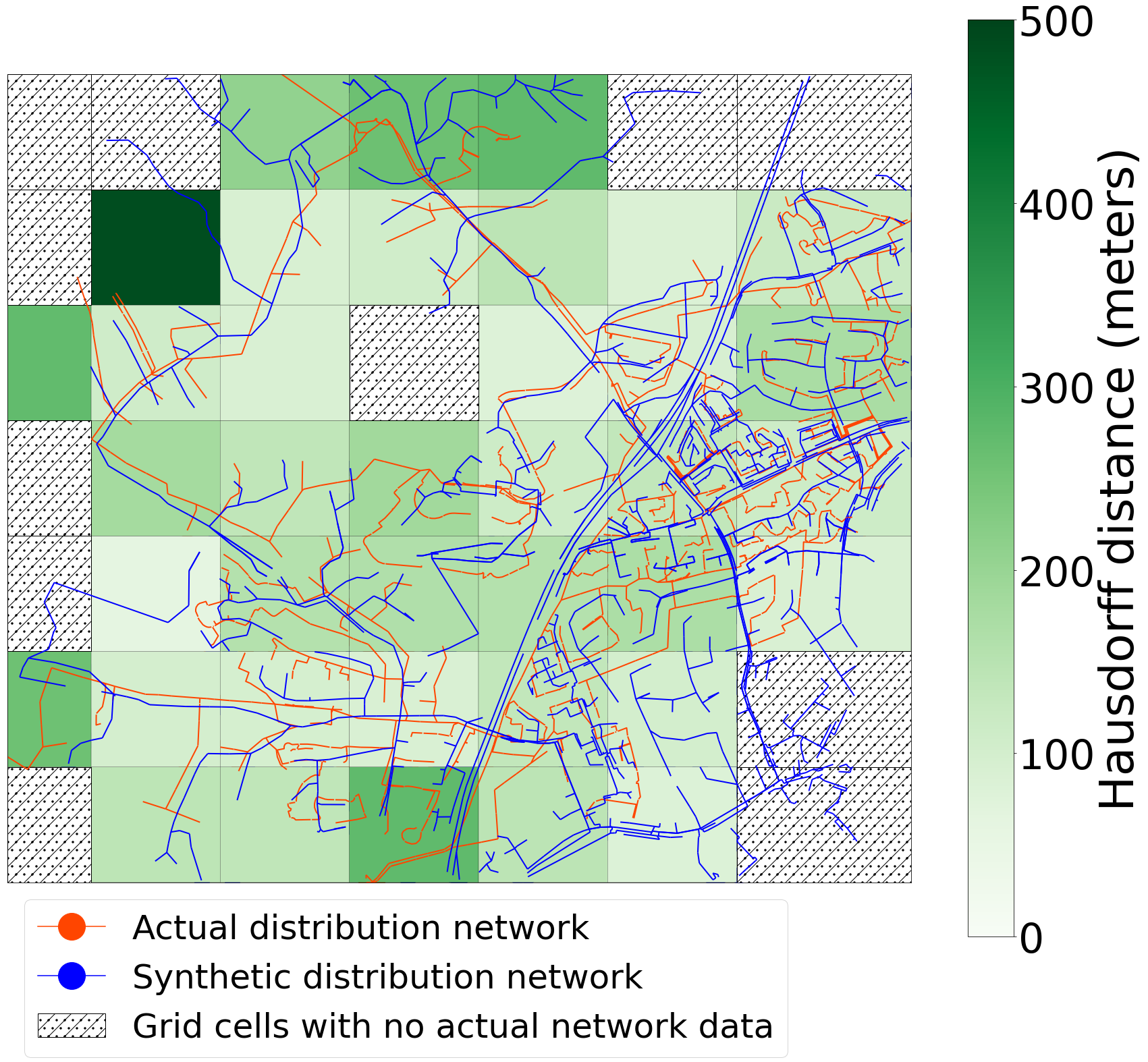}
    \includegraphics[width=0.32\textwidth]{figs/hauss-comparison-7-7.png}
    \includegraphics[width=0.32\textwidth]{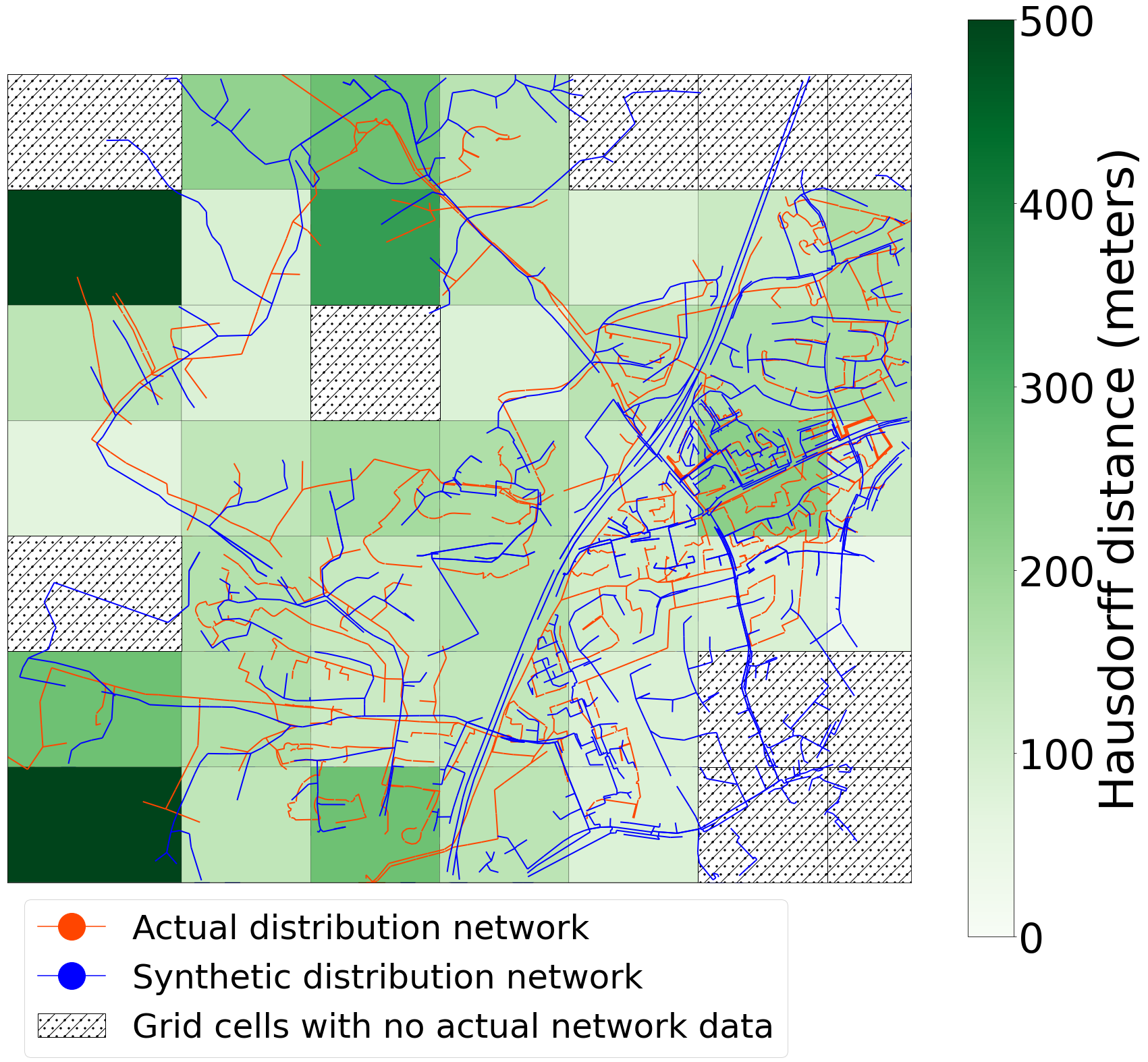}
    \caption{Plots comparing Hausdorff distance between edge geometries of actual and synthetic networks. The comparison is made for rectangular grid cells where data regarding the actual network is available. Two different grid resolutions are considered for the spatial distribution comparison. Also we compare the Hausdorff distance by moving the horizontal grid demarcations and observe varied results.}
    \label{fig:hausdorff-extra}
\end{figure}
\FloatBarrier

In the context of networks associated with geographic attribute, we define the metric space $\mathscr{M}$ to be the set of all points along edges in the network and the metric $d(x,y):=\mathsf{dist}(x,y)$ which is the geodesic distance between a pair of points $x,y\in\mathscr{M}$. We interpolate points along each edge of either network in order to evaluate an exact comparison of the network geometries. Let the set of interpolated points along the actual and synthetic networks be represented by $\mathscr{P}_{\textrm{act}}$ and $\mathscr{P}_{\textrm{syn}}$ respectively. Note that $\mathscr{P}_{\textrm{act}},\mathscr{P}_{\textrm{syn}}\subseteq\mathscr{M}$. We define the Hausdorff distance between networks $\mathscr{G}_{\textrm{act}}$ and $\mathscr{G}_{\textrm{syn}}$ as
\begin{equation}
    \mathsf{D}_{\textrm{H}}\left(\mathscr{G}_{\textrm{act}},\mathscr{G}_{\textrm{syn}}\right):=\max_{x\in\mathscr{P}_{\textrm{act}}}\min_{y\in\mathscr{P}_{\textrm{syn}}}\mathsf{dist}(x,y)
\end{equation}
Fig.~\ref{fig:hausdorff-extra} shows a Hausdorff distance-based network geometry comparison between actual and synthetic networks for uniform rectangular grid partitions of two different resolutions. This is accompanied by shifting the grid cell demarcations horizontally. We observe that as the cell demarcations are shifted horizontally, the Hausdorff distance between the edge geometries in each grid cell alters. Therefore, the grid resolution and grid demarcations play an important role in the geometry comparison as well.

\section*{Impact of photovoltaic penetration}
Installing photovoltaic (PV) generation on residence rooftops is an easy and viable option to improve the resilience of distribution systems. However, the continuous increase in PV penetration has led to serious operational issues such as overvoltage and unbalances in the distribution grid~\cite{pvhost_review}. Therefore, it is necessary to identify the maximum allowable PV penetration without violating operational and performance constraints.

It has been observed that a penetration as low as $2.5\%$ is capable of causing voltage violation when a large PV generation is installed at a single point in the medium voltage (MV) network. On the contrary, the low voltage (LV) network can withstand multiple PV generators with penetration levels up to $110\%$~\cite{pvhost_lvmv}. In this work, we focus on the impact of PV penetration in the LV networks through the installation of PV generation on residence rooftops. We aim to address the following question: does PV penetration provide any advantage to grid operation, and how much PV penetration is acceptable without violating operational constraints.

While planning a practical network, the LV feeders are designed with a suitable rating to supply a certain load to certain groups of residences without considering the impact of distributed generation (DGs). Most grid connected DGs needs to confer with the CAN/CSA-C22.2 No.257-06~\cite{csa2006} standard, which specifies the electrical requirements for the interconnection of inverter-based micro-distributed resource systems with grid-connected low-voltage systems. For single phase connection, it is considered normal operating conditions (NR-Normal Range) when the voltage level is between 0.917 and 1.042 pu. On extreme operation conditions (ER-Emergency Range), the under and over steady state voltage limits are 0.88 pu and 1.058 pu, respectively. It is worth mentioning that although networks are allowed to operate under extreme conditions, the utility would need to take a corrective action.

\begin{figure*}[tbhp]
    \centering
    \includegraphics[width=0.45\textwidth]{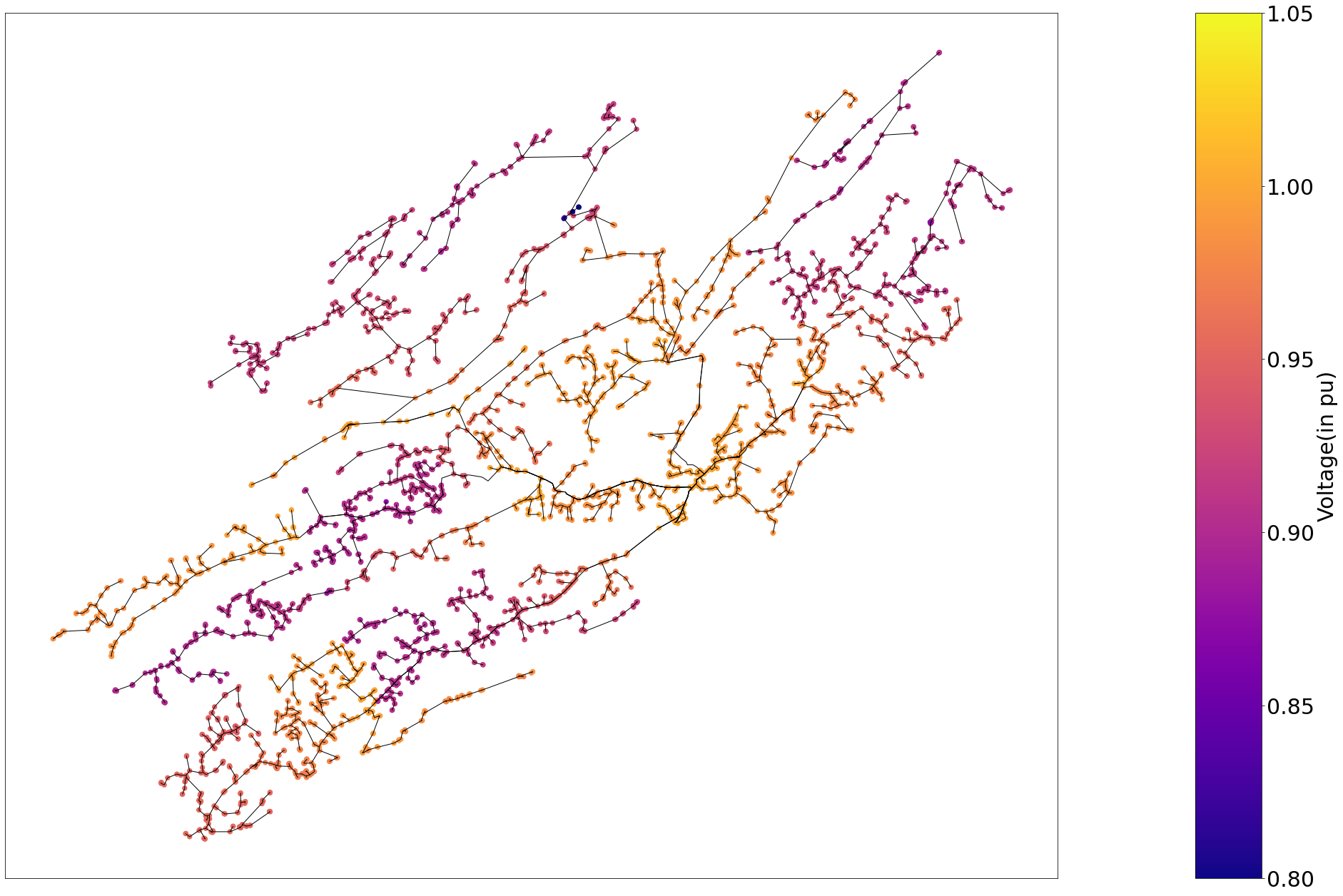}
    \includegraphics[width=0.45\textwidth]{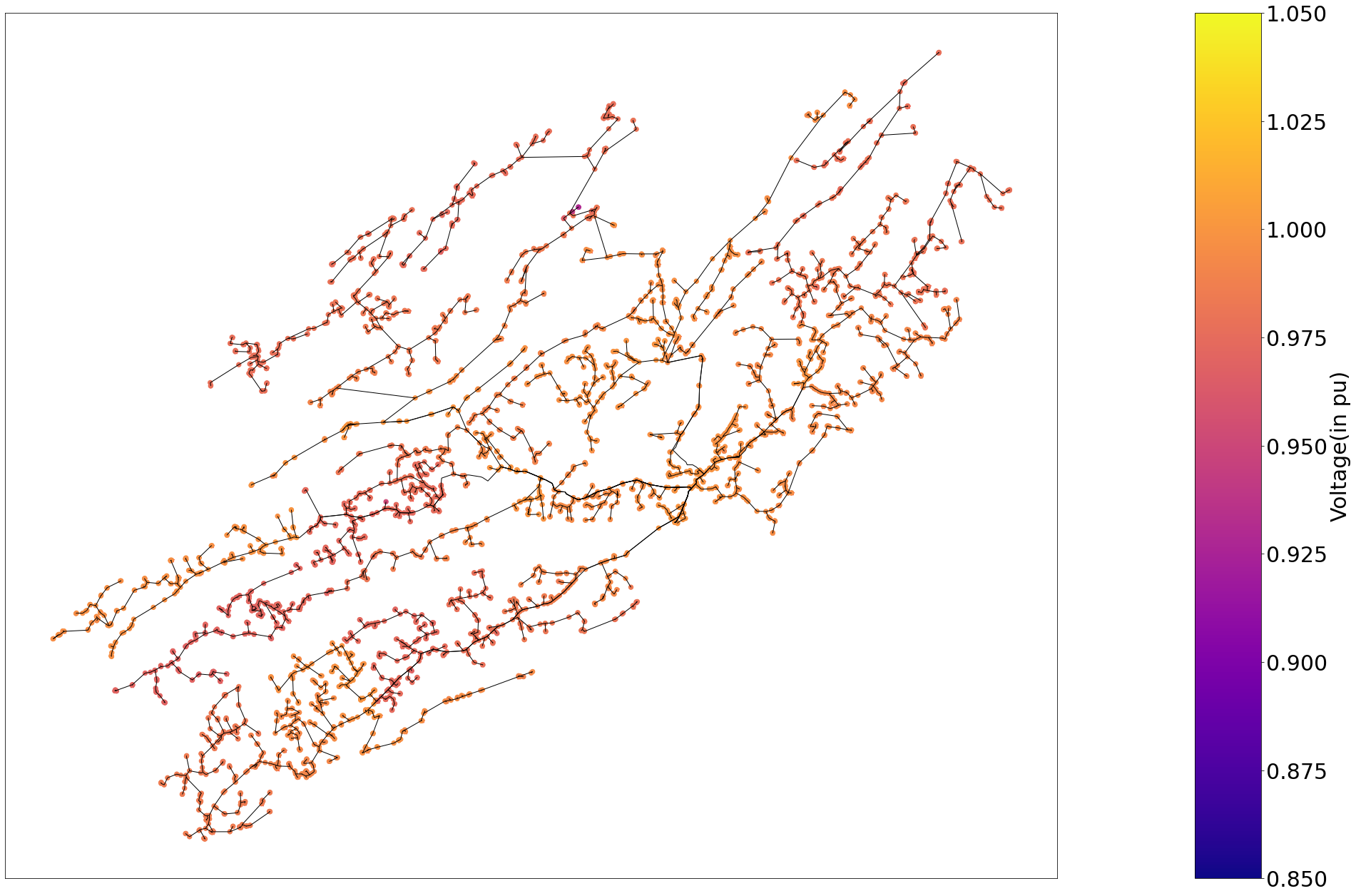}
    \caption{Comparison of voltage profile without and with PV penetration in the distribution network. With the introduction of rooftop PVs, the voltage profile is alleviated to 1 pu, which is recommended by engineering standards.}
    \label{fig:host_vs_nohost}
\end{figure*}
\FloatBarrier
If a distribution network is operated without any PV, the substation feeder is solely responsible for delivering power to all residences in the network. Therefore, the farthest connected residences suffer the maximum voltage drop and experience a voltage sag. This is denoted in the left figure of Fig.~\ref{fig:host_vs_nohost} where we observe that a significant percentage of nodes experience undervoltage (less than 0.9 pu). The introduction of rooftop PVs installed on randomly chosen $50\%$ of residences alleviates the voltage profile of the entire grid as shown in the right figure of Fig.~\ref{fig:host_vs_nohost}. In this example, we have considered a $30\%$ PV penetration , that is the total PV generation in the distribution network amounts to $30\%$ of the total load. Therefore, the substation feeders are responsible for delivering the remaining $70\%$ of the power.

\begin{figure*}[tbhp]
    \centering
    \includegraphics[width=\textwidth]{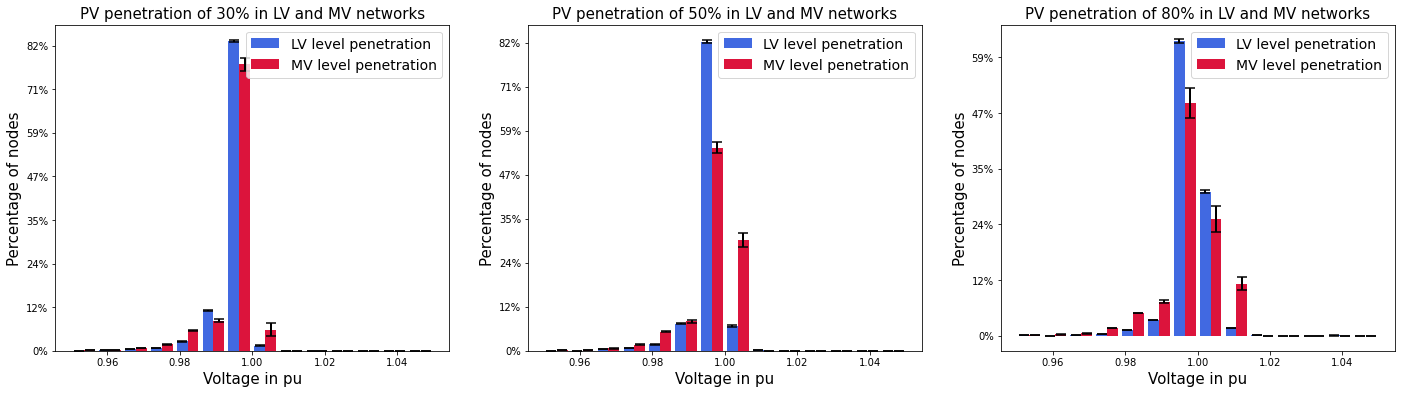}
    \includegraphics[width=\textwidth]{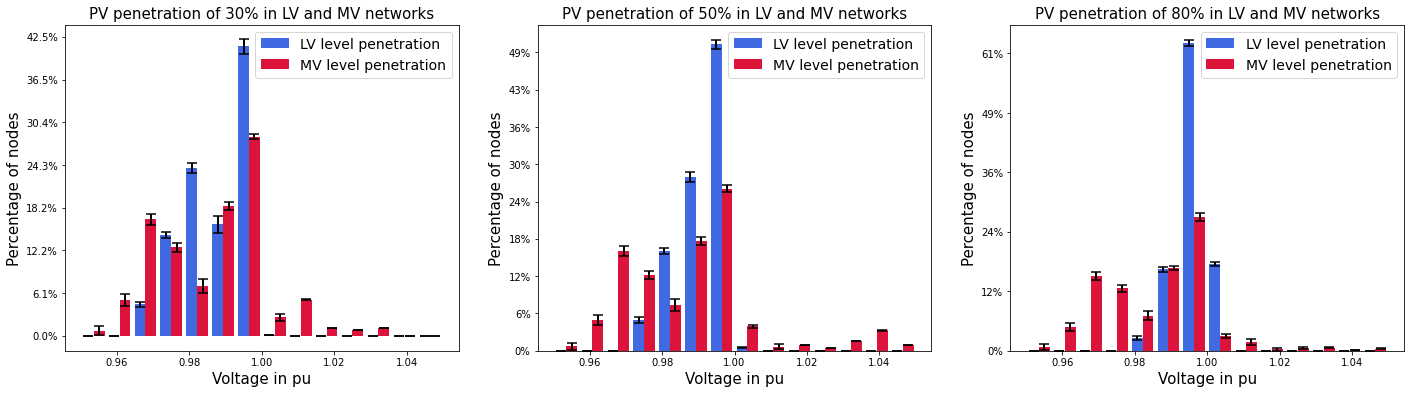}
    \caption{Comparing the impact of PV integration in an urban feeder (top figures) and rural feeder (bottom figures) for three different penetration levels. Colors depict PV addition at multiple locations in LV network (blue) and single node PV addition in MV network (red). The error bar shows the variation in impact when analyzed over an ensemble of $20$ networks. LV level penetration is less likely to cause overvoltage as compared to similar penetration in the MV network. Rural feeders are more prone to overvoltage issues.}
    \label{fig:urban-rural}
\end{figure*}

We perform the comparison on two different synthetic networks. Fig.~\ref{fig:urban-rural} shows the histogram of node voltages for comparing the impact of PV penetration in an urban feeder (top three figures) and a rural feeder (bottom three figures). The urban feeder network consists of shorter length lines. We observe that LV level PV generation is less likely to cause overvoltage issues as compared to a single node MV level PV integration. In the case of a rural feeder with longer lines, we observe a similar trend. However, the percentage of nodes experiencing severe overvoltage (around 1.05 pu which is the extreme limit of an acceptable overvoltage) is higher for rural networks as compared to urban feeder networks for all penetration levels when we consider a single node PV integration at the MV level. Therefore an optimal placement of PV generators is required for the rural feeders so that they do not suffer from overvoltage issues.
\FloatBarrier

\section*{Short Circuit Analysis}
\begin{figure*}[tbhp]
    \centering
    \includegraphics[width=0.45\textwidth]{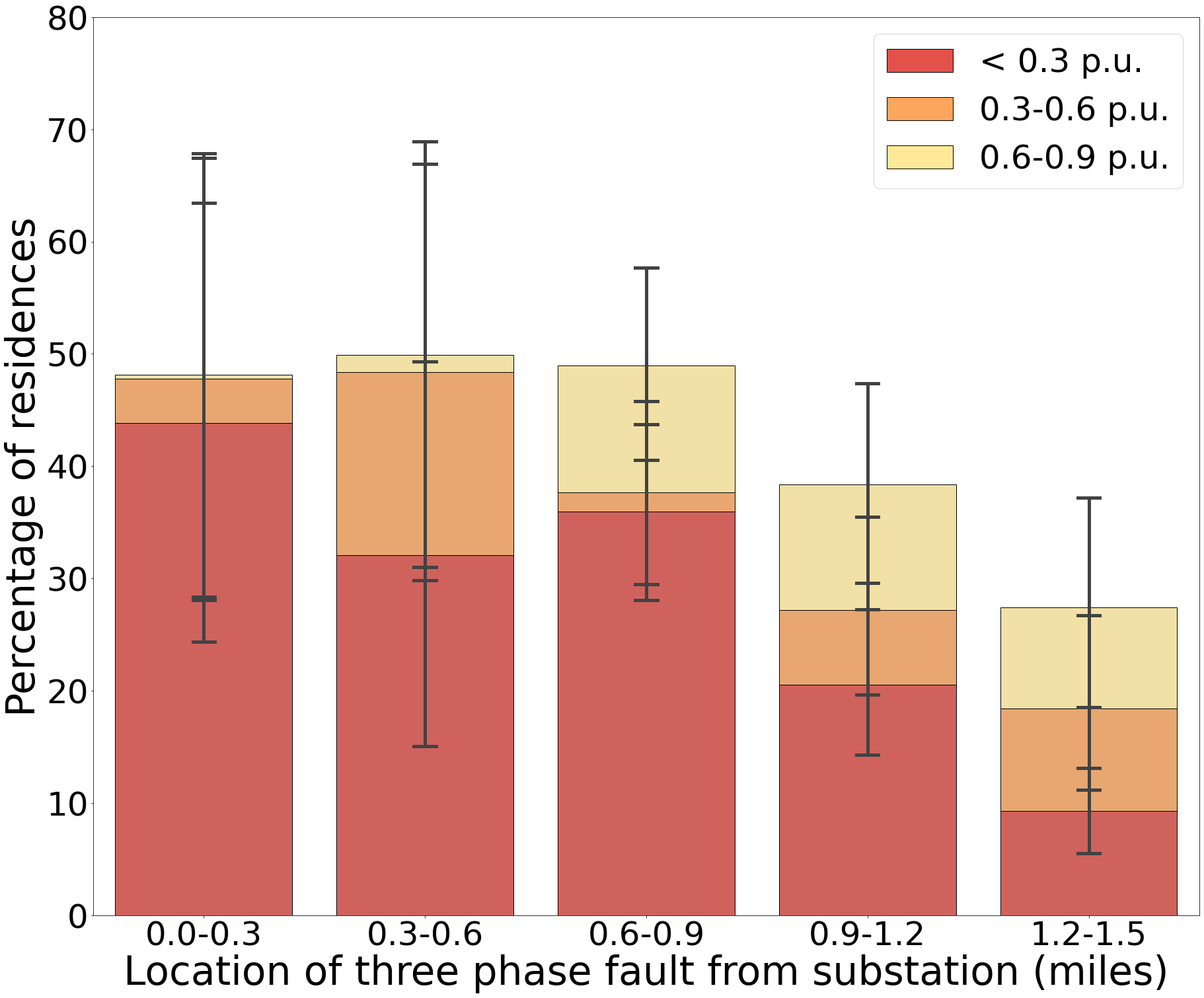}
    \includegraphics[width=0.45\textwidth]{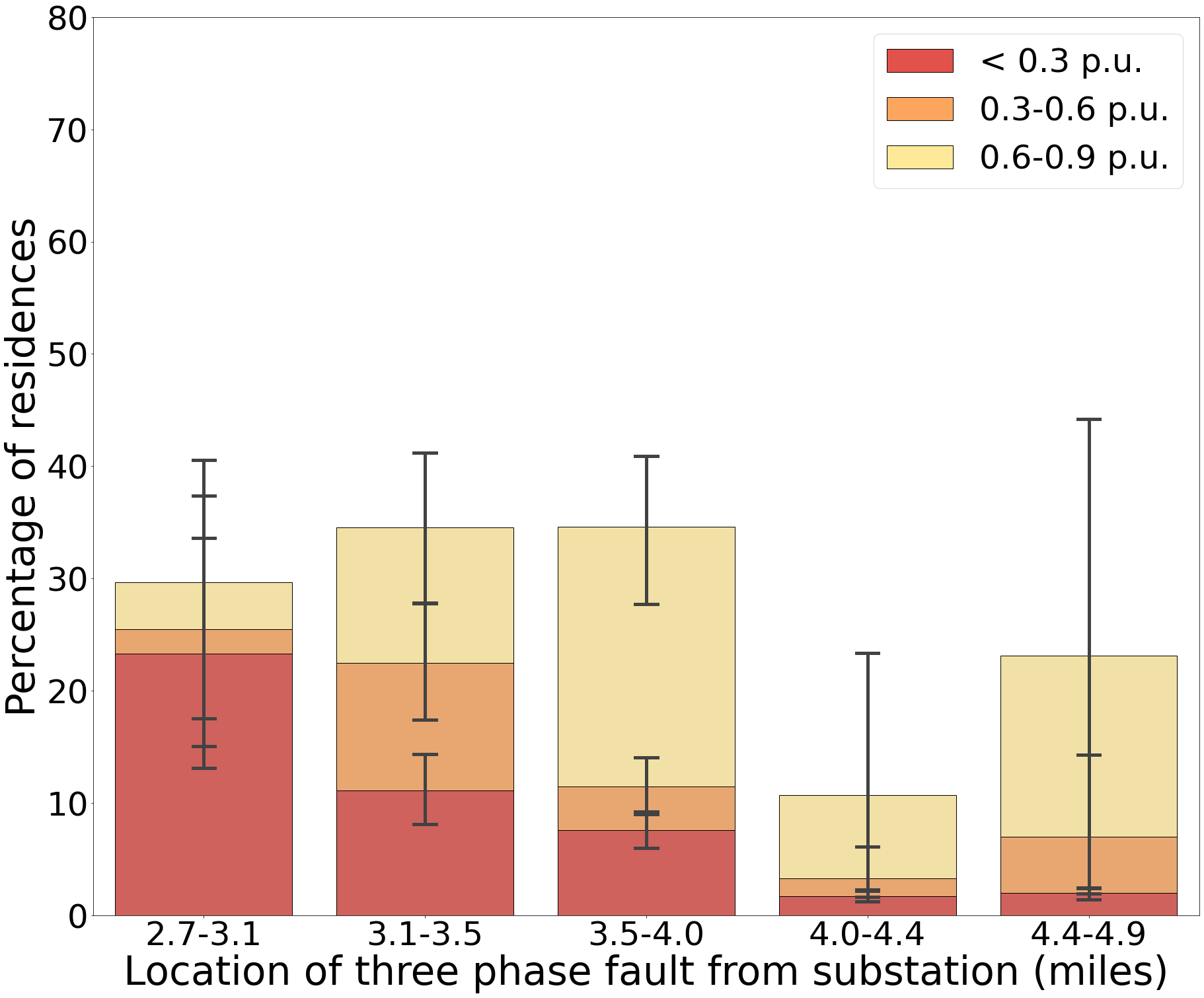}
    \caption{Plots showing short circuit analysis on two different feeders in one of the created synthetic networks. Three phase symmetrical faults are simulated at multiple locations along each feeder and post-fault voltages at residences are computed.}
    \label{fig:sca-resvolt}
\end{figure*}

We can use the created networks to perform short circuit analysis. Since we have created a positive sequence network for the distribution system, we can carry out a short circuit analysis for three phase symmetrical faults. To this end, we construct the bus impedance matrix $\mathbf{Z}_{\textrm{bus}}$ using the traditional methodology~\cite{stevenson}.Thereafter, we compute the post-fault voltage at bus $j$ for a three phase symmetrical fault at bus $k$ using (\ref{eq:fault}).
\begin{equation}
    V^{\textrm{post-fault}}_j = V^{\textrm{pre-fault}}_j - \frac{Z_{jk}}{Z_{kk}}V^{\textrm{pre-fault}}_k
    \label{eq:fault}
\end{equation}
Note that the pre-fault voltages are computed by solving the power flow problem using the LDF model~\cite{ldf} with each residence consuming average hourly demand. Fig.~\ref{fig:sca-resvolt} shows the percentage of residences experiencing undervoltage instantaneously after the occurrence of a symmetrical three phase fault at different locations in the network. We perform the short circuit analysis for two different feeders in a single synthetic network. We simulate symmetrical three phase fault at various locations along each feeder and observe the post-fault voltages at each residence nodes.

Though we have created positive sequence networks, we are able to perform short circuit analysis for unsymmetrical faults (line-ground and line-line). This can be performed under the assumption that all lines in the distribution network are overhead lines. Under such an assumption, the zero sequence impedance is three times the positive sequence impedance of the lines. Thereafter, the fault currents can be computed accordingly~\cite{Phadke-Horowitz,wadhwa,stevenson}.
\FloatBarrier







\bibliographystyle{unsrt}
\bibliography{references}